\documentclass[12pt]{article}

\usepackage[noadjust]{scicite}
\usepackage{times}
\usepackage{graphicx}
\usepackage{subcaption}
\usepackage{amssymb}
\usepackage{amsmath}
\usepackage{mathtools}
\usepackage{nicefrac}
\usepackage[colorlinks=true, linkcolor=purple]{hyperref}
\usepackage{cleveref}
\usepackage{soul}

\usepackage[final]{pdfpages}

\usepackage{algorithm}
\usepackage[noend]{algpseudocode}
\makeatletter
\renewcommand{\ALG@beginalgorithmic}{\small}
\makeatother
\algrenewcommand\alglinenumber[1]{\small #1:}
\algnewcommand{\IFor}[1]{\State\algorithmicfor\ #1\ \algorithmicdo}
\algnewcommand{\EndIFor}{\unskip\ \algorithmicend\ \algorithmicfor}
\usepackage{newfloat}
\DeclareFloatingEnvironment[
    fileext=loa,
    listname={List of Algorithms},
    name=Algorithm,
    placement=tbhp,
    within=section,
]{figalg}

\crefname{figalg}{Algorithm}{Algorithms}

\usepackage{xcolor}

\usepackage{xr}
\usepackage{mathrsfs}	

\newcommand{\eg}{\emph{e.g.,~}}
\newcommand{\ie}{\emph{i.e.,~}}

\newenvironment{sabstract}{%
\begin{quote} \bf}
{\end{quote}}

\title{Shining light on data: Geometric data analysis through quantum dynamics}

\author
{Akshat Kumar,$^{1\ast}$ Mohan Sarovar$^{2\ast}$\\
\\
\normalsize{$^{1}$Department of Mathematics, Clarkson University, Potsdam, NY 13699 USA}\\
\normalsize{Instituto de Telecomunica\c{c}\~{o}es, Lisbon, Portugal}\\
\normalsize{$^{2}$Quantum Algorithms and Applications Collaboratory}\\
\normalsize{Sandia National Laboratories, Livermore, California 94550 USA}\\
\\
\normalsize{$^\ast$E-mails:  akumar@clarkson.edu, mnsarov@sandia.gov}
}

\date{}

\begin{document}

\baselineskip24pt

\maketitle

\begin{sabstract}
Experimental sciences have come to depend heavily on our ability to organize and interpret high-dimensional datasets. Natural laws, conservation principles, and inter-dependencies among observed variables yield geometric structure, with fewer degrees of freedom, on the dataset.
We introduce the frameworks of semiclassical and microlocal analysis to data analysis and develop a novel, yet natural uncertainty principle for extracting fine-scale features of this geometric structure in data, crucially dependent on data-driven approximations to quantum mechanical processes underlying geometric optics.
This leads to the first tractable algorithm for approximation of wave dynamics and geodesics on data manifolds with rigorous probabilistic convergence rates under the manifold hypothesis.
We demonstrate our algorithm on real-world datasets, including an analysis of population mobility information during the COVID-19 pandemic to achieve four-fold improvement in dimensionality reduction over existing state-of-the-art and reveal anomalous behavior exhibited by less than 1.2\% of the entire dataset.
Our work initiates the study of data-driven quantum dynamics for analyzing datasets, and we outline several future directions for research.
\end{sabstract}

Nature is complex, yet organized -- this basic facet has become a cornerstone for how we analyze and interpret vast amounts of data ranging from the biological, physical, geological, meteorological, all the way to the astronomical.
In fact, we owe to this, nearly all of the significant advances in data analysis disciplines in the past two decades, including signals processing \cite{Damelin_Miller_2012, Wright_Ma_2022} and machine learning \cite{Bishop_2006,Candes_2019,Glielmo_2021}, where tractable models capture, interpret and reproduce complex natural data.
We can turn this observation on its head and ask: are there certain natural processes that are fundamental to understanding the structure in complex, yet organized data?

\begin{figure}[t]
\centering
\includegraphics[width=\textwidth]{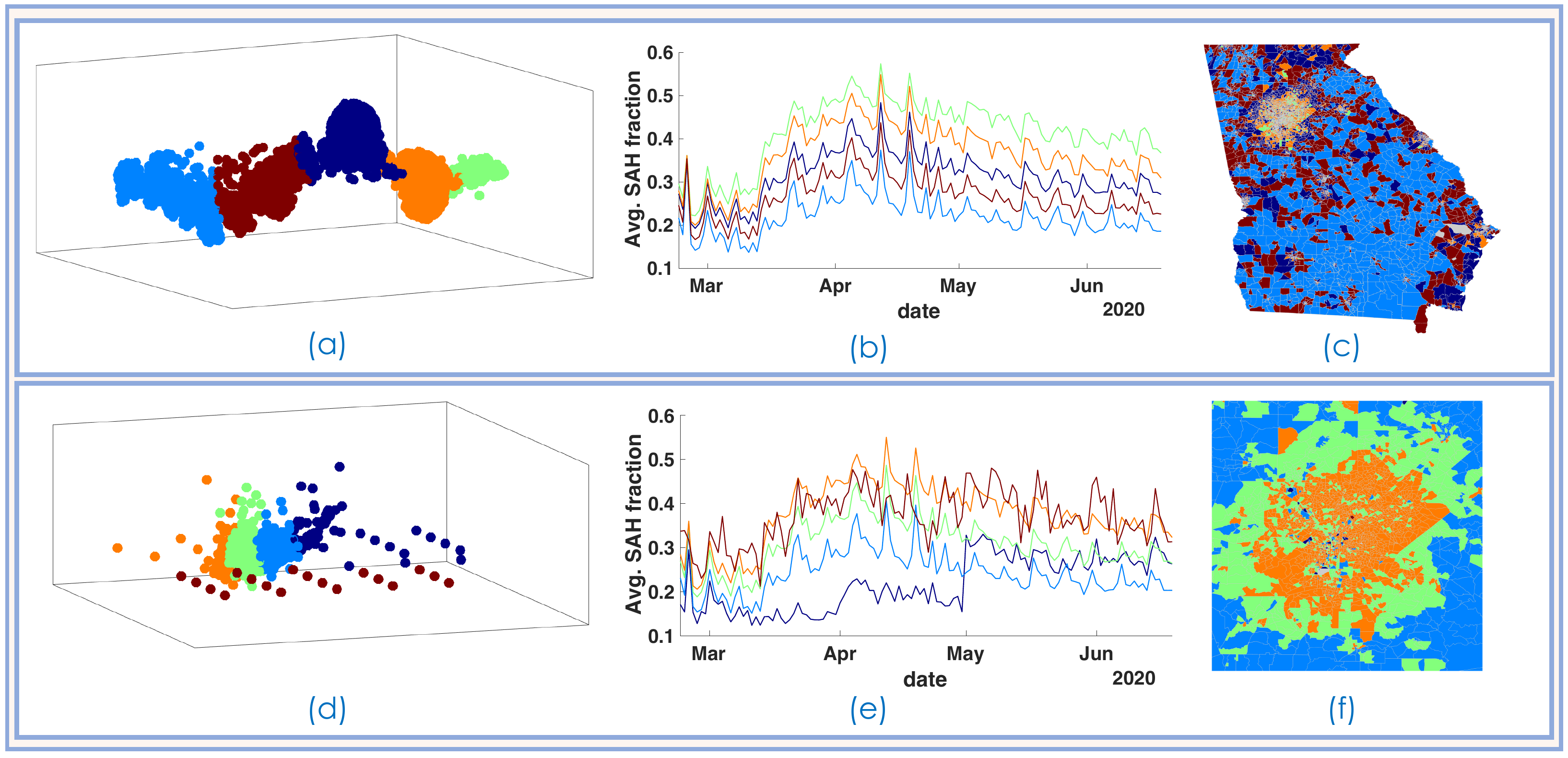}
\caption[LoF entry]{\scriptsize
Analysis of adherence to social distancing measures during the COVID-19 pandemic using the dataset \cite{safegraph}, which is a collection of geolocation information from mobile devices in the United States, aggregated at the census block group (CBG) level and recorded daily for a period of over a year. As previously studied in Ref. \cite{Levin_Chao_Wenger_Proctor_2020}, we compute the stay-at-home (SAH) fraction as a simple metric of adherence to social distancing from this data (see SM, \cref{SM-sec:app} for details). To compare with Ref. \cite{Levin_Chao_Wenger_Proctor_2020} we consider data for the state of Georgia (GA) and limit it to a 117-day time period, which provides a snapshot of mobility patterns during the first four months of the pandemic. We consider 5509 CBGs within the state, thus our dataset $X_N$ has $N = 5509$ samples, each of dimension 117.
Our workflow on this dataset creates associations between samples based on data-driven geometric optics and produces a sparse graph $\mathcal{X}_N$ that is amenable to $3$D embedding (see \Cref{fig:embeddings}).
We do $k$-means clustering on this embedding with $k=5$ set by Ref. \cite{Levin_Chao_Wenger_Proctor_2020}, for comparison.
One application involves using the \emph{expected position} $\bar{x}_t$ to associate a point distance $t$ away from a given sample point $x_j \in X_N$ and another involves using the \emph{maximum position} $\hat{x}_t$; the details of construction, theory, interpretation and convergence of this framework are discussed below.
\textbf{Top (a,b,c)}: Embedding of $\mathcal{X}_N$ and clustering using $\bar{x}_t$.
\textbf{Bottom (d,e,f)}: Embedding and clustering using $\hat{x}_t$.
\textbf{(a)} and \textbf{(d)} show $3$D embeddings of $\mathcal{X}_N$ and clusters of the resulting points coded by color.
\textbf{(b)} and \textbf{(e)} show the average SAH fraction time series for each cluster.
In \textbf{(b)} we see a clear separation of clusters by their SAH behavior, and in \textbf{(e)} we have identified an anomalous SAH pattern (in dark blue) in one of the clusters.
\textbf{(c)} and (\textbf{f}) color code the CBGs in GA according to the cluster they belong to.
\textbf{(c)} shows a clear rural-urban divide in degree of social distancing behavior, and (\textbf{f}) is a magnification of the Atlanta metropolitan region, because many of the outlier CBGs identified are located in this area. See SM, \cref{SM-sec:app} for further details of analysis, including parameter values.
\label{fig:covid}}
\end{figure}

We propose that, much like how quantum mechanics models nature at fine scales, the fine-scaled resolution of the organization and structure of data is also best characterized using quantum mechanical processes.
A real-world application of our algorithms (see SM, \Cref{SM-sec:app} and \Cref{SM-app:code}) that realize this proposal, is shown in \Cref{fig:covid}, in which we analyze a publicly available dataset \cite{safegraph} that collected population mobility information (based on GPS data from mobile phones) across the US during the initial stages of the COVID-19 pandemic, when social distancing measures were being placed to thwart the spread of the virus.
We consider daily mobility information over a roughly $4$ month period ($117$ days: February 23, 2020 - June 19, 2020) and focus on the state of Georgia (GA). The mobility information is aggregated in $5509$ geographic regions called census block groups (CBGs)
that partition the state. We wish to organize the CBGs according to the patterns of mobility among their populations. To do so, we compute a daily stay-at-home (SAH) statistic for each CBG that captures the fraction of devices that stayed within their home CBG for the entire day.
This defines $5509$ data points, each of $117$ dimensions, whose components give the time-series of stay-at-home adherence.
Our quantum mechanical framework gives organizations of this dataset in just $3$ dimensions, (see \Cref{fig:covid}(a,d)).
Based on one application of the framework, we have the organization in \Cref{fig:covid}(a) that naturally separates into five distinct clusters of regions, whose average SAH fractions (in \Cref{fig:covid}(b)) show that the mobility patterns are statistically quite similar,
roughly differing from each other only by a small \emph{shift} of the baseline.
Upon color coding the CBGs in GA based on their cluster (\Cref{fig:covid}(c)) we see that the CBGs in clusters with lower baselines are more rural, while the relatively fewer CBGs in the clusters with higher stay-at-home adherence baseline are in urban, metropolitan areas.
A second application of our framework yields the organization in \Cref{fig:covid}(d), which again naturally separates into five distinct clusters of regions, as before, except for a key difference: we find that the CBGs in the clusters colored in brown and dark blue have very different average patterns of mobility than the rest (see \Cref{fig:covid}(e)).
Both clusters are concentrated in urban areas, thus offering an even finer-grained perspective of the dataset than \Cref{fig:covid}(a-c) and in \Cref{fig:covid}(f) we show a zoom-in of the city of Atlanta, where many of the CBGs with the anomalous SAH time series are present.
Particularly notable are the CBGs in the dark blue cluster, whose average SAH time series exhibits a large increase in SAH fraction in May.
On inspection, we find that most of the CBGs in this cluster are home to
the major universities of Georgia.
As explained in SM, \Cref{SM-sec:app}, we identified that these universities implemented initial distance-learning mandates starting late-March (following which, the statistics show a slight rise in early-April) and full stay-at-home measures, with canceled or virtual final examinations and commencement ceremonies, beginning in May, which provides a possible explanation for the distinct behavior in average SAH fraction for this cluster.
We note that the same dataset was analyzed with state-of-the-art methods in \cite{Levin_Chao_Wenger_Proctor_2020} and we chose the same number of clusters in order to compare our results. They required $14$ dimensions to identify $5$ clusters similar to \Cref{fig:covid}(b), which is more than a four-fold increase from our reduction to $3$ dimensions.
Strikingly, ours is the only approach to date that identifies the anomalous, weak signals in \Cref{fig:covid}(e), which contain only $62$ ($18$) of the $5509$ data points in the dark blue (brown) clusters. In the remainder of the article we explain the quantum mechanical framework for geometric data analysis and associated algorithms that produced the results in \Cref{fig:covid}.

The dataset just considered is an example of the following more general scenario: an experimentalist obtains a sequence of measurements, each consisting of a value for $D$ variables. Each measurement is thus a point in $\mathbb{R}^D$. Of course, measurements coming from nature are bound by physical laws, so in fact one must imagine that the \emph{true} ambient space is some other, nonlinear structure $\mathcal{M}$, residing in $\mathbb{R}^D$. Moreover, if the experiment is governed by a number of parameters, then these give $\mathcal{M}$ its local degrees of freedom around any given measurement. This principle is in fact often studied and known as the \emph{manifold hypothesis} (MH) \cite{fefferman_testing_2016}: measurements of arbitrarily high dimensions (residing in $\mathbb{R}^D$) arising from natural processes are confined to low-dimensional manifolds (\ie $\mathcal{M} \subset \mathbb{R}^D$ with $\dim \mathcal{M} \ll D$). Due to the MH, the experimentalist's aim of connecting model parameters to observations through the relationships among measurements is just the analysis of the organizational structure, or \emph{geometry} of data. In the natural sciences, such analyses are commonly performed through dimensionality reduction, classification, \emph{etc.} using techniques such as principal component analysis (PCA), t-distributed stochastic neighbor embedding (t-SNE), or variants of Laplacian eigenmaps, that indirectly probe the geometry of data, \eg  \cite{coifman_geometric_2005, Maggioni_2015, Kobak_Berens_2019}. In addition, it has recently been appreciated that understanding and exploiting the structure of data, and sometimes reorganizing/reparameterizing it, has advantages in learning frameworks such as convolutional neural networks, \eg \cite{Zhu_2017, Bronstein_2021}.
While the state-of-the-art in learning the structure and organization of data is founded on principles of Markovian dynamics \cite{belkin_laplacian_2003, coifman_diffusion_2006} -- primarily because of their \emph{local} nature that forms an accessible link between random walks and Markov processes -- in practice, these are suitable only for accessing coarse features of the data due to the slowly varying nature of the steady-states that form the basis of these techniques.
We have found -- as in \Cref{fig:covid} and the applications in the SM, \Cref{SM-sec:eg} -- that to perform a fine-scaled data analysis, it is advantageous to shift the paradigm to quantum mechanics, which provides a natural way to formulate dynamics on the data that respects limits to resolution of its geometric structure set by data sampling, through the introduction of a characteristically quantum mechanical uncertainty principle.

\begin{figalg}[t]
\centering
\fbox{
  \begin{minipage}[c]{0.55\textwidth}
\begin{algorithmic}[1]
	\State {\bf Inputs:} $X_N=\{v_1, ..., v_N\}, v^*, \epsilon>0, \alpha\geq 1, t>0$
	\State {\bf Output:} Propagated state $[\psi_h^\zeta](t)$

    \Procedure{Propagate}{}
	    \IFor{$i,j = 1:N$} $\left[T_{\epsilon}\right]_{i,j} \gets k(\nicefrac{||v_i - v_j||^2}{\epsilon})$	\label{alg:kernelmat}
	    \State $D_{\epsilon} \gets \operatorname{diagonal \, matrix}\left( \sum_{j=1}^N [T_{\epsilon}]_{i,j} \right)_{1 \leq i \leq N}$	\label{alg:normalizer}
    	\State $\Delta_{\epsilon,N} \gets \frac{4(I_N-D_{\epsilon}^{-1} T_{\epsilon})}{\epsilon}$
		\State $U_{\epsilon,N}^{t} \leftarrow  \exp(-i t \sqrt{\Delta_{\epsilon,N}})$	\label{alg:propagator}
      \State $h \gets \epsilon^{\frac{1}{(2+\alpha)}}$
      \State $p_0 \gets v_j - v^*$ for $v_j$ closest to point $v^*$
\While {$1 \leq \ell \leq N$}
      \State $[\psi^\zeta_h]_\ell \leftarrow e^{-\nicefrac{\vert\vert v_\ell - v^*\vert\vert^2}{2h}} e^{\frac{i}{h} (v_\ell - v^*)^{\sf T} \nicefrac{p_0}{||p_0||}}$
\EndWhile
      \State \textbf{return} $[\psi^\zeta_h](t) = U^t_{\epsilon, N} [\psi^\zeta_h]$
    \EndProcedure
    \end{algorithmic}

  \end{minipage}\hfill
}
  \begin{minipage}[c]{0.42\textwidth}
\caption[LoF entry]{\footnotesize Pseudocode for algorithm that performs data-driven propagation of coherent states on a data graph. The inputs are the dataset $X_N \subset \mathbb{R}^D$, an initial data point to propagate from, $v^*$, parameters $\epsilon>0, \alpha\geq1$, and a time to propagate for, $t>0$.
Lines 4-7 construct a data-driven quantum propagator, $U^t_{\epsilon,N}$ using a graph Laplacian $\Delta_{\epsilon,N}$ computed from the data.
In line 4, $k(\cdot)$ is an exponentially decaying function of the argument and in line 6, $I_N$ is the $N\times N$ identify matrix.
Lines 10-11 form the $N \times 1$ vector that approximates a coherent state at phase space point $\zeta := (v^*, \nicefrac{p_0}{\vert\vert p_0\vert\vert})$.
A more detailed description of this algorithm, and extensions of it, are presented in the SM, \cref{SM-app:code}.
\label{fig:alg}
}
  \end{minipage}
\end{figalg}

\begin{figure}[t]
\centering
\includegraphics[width=0.9\textwidth]{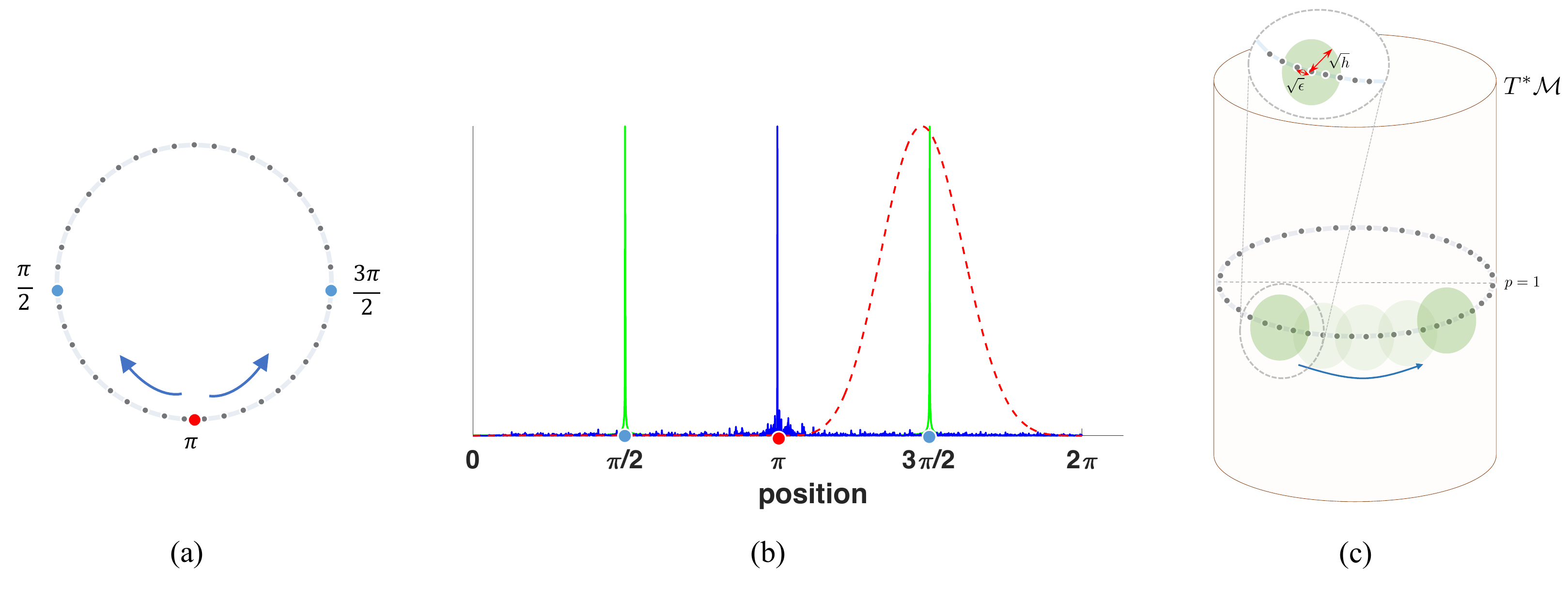}
\caption[LoF entry]{\scriptsize \textbf{(a)} Data is $N=2500$ regularly spaced samples from the unit circle: $x_j = (\cos \theta_j, \sin \theta_j)$ for $\theta_j \in [0, 2\pi)$.
\textbf{(b)} We observe the squared magnitude of: (green) an undirected optical ray from $x^* = \pi$ at time $t = \frac{\pi}{2}$, \ie $|U^t[\delta_{\pi}]|^2$, and (blue) the corresponding data-driven quantum propagation, \ie $|U_{\epsilon,N}^t[\delta_{\pi}]|^2$. While the former is concentrated at $x_{-}=\pi-t = -\nicefrac{\pi}{2}, x_{+}=\pi+t=\nicefrac{3\pi}{2}$, which are points that minimize $|d_g(x^*, x) - t|$ ($d_g$ is geodesic distance), the latter state bears no resemblance to geodesic propagation, with significant attenuation over the green signal, and an undiminished peak at the source, $x^*$. We also show (dashed red) the data-driven propagation of a coherent state centered at $x^*$ with uncertainty parameter $h$, \ie $|U_{\epsilon,N}^t[\psi^\zeta_h]|^2$, with $\zeta=(\pi,1)$. This state is approximately centered at the point $x_+$, and in fact, its expected position, $\langle x \rangle$, satisfies $|\langle x\rangle - x_+|\leq h$. Note that all curves are normalized to have the same maximum value in order to plot them on the same scale. \textbf{(c)} Depiction of coherent state propagation in \emph{phase space}, $T^*\mathcal{M}$. The zoom-in schematically shows $\sqrt{\epsilon}$ and $\sqrt{h}$, the data-determined scale and uncertainty parameters, respectively.
\label{fig:circle}}
\end{figure}

Our implementation is built upon the simulation of quantum dynamics on data as sketched in \Cref{fig:alg}.
The fundamental relationship connecting this dynamics to the structure of data is given by a \emph{discrete} quantum-classical correspondence principle (QCC), which is established in Ref. \cite{kumar_math} and summarized in the SM, \cref{SM-sec:method-theory}, and applies when the MH holds for the data in \(\mathbb{R}^D\), confined to \(\mathcal{M}\) a smooth, compact, boundaryless submanifold.
This relationship is supported by the (traditional) continuum form of the QCC, which physically in our geometric setting connects the propagation of wavefunctions of photons (quantized excitations) on curved space ($\mathcal{M}$) with geodesic propagation of light rays \cite{Gloge_Marcuse_1969}.
Such a propagation is illustrated in green in \Cref{fig:circle}.
More formally, it is a well-known result in microlocal analysis \cite{zelditch_2017, zworski_semiclassical_2012} that the trajectory of an impulse \(\delta_{x^*}\) at \(x^* \in \mathcal{M}\) with respect to the propagation $U^t[\delta_{x^*}] := e^{-i t\sqrt{\Delta}}[\delta_{x^*}]$, where \(\Delta\) is the Laplace-Beltrami operator of \(\mathcal{M}\), has singular support along geodesics in all directions at distance \(|t|<T\) from \(x^*\), for a bounded time \(T\).
This is the continuum QCC in action: it connects the quantum mechanical propagation $U^t[\delta_{x^*}]$ generated by the quantization $\sqrt{\Delta}$ of the pure kinetic energy classical Hamiltonian, to the geodesic flow, by moving the energy concentration of the singular state $\delta_{x^*}$ to time $t$ along the classical flow in directions of the initial state's asymptotically large momenta \cite{zelditch_2017}. The energy concentration propagates to all geodesics at time $t$ emanating from $x^*$ because the initial state  $\delta_{x^*}$ has asymptotically large momenta\footnote{By wave-particle duality, the momentum of a particle corresponds to the frequency of its wavefunction.} in all directions.

Coming back to quantum dynamics on the data, we couple the above continuum QCC with the now well-known result that, assuming the MH, we can approximate from \(N\) measurement samples \(X_N := \{ v_1, \ldots, v_N \} \subset \mathbb{R}^D\) the Laplace-Beltrami operator of their underlying manifold through the graph Laplacian \(\Delta_{\epsilon,N}\) of a \(\sqrt{\epsilon}\)-nearest-neighbour (\(\sqrt{\epsilon}\)-n.n.) graph on \(X_N\) with high probability.
This is founded on the above-mentioned Markovian methods: in fact, \(\Delta_{\epsilon,N}\) is a first-order approximation, in \(\epsilon\), to the generator of a Markov process on \(X_N\), which almost surely limits to a continuum Markov process on \(\mathcal{M}\) as \(N \to \infty\) \cite{belkin_laplacian_2003, hein2005graphs, coifman_diffusion_2006}. Combined with some essential new ingredients, a data-driven approximation to the quantum mechanical propagator \(U^t\) is given with high probability, in Ref. \cite{kumar_math}, by the \(N \times N\) matrix $U_{\epsilon,N}^t := e^{-i t \sqrt{\Delta_{\epsilon,N}}}$, see line 7 of \Cref{fig:alg}.

Given this construction, it is tempting to reproduce the geodesic propagation of light rays given through \(U^t[\delta_{x^*}]\) with \(U_{\epsilon,N}^t[\delta_{x^*}]\) \footnote{We denote a discrete approximation to an impulse, an $N\times 1$ vector with zeros everywhere but one entry, as $\delta_{x^*}$ also, as will be clear from the context in which it is used.}.
However, as shown in the blue curve in \Cref{fig:circle}(b), this data-driven propagation bears no resemblance to the continuum signal \(U^t[\delta_{x^*}]\) (green curve).
To see why this is, we go back to the Markov process roots of $U^t_{\epsilon,N}$: namely, the Markov process defining $\Delta_{\epsilon,N}$ is discretized to $\sqrt{\epsilon}$-balls%
\footnote{
When we refer to $\sqrt{\epsilon}$- and $\sqrt{h}$-balls, these are the radii of the balls up to constant factors; \ie the balls are $O(\sqrt{\epsilon})$ or $O(\sqrt{h})$, however for conciseness we omit the order notation.
}
on the manifold.
Then, based on the uncertainty principle, the analysis in \cite{kumar_math} shows that this Markov process acts like a low-pass filter%
\footnote{Specifically, this follows from the semiclassical analysis in \cite{kumar_math} specified to $h = 1$ for the continuum Markov process, combined with convergence to bring this to the discrete setting. The continuum Markov process is an integral kernel operator with $C^{\infty}$ kernel and expressed in phase space as in \cite{kumar_math}, this operator attenuates spatial frequencies $\xi$ with $|\xi| \gtrsim 1/\sqrt{\epsilon}$ at the rate $O(|\xi|^{-M})$ for all $M > 0$.}
by attenuating spatial frequencies with magnitude $\gtrsim 1/\sqrt{\epsilon}$.
Therefore, $\Delta_{\epsilon,N}$ approximates a scalar operator at bandwidth $\gtrsim 1/\sqrt{\epsilon}$ and by virtue of being its spectral function, so does $U_{\epsilon,N}^t$.
This explains the concentration of the blue curve in \Cref{fig:circle}(b) about $x^*$ and predicts that this behaviour will persist even in the limit $N \to \infty$ while $\epsilon >0$ remains bounded away from zero.

Understanding this issue is the fulcrum of our discrete QCC, which recovers light rays from data-driven simulation of quantum dynamics. As we have seen, the primary obstruction to approximating the continuum QCC with $U_{\epsilon,N}^t[\delta_{x^*}]$ is the over-concentration of the initial state $\delta_{x^*}$, which gives frequency content above the data-determined threshold $\sim 1/\sqrt{\epsilon}$.
To control this, we introduce a state $\psi_h \in C^{\infty}(\mathcal{M} \times (0, 1]_h)$ surrogate to $\delta_{x^*}$, whose bandwidth we can control to scale as $\sim 1/h$ and whose continuum propagation $U^t[\psi_h]$ follows the light ray emanating from $x^*$ until time $t$, to within a $\sqrt{h}$-ball throughout the propagation.
This is satisfied by \emph{coherent states}, which in the continuum have position-space representation: $\psi_h^{\zeta}(v) := e^{-\frac{1}{2h} ||v - x^*||^2} e^{\frac{i}{h} \langle v - x^*, p \rangle}$ with position \(x^* \in \mathcal{M}\) and momentum $p \in T^*_{x^*}\mathcal{M}$.
In particular, $\psi_h^{\zeta}$ (time $t = 0$) is localized in phase space in a $\sim \sqrt{h}$-ball about the point $\zeta := (x^*,p/h) \in T^*\mathcal{M}$. The wave-particle duality is apparent in this state: its spatial oscillation frequency is set by $p/h$, where $p$ is the classical momentum vector which sets the direction of maximum variation.
Thus, the \emph{uncertainty parameter}\footnote{In mathematical physics terminology, $h$ is more commonly called a \emph{semiclassical parameter}.}
 $0 < h \leq 1$ governing the quantum mechanical properties of coherent states gives us the desired control to establish our discrete QCC.

The finite discretization of $\psi_h^\zeta$ can be defined
on the sampled dataset \(X_N\) as a vector \([\psi_h^{\zeta}]\) of dimension \(N\times 1\) as per line 10 of \Cref{fig:alg}, and its extensions in SM, \cref{SM-app:code}.
We approximate the momentum $p$ from the data $X_N$ using various techniques that approximate the tangent space at a given $x^*\in X_N$, and show in SM, \cref{SM-sec:geodesic-measurements}.
In order for the data-driven propagation $U_{\epsilon,N}^t[\psi_h^{\zeta}]$ to approximate $U^t[\psi_h^{\zeta}]$ and therefore, to follow the light rays on data, the bandwidth $1/h$ must be kept $\ll 1/\sqrt{\epsilon}$.
Hence, crucially we scale $h \gg \sqrt{\epsilon}$. The convergence rates in [\cite{kumar_math} and SM, \cref{SM-sec:data_dyn}] that take into account the finite size $N$ of the dataset, establish that the parameter $h(N)$ dictates how the uncertainty in the fine-scaled resolution of the geometry of the dataset scales with $N$.

In \Cref{fig:circle}(b) the dashed red line shows the result of propagating a coherent state using a data-driven propagator for the circle example. By incorporating the intrinsic uncertainty induced by finite sampling into the formulation of quantum dynamics, we recover accurate propagation. In \Cref{fig:circle}(c) we also depict coherent state propagation in phase space for this example, and graphically show the relationship between $\epsilon$ and $h$. We stress that this uncertainty relation crucially depends on the quantum formulation that realizes the graph Laplacian at length scale $\epsilon$ as a quantization of the geodesic Hamiltonian up to wavelength $h$ (this is further explained with detail in \cite{kumar_math}).

The coherent states just considered are meaningful for the data: their localization and momentum properties jointly encode the local density and variation of the data samples.
The MH can be interpreted as a combined, local and global hypothesis on these properties of the dataset: it dictates that as more data samples are acquired from measurements of a physical system, their variation and concentration in ambient space are subject to geometric constraints, which are fixed by the natural process and thus independent of the number of measurements.
The discrete QCC realized by $U_{\epsilon,N}^t[\psi_h^{\zeta}]$ extends the local information encoded in $\psi_h^{\zeta}$ non-locally to find optimal paths and lengths between data points at a sufficiently large $N$, with respect to the constraints governed by the system being measured.
It is remarkable that this simulated quantum dynamics on the data, with finite samples, traverses the optimal path by forecasting new measurements that are feasible for the underlying physical system.

In the SM we derive the convergence results to establish the following: given $N$ data samples, $X_N$, from a smooth density on $\mathcal{M}$,
the data-driven
finite-dimensional matrix propagation $[\psi_h^{\zeta}](t) := U_{\epsilon,N}^t[\psi_h^{\zeta}]$ returned by \Cref{fig:alg}, with $h \propto \epsilon^{\frac{1}{2 + \alpha}}$ for $\alpha \geq 1$ agrees with $U^t[\psi_h^{\zeta}]$ up to uniform error $O(h)$ \emph{w.h.p.}, provided $h \gtrsim N^{-\frac{1}{\gamma}}$ for $\gamma > 0$ a constant depending only on $\alpha$ and $\dim \mathcal{M}$. Thus,
$[\psi_h^{\zeta}](t)$ traverses within an $O(\sqrt{h})$ radius of the geodesic beam emanating from $v^*$ in the direction $p$ to time $t$. Moreover, the point $\bar{x}_t := ||\psi_h^{\zeta}(t)||^{-2} \sum_{j=1}^N \begin{psmallmatrix} v_{j,1} & \cdots & v_{j,D} \end{psmallmatrix} |[\psi_h^{\zeta}](t)|^2$, which is the expected position of the propagated state, is \emph{w.h.p.} within geodesic distance $O(h)$ to the point $x^*_t$, that is geodesic distance $t$ from the initial position $v^*$ in the direction $p$.
While this expected value of the position coordinate, $\bar{x}_t$, is the best estimate of position along the geodesic path, due to the localization of the propagated state the maximum of the wavepacket distribution, $\hat{x}_t := \arg\max |[\psi_h^{\zeta}](t)|^2$ is also \emph{w.h.p.} within geodesic distance $O(h)$ from $x^*_t$ \cite{kumar_math}.

The fine-scaled resolution properties of quantum dynamics on data are now apparent: geodesics are inherently \emph{high-frequency} features of the data \cite{zelditch_2017}. While Markovian methods have introduced a crucial $\epsilon$ dependence in data analysis, which signifies that the quality of approximation of $\Delta_{\epsilon,N}$ is tied to a random walker's traversibility of the underlying $\sqrt{\epsilon}$-n.n. graph \cite{hein2005graphs}, we have switched the perspective to a quantum mechanical \emph{wavelength} $h$ at which the dataset's fine-scaled geometry can be resolved.

The uncertainty principles utilized above to study the geometry of data through the crucial $h(\epsilon)$ relationship, can also be viewed from a signal processing perspective.  In fact, quantum mechanics in the continuum and time-frequency analysis are two sides of the same coin, and share common notions of uncertainty principles \cite{Grochenig_2001}.
Particularly relevant to this work is the short-time Fourier transform (STFT) and its relation to localized wavepackets.

To bring our results firmly into the signals processing setting, we begin by noting that our initial coherent states $\psi_h^{\zeta}$ are actually Gabor wavelets \cite{Antoine_2015}.
These wavelets define a form of the STFT, namely $T_h:\mathcal{D}(\mathcal{M}) \to C^{\infty}(T^*\mathcal{M})$ that maps a distribution $f \in \mathcal{D}$ to its time-frequency%
\footnote{The terminology \emph{time-frequency} is borrowed from the conventional signal processing setting, but it is important to note that in our geometric context, this is the \emph{position-momentum} or \emph{position-frequency} representation of a function over phase space and thus, \emph{time} in time-frequency refers to a point of $\mathcal{M}$ rather than the temporal parameter $t$ in the propagators $U^t$ or $U_{\epsilon,N}^t$ (see SM, \cref{SM-sec:dual}).}
(or phase space) representation by integrating%
\footnote{See SM, \cref{SM-sec:dual} for explicit definition of $T_h$. In the context of manifolds, this is commonly called the FBI transform and in quantum mechanics, its result is known as the Husimi phase space distribution.} against $\overline{\psi^{\zeta}_h}$.
Since this STFT's window function saturates the time-frequency uncertainty, this is just the Gabor transform adapted to a manifold and frequency scaled by a factor of $1/h$.
The Gabor spectrogram $|T_h[f](\zeta)|^2$ is an essential tool for signal analysis and especially detection, since it gives a picture of the frequency content of $f$ occurring across spatial windows with equal order of resolution in space (position) and frequency, which in our case is a $\sqrt{h}$-ball in phase space.
In practice, we realize this spectrogram from just the measurement samples $X_N = \{ v_1, \ldots, v_N \} \subset \mathbb{R}^D$ by taking directions $p$ from an approximate tangent space at each $v_j$ and computing $\zeta := (x_j, p) \mapsto T_{h,N}[f](\zeta) := [\psi_h^{\zeta}]^{\dagger} [f]$.
We show in the SM, \cref{SM-sec:dual} that with $\zeta$ taken in this way,
$\mathscr{T}_h^t : \zeta \mapsto |T_{h,N}[U_{\epsilon,N}^{-t}[\delta_{x^*}]](\zeta)|^2$ approximates \emph{w.h.p.}, to within $O(h)$ error, the spectrogram of $U^{-t}[\delta_{x^*}]$ \emph{uniformly} over any bounded region $\mathcal{B}$ of phase space at bandwidths $O(1/h)$, provided that $h \gg \sqrt{\epsilon}$.
Since as discussed above, $U^{-t}[\delta_{x^*}]$ has light rays emanating from $x^*$ in \emph{all} possible directions on $\mathcal{M}$, these can be computationally recovered from $\mathscr{T}_h^t$ evaluated on a region $\mathcal{B}$ containing all unit speed phase space points.
More specifically, our use of the Gabor transform and the relationship between uncertainty and sampling density given by $h(\epsilon)$ show that the high-frequency content of $U^{-t}[\delta_{x^*}]$ is attenuated to within frequency band $1/h \ll 1/\sqrt{\epsilon}$ and spread across phase space windows with radius of order $\sqrt{h} \gg \epsilon^{\frac{1}{4}}$.
For this reason,
we view $U^{-t}[\delta_{x^*}]$ as a signal with arbitrarily high-frequency content that is attenuated to $U_{\epsilon,N}^{-t}[\delta_{x^*}]$ by the finite sampling that gives the dataset $X_N$.

\begin{figure}[t]
\centering
\includegraphics[width=\textwidth]{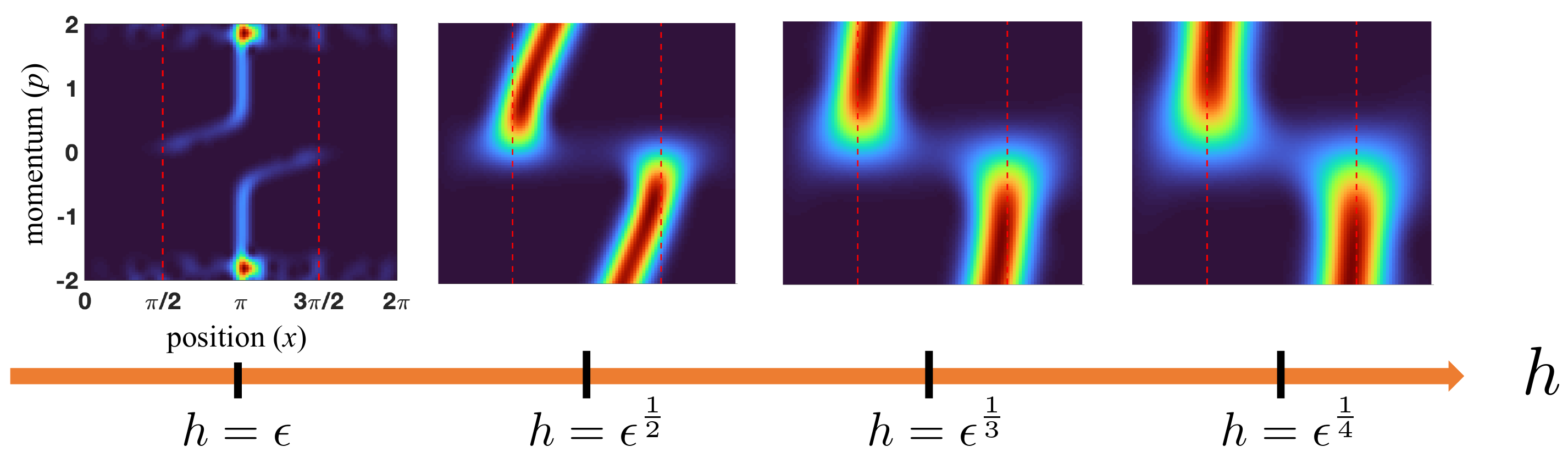}
\caption[LoF entry]{\scriptsize
Illustrating the dual picture of resolving time evolved states at scales set by $h$. We propagate an initial impulse $\delta_{\pi}$ on sampled circle (as in \Cref{fig:circle}, with $N=2500$) with the data-driven propagator $U_{\epsilon, N}^{t}$ for $t=\nicefrac{\pi}{2}$. Each plot shows the discrete approximation $\mathscr{T}_h^t(x,p)$ to the spectrogram of this state at varying resolution $h$ as a function of $\epsilon$, the scale parameter set by the data sampling.
At resolutions $h \lesssim \sqrt{\epsilon}$ the spectrogram's localization is inaccurate in terms of the \emph{high-frequency content} $U^t[\delta_{\pi}]$ (red, dashed lines), while at appropriate resolutions, $h=\epsilon^{\nicefrac{1}{(2+\alpha)}}$, for $\alpha>1$, we see a state localized at $x^* \pm t = \pi \pm \nicefrac{\pi}{2}$ and delocalized in momentum, as we would expect from a propagating a state perfectly localized in position space.
This illustrates (1) the interpretation of $h$ as setting the scale at which data-driven dynamics can be accurately resolved, and (2) that the spectrogram reveals at once, the position at time $t$ of light-rays emanating from $x^*$ in \emph{all} possible directions.
\label{fig:phase_space}}
\end{figure}

To illustrate this dual picture, we present spectrograms for the circle example from \Cref{fig:circle} at a fixed propagation time and various resolutions, $h$, in \Cref{fig:phase_space}. When $h\lesssim \sqrt{\epsilon}$ the inaccuracy of the propagation is evident. This is another perspective on the observation from \Cref{fig:circle}, where the direct application $U_{\epsilon,N}^t[\delta_{x^*}]$ is unable to recover the propagated state due to the attenuation of high-frequency content in the signal discussed above in terms of quantum dynamics. Whereas, when the resolution of the spectrogram is at $h\gg\sqrt{\epsilon}$ the signal $U^t[\delta_{x^*}]$ is recovered.

The data-driven quantum dynamics formulated above enables estimation of intrinsic distances between points in a dataset: if $\bar{x}_t(j ; p_j) \in X_N$ denotes the data point closest to $\bar{x}_t$ (or $\hat{x}_t$) computed as above with respect to an initial state $\psi_h^{\zeta}$ localized at $\zeta := (v_j, p_j)$, then $t$ gives the propagation time of a coherent state following approximately a ray emanating in direction $p_j$ from $v_j$. When the data is sampled from a smooth density on a smooth, compact, boundaryless manifold, we have established that $t$ is \emph{w.h.p.}, within $O(h)$ of the geodesic distance . In fact, by using local PCA to define $\psi^{\zeta}_h$ and hence, $p_j$, we can \emph{chart} the dataset $X_N$ with \emph{geodesic polar coordinates} (GPC), or equivalently, normal coordinates (SM \cref{SM-sec:state-prep}).
We pause to emphasize that the procedure we have described gives access to geodesics and GPC on a manifold, which are inherently described by non-linear dynamical equations, through linear, matrix computations. Computing such quantities, even when much more is known about the manifold, is generally computationally difficult since typical approximation methods are insufficient: polyhedral approximations are not a faithful model of manifolds with curvature restrictions \cite{petrunin2003polyhedral} and even forward marching type approximations are known to be prone to failures \cite{peyre_geodesic_2010}. Furthermore, our result establishes the first general convergence result for geodesics and GPC from data.

\begin{figure}[t]
\centering
\includegraphics[width=\textwidth]{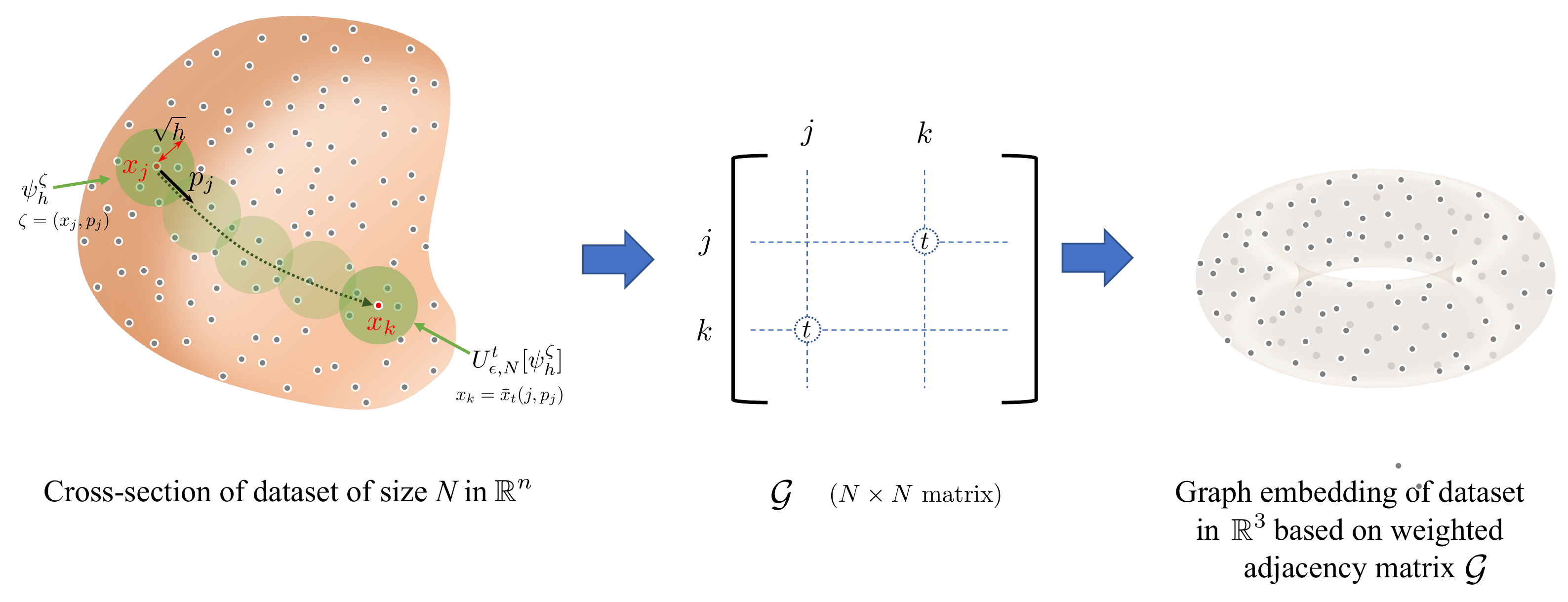}
\caption[LoF entry]{\scriptsize Illustration of how data-driven extraction of geodesic distances between data points establishes a similarity metric on the dataset that can be used to perform low-dimensional embedding of the data. By propagating a coherent state centered on $x_j$, with initial momentum $p_j$ ($|p_j|=1$), with $U_{\epsilon,N}^t$, and extracting the point closest to the expected position of the resulting state, $x_k \equiv \bar{x}_t(j,p_j)$, we establish a distance $t$ between these points (left). Repeating this process for a set of initial point and times allows construction of a matrix $\mathcal{G}$, that approximates geodesic distances between data points (middle). This matrix can be utilized as a weighted adjacency matrix for a  (low-dimensional) graph embedding of the dataset (right).
\label{fig:embeddings}}
\end{figure}

More broadly, the assignment of a data point $\bar{x}_t(j, p_j) = v_k$ to a given data point $v_j$ is, in itself, independent of further structural assumptions on the dataset. Even when the data $X_N$ is not guaranteed to be sampled from a manifold, this defines a distance relationship between two points based on the \emph{quantum walk} $U_{\epsilon,N}^t[\psi_h^{\zeta}]$.
Based on this distance relationship we can build an $N\times N$ adjacency matrix, $\mathcal{G}$, for a graph $\mathcal{X}_N$ on $X_N$, with elements $\mathcal{G}_{j,k} = \mathcal{G}_{k,j}=t$; see \Cref{fig:embeddings}. Repeating this process for a collection of initial points $v_j$ and time-steps, $t_1, \ldots, t_m$ populates this adjacency matrix, which captures a notion of distance between the data points in $X_N$ given by the quantum propagation times of coherent states.
We can perform emebeddings of $\mathcal{X}_N$ to achieve tasks such as recovering reduced-dimensional coordinates, clustering, classification, \emph{etc.} We find that even the most classical embedding of $\mathcal{X}_N$ in few dimensions, such as the Fruchterman-Reingold (FR) method of springs and electrostatic forces \cite{Fruchterman_Reingold_1991}, recovers salient features of complex datasets.
For example, \Cref{fig:covid} shows examples of FR embedding into 3 dimensions followed by $k$-means clustering of the COVID-19 mobility information dataset.

We have argued that, similar to the Fourier transform being fundamental for signal analysis, quantum propagation and sensing are fundamental for \emph{structure} analysis in sampled data.
\Cref{fig:covid} exhibits the ability of this approach to organize and discover previously unseen anomalous behavior in social distancing data collected during the COVID-19 crisis, affected by irregularity and noise and generated by complex, intractable mobility dynamics.
As seen from \Cref{fig:alg} the basic transformation of data realizing this approach is algorithmically quite simple, involving only \emph{linear} operations, and moreover, we have rigorous justification with rates of convergence and mild assumptions that are supported from our novel inclusion of graph Laplacians in the framework of semiclassical analysis and probabilistic reasoning.
Furthermore, we have shown both sides of a \emph{dual picture}, providing interpretations through the QCC on the one hand and signals processing on the other, as exhibited in \Cref{fig:circle,fig:phase_space}, respectively, which are again firmly grounded mathematically through probabilistic convergence rates.
In the SM we present several additional applications -- combined with the graph embedding method depicted in \Cref{fig:embeddings} -- even for meaningful organization of very small datasets, and a statistical analysis of the quality of geodesic extraction on model manifolds, with the corresponding algorithms.
Our approach to data analysis combines methods from a wide array of fields, including dynamics, geometric inverse problems, statistics and physics, and has relationships to yet more disciplines, which we discuss in some detail in the SM, \cref{SM-sec:disc}.

\section*{Acknowledgments}
This work was supported by the Laboratory Directed Research and Development program at Sandia National Laboratories, a multimission laboratory managed and operated by National Technology and Engineering Solutions of Sandia, LLC., a wholly owned subsidiary of Honeywell International, Inc., for the U.S. Department of Energy's National Nuclear Security Administration under contract {DE-NA-0003525}. MS was also supported by the U.S. Department of Energy, Office of Science, National Quantum Information Science Research Centers.

\bibliography{bib}

\bibliographystyle{Science}

\clearpage



\section*{Supplementary material}

# Supplementary Material for "Shining light on data: Geometric data analysis through quantum dynamics"

Akshat Kumar[1, 2, a)] and Mohan Sarovar[3, b)]

[1)]*Department of Mathematics, Clarkson University, Potsdam, NY 13699 USA*
[2)]*Instituto de Telecomunicações, Lisbon, Portugal*
[3)]*Quantum Algorithms and Applications Collaboratory, Sandia National Laboratories, Livermore, California 94550, USA*

CONTENTS

---

[a)]Electronic mail: akumar@clarkson.edu
[b)]Electronic mail: mnsarov@sandia.gov

# I. OVERVIEW OF SUPPLEMENTARY MATERIALS

In the main text we introduced a new approach to learning the geometry of datasets through linear dynamics generated by the square-root of the data-driven graph Laplacian that evolve data-driven analogues of coherent states in imaginary time. As explained there, when the datasets satisfy the manifold hypothesis (MH), these finite-dimensional dynamics – along with the induced dynamics of measurements on these propagated states (*e.g.,* taking inner products or identifying maximizers) – are, with high probability, small perturbations of quantum dynamics on the underlying manifold carrying the data. In turn, these quantum dynamics on the manifold satisfy a *quantum-classical correspondence* (QCC) that recovers the geodesic flow through the propagation of coherent states. Consequently, the data-driven linear dynamics we propose, give rise to a *discrete* QCC that recovers the geodesic paths between data samples. From a dual perspective, the linear measurements we evolve are approximations to short-time Fourier transforms of the data-driven propagation of point particles on the manifold. Thus, the corresponding spectrogram is concentrated about the geodesic sphere centered at the position of the initial point particle, of radius given by the length of time of propagation; this, we view as (the *high-frequency content* in) the geometric analogue to *signal* in the data.

The following supplementary sections serve the purposes of: (1) giving the mathematical background and model for this learning problem and for our approach, (2) giving a rigorous treatment to the connections with quantum dynamics through the mathematical framework of semiclassical analysis and to the problems of convergence from finite data, (3) providing further details on the applications discussed in the main text as well as more real-world applications and (4) detailing algorithms to implement our proposed approaches on actual data. In Section II, we discuss the mathematical details and give a more general treatment than explained in the main text: namely, we develop the discrete QCC for a (commonly used) parameterized family of normalized graph Laplacians and for initial states and measurements taken in a variety of ways amenable to computation. Much of the theory and convergence results are developed in [1] and here we mainly extend these results to the (more) practical class of initial states we consider in this work. Along the way, we discuss more on the semiclassical parameter $h$ and the data-driven method for choosing $h$ and $\epsilon$, based on a computable *deviation measure*. In Section III we provide details on our numerical experiments as well as real-world applications and in Section IV and Section V we give further details on our computations for the model test cases that serve as controls for our numerical experiments. The choices of $h$ and $\epsilon$ parameters have been important to achieve effective results in all applications and in Section VI we show that the useful selections come from near the specified regions of our data-determined *deviation measure*. Further, in Section VII we provide algorithms along with explanations of all input parameters, which we have used to implement our approaches in all of the applications. Finally, our methodology connects various fields of active research and we give some perspectives on this in Section VIII.

# II. METHODS AND THEORY

In this section we present background material and details of our proposed approach, including the derivation of theorems justifying the statements in the main text, and details of a general algorithm for manifold learning based on our new approach. We also show how our data-driven approach to extracting geodesics on data manifolds can be coupled with embedding techniques to achieve non-linear dimensionality reduction.

The key ingredients to our approach are (i) the approximation of a time evolution operator (propagator) for a quantization of Hamiltonian dynamics on the data manifold, (ii) propagation of states that approximate classical point particle trajectories, and (iii) a rescaling of the diffusion approximation parameter to support the quantized dynamics. Quantization of the dynamics decomposes into linear operations the classically non-linear problem of solving for geodesics on a manifold, and moreover, our construction of a discrete quantum classical correspondence allows us to prove strong asymptotic connections between the dynamics induced by quantum Hamiltonians and geodesic distances on the underlying manifold.

We begin with defining notation in Section II.A, provide background relating quantum and classical dynamics in the continuum setting in Section II.B, and then developing data-driven constructions in Sections II.C and II.D.

## II.A. Notation

Throughout this work, we will connect our discussion to concepts in physics. To aid with this we use the following Dirac notation and physics terminology: operators and quantum states are denoted as $\hat{A}$ and $|\psi\rangle$, respectively, and complex inner products are written as $\langle\phi|\psi\rangle$, where $\langle\phi|$ is the dual of $|\phi\rangle$, *i.e.,* this inner product is anti-linear in the first component. Further, operator expectations are denoted $\langle\hat{A}\rangle := \langle\psi|\hat{A}|\psi\rangle = \langle\psi|\hat{A}\psi\rangle$, and we often use Schwartz

distributions so that we may represent states in the position basis as $\psi(\mathbf{x}) := \langle \delta_\mathbf{x} | \psi \rangle$, where $|\delta_\mathbf{x}\rangle$ a point mass at $\mathbf{x} \in \mathcal{M}$, hence allowing us to think of them as $L^2$ functions on the manifold. Finite-dimensional matrices, especially those corresponding to discrete approximations of infinite-dimensional operators are denoted by $\breve{A}$. A complete table of notation is provided in Table I.

### II.B. Continuum quantum and classical dynamics

As in the main text, we denote the dataset as $X_N = \{v_1, v_2, ..., v_N\}$, which consists of $N$ samples from the underlying compact, smooth, and boundaryless Riemannian manifold $\mathcal{M}$ with dimension $\nu$ and metric tensor $g_{ij}$. The manifold is assumed to be isometrically and smoothly ($C^\infty$) embedded in $\mathbb{R}^n$ with a bounded second fundamental form. We view the index $\ell$ to correspond to a sample on $\mathcal{M}$ and $v_\ell \in \mathbb{R}^n$ to specify its *extrinsic coordinates*. The samples in $X_N$ are drawn according to a fixed distribution on $\mathbb{R}^n$ that is supported on the embedding of $\mathcal{M}$, and has smooth density p with respect to its induced volume form.

Often, in practice, $n \gg \nu$ and this renders resolving many geometric properties of the underlying manifold, including geodesics, computationally intractable. The geodesic equation, from a Hamiltonian dynamics point of view, models a "test particle" moving freely on phase space, which is defined to be the cotangent bundle $T^*\mathcal{M}$. Upon projection onto the *configuration space* $\mathcal{M}$, the motion of this particle traces out a geodesic path with starting position and direction fixed by the initial conditions to the Hamiltonian flow. Simulations of this Hamiltonian flow are difficult to achieve directly since there is no straightforward way to use samples only from configuration space, $\mathcal{M}$, to simulate dynamics that are naturally described in terms of phase space variables, *i.e.*, $\mathbf{x} = (x^1, ...x^\nu)$ and $\mathbf{p} = (p_1, ...p_\nu)$, the canonical configuration (or position) and momentum coordinates, respectively. Another perspective, coming from geometric optics and modeled by the eikonal equation, is that the phase space dynamics simulate the propagation of a light ray in curved space. To obtain geodesics using samples from the manifold the conventional approach involves solving a discrete approximation to the eikonal equation [2], usually after a triangulation step that approximates a surface from the point cloud $X_N$.

Our approach to obtaining geodesics does not require solving a nonlinear equation, or triangulation of the point cloud, and instead requires only simple matrix operations on the dataset $X_N$. We proceed through a *quantization* of the dynamical system, which is a linear (albeit infinite-dimensional) system, and propagation of wavepackets defined on $\mathcal{M}$, but localized to a neighborhood of a point in $T^*\mathcal{M} \setminus 0$ (while still respecting the uncertainty principle) that quantum mechanically approximate the classical test particles. We then assess certain *statistics* of these propagated wavepackets, which constitutes an effective *dequantization* that enables access to the underlying classical – and non-linear! – dynamics, tracing out the geodesic path with initial position and direction fixed by the localization of the initial wavepacket.

In general terms, the quantization we are concerned with connects certain dynamics of operators on $L^2(\mathcal{M})$ with Hamiltonian mechanics, or in the functional form, Liouvillian dynamics on $T^*\mathcal{M}$. The operator dynamics are generated by a linear operator $\hat{Q}_h : C^\infty(\mathcal{M}) \to C^\infty(\mathcal{M})$ with parameter $h \in (0, h_0]$ for some $h_0 > 0$ and driven by the operator solution $\hat{U}_t$ to $h \partial_t \hat{U}_t = -i \hat{Q}_h \hat{U}_t$ with initial condition, $\hat{U}_{t=0} = \hat{I}$. Then, the evolution of $\hat{A} : L^2(\mathcal{M}) \to L^2(\mathcal{M})$ is given by $\hat{A}_t := \hat{U}_t^\dagger \hat{A} \hat{U}_t$, which solves $\partial_t \hat{A}_t = \frac{i}{h}[\hat{Q}_h, \hat{A}_t]$ with initial condition $\hat{A}_{t=0} = \hat{A}$. To bring this to *classical dynamics* on phase space, a connective link is Egorov's theorem [3], which states roughly that if $\hat{Q}_h$ and $\hat{A}$ are the *quantizations* of $q, a \in C^\infty(T^*\mathcal{M})$, respectively satisfying certain properties, then $\hat{A}_t$ is the quantization of $a \circ \Phi_q^t$ with $\Phi_q^t$ the Hamiltonian flow given by $q$. A graphical description of our approach is shown in Fig. 1.

This first step in our approach requires the notion of *quantization*: as alluded to, this is a map that takes certain classes of smooth functions on phase space, called *symbols*, to linear operators. The resulting operators are then called *pseudodifferential operators* (or shortly, $\Psi$DOs). We refer to [3] for details, but to give a sense of symbols and $\Psi$DOs, consider the situation in $\mathbb{R}^d$. On identifying $T^*\mathbb{R}^d \cong \mathbb{R}^{2d}$ for the purposes of analogy, consider a symbol $a \in C^\infty(\mathbb{R}^{2d} \times [0, h_0))$ for some $h_0 > 0$ that is Schwartz class in $\mathbb{R}^{2d}$. Then, its *quantization* is any of the *continuous maps*

$$\mathrm{Op}_{h,\tau}(a) : \mathscr{S}' \ni u \to \frac{1}{(2\pi h)^d} \int_{\mathbb{R}^{2d}} e^{\frac{i}{h}\langle x-y, \xi \rangle} a(\tau x + (1-\tau)y, \xi; h) u(y) \, dy d\xi \in \mathscr{S} \tag{1}$$

for $\mathscr{S}$ the Schwartz space of functions and $\mathscr{S}'$ the space of tempered distributions on $\mathbb{R}^d$ and $\tau \in [0, 1]$. Particular cases of interest are the *standard quantization* with $\tau = 0$, *Weyl quantization* with $\tau = 1/2$ and *right quantization* with $\tau = 1$. The Schwartz class of symbols is too restrictive, since we lack even the differential operators here upon quantization; to remedy this, one defines a more general *symbol class of order* $(k, m)$, which we denote by $h^k S^m$, to be given by functions $a \in C^\infty(\mathbb{R}^{2d})$ such that for all $\beta, \gamma \in \mathbb{N}_0^d$, there is a $C_{\beta,\gamma} \geq 0$ so that for all $h \in [0, h_0]$ and $(\mathbf{x}, \mathbf{p}) \in \mathbb{R}^{2d}$, $|\partial_\mathbf{x}^\gamma \partial_\mathbf{p}^\beta a| \leq C_{\beta,\gamma} h^{-k} (1 + |\mathbf{p}|^2)^{(m-|\beta|)/2}$. This symbol class is defined by a certain decay of higher order

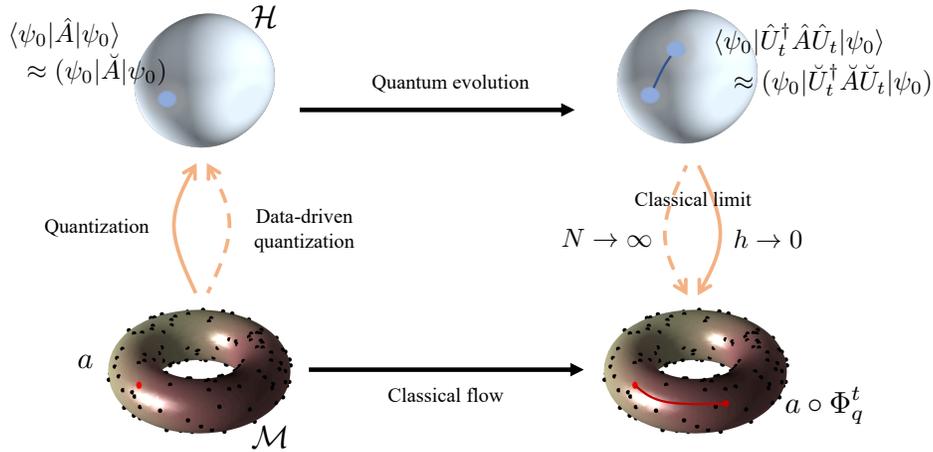

FIG. 1: A schematic representation of our general approach. Through Egorov's theorem a quantum-classical correspondence can be established where the classical flow on a manifold ($\mathcal{M}$) is related to a quantized evolution in Hilbert space ($\mathcal{H}$). Expectations of time-propagated quantized observables, $\langle\psi_0|\hat{U}_t^\dagger \hat{A}\hat{U}_t|\psi_0\rangle$, can recover corresponding classical observables under geodesic flow, $a \circ \Phi_q^t$, in the classical limit, $h \to 0$. We show that quantized operators in Hilbert space can be approximated from sample points on the manifold (black dots on the torus above), and construct discrete quantum propagators ($\check{U}$), states ($|\psi_0\rangle$) and observables ($\check{A}$) to approximate the time-dependent observable expectation by $(\psi_0|\check{U}_t^\dagger \check{A}\check{U}_t|\psi_0)$. This approximation converges to the classical flow in the infinite sample limit, $N \to \infty$.

derivatives with respect to $\mathbf{p}$, which ensures that the symbols have an asymptotic expansion at large $|\mathbf{p}|$ that allows us to distinguish dominant parts of the symbol. The corresponding image of $\text{Op}_{h,\tau}$ is the space of $\Psi$DOs *of order* $(k, m)$, which we write as $h^k \Psi^m$.

The agreement of two $\Psi$DOs up to *a lower order* signifies that the symbol of their residual decays an order faster *both*, in $h \to 0$ and in $|\mathbf{p}| \to \infty$; so *e.g.*, $\Psi$DOs from two different quantizations, *viz.*, different values of $\tau$, of a common symbol are equal up to a lower order $\Psi$DO. Now, indeed, we have the polynomials in the symbol classes and in particular, the Laplacian (up to semi-classical scaling), $h^2\Delta \in h^0\Psi^2$.

A particularly useful concept is that of the *principal symbol* of a $\Psi$DO $\hat{A} \in h^k\Psi^m$, which is the equivalence class $[a_0] \in h^k S^m / h^{k+1} S^{m-1}$ such that for any $a \in [a_0]$, $\hat{A} - \text{Op}_{h,\tau}(a) \in h^{k+1}\Psi^{m-1}$. Roughly, the principal symbol captures the leading order behavior in the limit $h \to 0$, which is the definition of the *classical limit* in the *quantum-classical correspondence*. This can be understood as the operator's *high-frequency* component in the following way: if $\phi \in C^\infty(\mathbb{R}^d)$ has derivative $d\phi|_{\mathbf{x}_0} = -\mathbf{p}_0$ and $q(\mathbf{x}, \mathbf{p}) = \sum_{|\beta|\le D} q_\beta(\mathbf{x}) h^{D-|\beta|} \mathbf{p}^\beta$ is a symbol, then application of its quantization to a *wave-packet* with phase $\phi$ leads to, $e^{-\frac{i}{h}\phi} \text{Op}_{h,\tau}[q(\mathbf{x},\mathbf{p})](e^{\frac{i}{h}\phi})(\mathbf{x}_0) \to \sum_{|\beta|=D} q_\beta(\mathbf{x}_0) \mathbf{p}_0^\beta$ as $h \to 0$. Another key part of the relationship between $\Psi$DOs and symbols is that in as far as $\Psi$DOs commute up to a lower order operator, the principal symbol resulting from a product of $\Psi$DOs is the product of their principal symbols: $\hat{C} := \text{Op}_h(a) \text{Op}_h(b)$ has principal symbol $a_0 b_0$ with $a_0$ and $b_0$ the principal parts of $a$ and $b$, respectively. These properties of the principal symbol of a $\Psi$DO underlie the key relationship in the quantum-classical correspondence: principal symbols are the *classical observables* corresponding to $\Psi$DOs that are interpreted as *quantum observables*.

The semiclassical framework can be ported to $C^\infty$ boundaryless Riemannian manifolds by localizing the definition from Euclidean space to small balls, wherein the manifolds are roughly flat themselves and then patching them together via partitions of unity. A key aspect of this is that the symbol classes $h^k S^m$ are invariant to coordinate transformations [3, §14]. Likewise, the concept of principal symbol can be realized by taking *localized* wavepackets on the manifold and applying $\Psi$DOs to them as was sketched in the previous paragraph for general wavepackets in $\mathbb{R}^d$. This is expected since $\mathcal{M}$ locally resembles $\mathbb{R}^d$. Concretely, we will use *coherent states* to this effect: given $\zeta_0 \in T^*\mathcal{M}$, a *coherent state* localized at $\zeta_0 = (\mathbf{x}_0, \mathbf{p}_0)$ is the function

$$\psi_h^{\zeta_0}(\mathbf{x}) = \frac{1}{h^{\nu/4}} e^{-\frac{i}{h}\overline{\varphi}(\mathbf{x},\zeta_0)} \qquad (2)$$

with $\varphi \in C^\infty(\mathcal{M})$ the restriction to $\zeta_0$ of an *admissible phase* [1] $\phi \in C^\infty(\mathcal{M} \times T^*\mathcal{M})$. This admissibility condition entails roughly that in a neighbourhood of the diagonal $\{(\mathbf{x}; \mathbf{x}, \mathbf{p}) \in \mathcal{M} \times T^*\mathcal{M}\}$, $\phi$ can be approximated in local coordinates by, $\langle p, x-y\rangle_{\mathbb{R}^\nu} + i\|x-y\|_{\mathbb{R}^\nu}^2$. In the context of its interactions with $\Psi$DOs, the *phase space* properties

| Symbol | Description |
|---|---|
| $\mathcal{M}$ | Manifold that data is sampled from. |
| $g$ | Metric of manifold $\mathcal{M}$. |
| $\nu$ | Intrinsic dimension of $\mathcal{M}$. |
| $d_g(\mathbf{x}_1, \mathbf{x}_2)$ | Intrinsic, geodesic distance between points $\mathbf{x}_1, \mathbf{x}_2 \in \mathcal{M}$. |
| $\Delta_g$ | Laplace-Beltrami operator on $\mathcal{M}$. |
| $n$ | Dimension of extrinsic coordinates for points sampled from $\mathcal{M}$. |
| $X_N = \{v_i\}_{i=1}^N$ | Dataset, consisting of $N$ vectors in $v_i \in \mathbb{R}^n$. |
| $\mathsf{p}$ | Sampling distribution dictating sampling from $\mathcal{M}$. |
| $\iota(\mathbf{x})$ | Extrinsic coordinates for a point $\mathbf{x} \in \mathcal{M}$. |
| $\eta^{\mathbf{x}_0}(\mathbf{x})$ | Normal coordinates centered at $\mathbf{x}_0$ for a point $\mathbf{x} \in \mathcal{M}$. |
| $\mathscr{S}$ | Schwartz space of functions on $\mathbb{R}^d$. |
| $\mathscr{S}'$ | Space of tempered distributions on $\mathbb{R}^d$. |
| $\mathbf{x} = (x^1, ...x^\nu), \mathbf{p} = (p_1, ..., p_\nu)$ | Position and momentum phase space variables defined on $T^*\mathcal{M}$. |
| $\zeta = (\mathbf{x}, \mathbf{p}) \in T^*\mathcal{M}$ | A point in phase space. |
| $\Phi_q^t$ | Hamiltonian flow representing time evolution associated with symbol $q$. |
| $a \in C^\infty(T^*\mathcal{M})$ | A symbol representing a physical observable in phase space. |
| $h^k S^m$ | Symbol class of order $(k, m)$. |
| $h^k \Psi^m$ | Space of pseudo-differential operators of order $(k, m)$. |
| $\hat{A}: C^\infty(\mathcal{M}) \to C^\infty(\mathcal{M})$ | Operator representing a quantum observable. |
| $\text{Op}_h(a)$ | Quantization of a symbol $a$ to an operator $\hat{A}$. |
| $\check{A} \in \mathbb{C}^{N \times N}$ | An $N \times N$ complex matrix that depends on the dataset $X_N$. If a corresponding $\hat{A}$ is defined, $\check{A}$ is a finite-dimensional approximation of $\hat{A}$. |
| $\hat{U}_t: C^\infty(\mathcal{M}) \to C^\infty(\mathcal{M})$ | Time evolution map governed by the quantum Hamiltonian $\sqrt{\Delta_g}$: namely, $\hat{U}_t := e^{-it\sqrt{\Delta_g}}$. Denoted $U^t$ in the main text. |
| $\check{\mathcal{L}}_{\epsilon,\lambda} \in \mathbb{R}^{N \times N}$ | Graph Laplacian matrix with scale parameter $\epsilon$ and $\lambda$-renormalization; in terms of the main text, $\Delta_{\epsilon,N} := \check{\mathcal{L}}_{\epsilon,0}$. |
| $\hat{\mathcal{L}}_{\epsilon,\lambda}: C^\infty(\mathcal{M}) \to C^\infty(\mathcal{M})$ | Continuum graph Laplacian operator ($N \to \infty$ limit of $\check{\mathcal{L}}_{\epsilon,\lambda}$). |
| $\hat{V}_t: C^\infty(\mathcal{M}) \to C^\infty(\mathcal{M})$ | Propagator solving Schrödinger equation with generator $\hat{\mathcal{L}}_{\epsilon,\lambda}$: namely, $\hat{V}_t := e^{-it\hat{\mathcal{L}}_{\epsilon,\lambda}}$. |
| $\check{V}_t \in \mathbb{C}^{N \times N}$ | Matrix approximation to $\hat{V}_t$: namely, $\check{V}_t := e^{-it\check{\mathcal{L}}_{\epsilon,\lambda}}$. Denoted $U_{\epsilon,N}^t$ in the main text. |
| $|\psi\rangle \in C^\infty(\mathcal{M})$ | An (infinite-dimensional) quantum state. Denoted $\psi_h$ in the main text. |
| $|\psi_{\zeta_0}^h\rangle \in C^\infty(\mathcal{M})$ | A coherent state centered at phase space point $\zeta_0$. Denoted $\psi_h^\zeta$ in the main text. |
| $|\psi\rangle \in \mathbb{C}^N$ | An $N$-dimensional complex vector that depends on the dataset $X_N$. If a corresponding $|\psi\rangle$ is defined, $|\psi\rangle$ is the restriction of $|\psi\rangle$ to $X_N$. Sometimes we will abuse this notation and denote both the vector and a $C^\infty$ extension by $|\psi\rangle$. Denoted $[\psi_h]$ in the main text. |
| $\text{inj}(\mathbf{x}_0)$ | Injectivity radius of point $\mathbf{x}_0 \in \mathcal{M}$. |
| $\tilde{\nu}$ | Dimension for local principal component analysis. |
| $\check{O}: \mathbb{R}^n \to \mathbb{R}^{\tilde{\nu}}$ | Local principal component analysis (LPCA) linear map |
| $\vartheta_\ell^{v_0} \in \mathbb{R}^{\tilde{\nu}}$ | LPCA coordinates for data point $\ell$, with LPCA performed at point $v_0$. |
| $\epsilon > 0$ | Scale parameter dictating neighborhood size for sampled data points on $\mathcal{M}$. |
| $0 \leq \lambda \leq 1$ | Normalization parameter dictating convergence properties of data-driven Markov generator. |
| $\alpha > 0$ | Scale parameter dictating uncertainty of coherent states constructed from dataset $X_N$. |
| $\Delta t > 0$ | Time step taken by quantum propagator. |
| $n_{\text{prop}} \in \mathbb{Z}$ | Number of $\Delta t$ time steps to propagate. |
| $n_{\text{coll}} \in \mathbb{Z}$ | Number of initial states to propagate. |
| $\delta_{\text{PCA}} > 0$ | Radius of Euclidean ball that dictates the samples points over which LPCA is performed. |

TABLE I: Notation used in this SM

of this function can be understood from its *Husimi function*, as described in [1]. These states satisfy the properties of coherent states usually encountered in physics [4]; *i.e.*, they have minimum uncertainty volume in quantum phase space, satisfying the Heisenberg uncertainty principle, which is equally distributed between position and momentum degrees of freedom, defined through the $\text{Op}_h(\cdot)$ quantization.

Connecting the semiclassical framework and Egorov's theorem with coherent states, the form of quantum-classical correspondence we will utilize is given by,

**Theorem 1** (Ref. [1]). *Let $\hat{Q} = \text{Op}_h(q) \in h^0 \Psi^{-\infty} := \cap_{m \in \mathbb{Z}} h^0 \Psi^m$ with $q \in C_c^\infty(T^*\mathcal{M})$ having real principal symbol $q_0$ and $a \in h^0 S^m$ over $h \in (0, h_0]$ for some $h_0 > 0$. Then, given a coherent state $\psi_h$ localized at $\zeta_0 \in T^*\mathcal{M}$ and a fixed $T \geq 0$, we have for all $h \in (0, h_0]$ and $|t| \leq T$,*

$$|\langle \psi_h | e^{\frac{i}{\hbar} \hat{Q} t} \text{Op}_h(a) e^{-\frac{i}{\hbar} \hat{Q} t} | \psi_h \rangle - a \circ \Phi_{q_0}^t(\zeta_0)| = \mathcal{O}(h), \tag{3}$$

*where $\Phi_{q_0}^t$ is the Hamiltonian flow on $T^*\mathcal{M}$ generated by $q_0$.*

We often call $\hat{Q}$ the *quantized Hamiltonian*, or even just the *Hamiltonian* when the context is clear with regards to the classical case, in as far as it generates the dynamics on *quantum observables* $\hat{A} \in h^0 \Psi^m$ through the propagator $\hat{U}_t := e^{-\frac{i}{\hbar} \hat{Q} t}$. The expression Eq. (3) states that the time evolution of any observable of a quantized system approximates the classical Hamiltonian flow of the corresponding classical observable generated by the principal symbol of the quantized Hamiltonian $\hat{Q}$. This approximation improves as $h \to 0$. The validity of the relationship between the time evolution of the quantized observable and the related classical flow given by Eq. (3) is only valid until a fixed amount of time $T > 0$. This suffices for our application, since we will propagate only for a time smaller than the *injectivity radius* at a given point on $\mathcal{M}$.

The statement in Eq. (3) can be roughly understood by expanding the propagated quantum observable by application of the Baker-Campbell-Hausdorff formula:

$$\hat{A}(t) = e^{\frac{i}{\hbar} \hat{Q} t} \hat{A} e^{-\frac{i}{\hbar} \hat{Q} t} = \hat{A} + \frac{it}{\hbar}[\hat{Q}, \hat{A}] - \frac{t^2}{2\hbar^2}[\hat{Q}, [\hat{Q}, \hat{A}]] + \ldots, \tag{4}$$

with $[\hat{Q}, \hat{B}] \in h^{k+1} \Psi^{m-1}$ having principal symbol $\frac{h}{i}\{q_0, b\}$, whenever $\hat{B} = \text{Op}_h(b) \in h^k \Psi^m$ due to the properties of quantization discussed above. If we dequantize this operator to recover the corresponding propagated symbol, then the principal part is given by the series $a + t\{q_0, a\} + \frac{t^2}{2}\{q_0, \{q_0, a\}\} + \cdots$, which agrees with the Taylor series expansion at $t = 0$ of $a \circ \Phi^t$. This formal computation indicates the correspondence, given in Eq. (3), between the quantum propagation of quantum observables and the classical flow of classical observables and from this perspective, it is clear that even if $a$ has no dependence on $h$, the symbol at $t > 0$ will due to the sub-principal $O(h)$ terms in this expansion. These are quantum corrections that decay as $h \to 0$ to yield the flow of the classical observable, $a \circ \Phi_{q_0}^t$.

### II.C. Quantization for geometric data anaysis

We recap some notation that is heavily used in the following. Finite-dimensional, data-dependent representations or approximations of operators, $\hat{A}$, and states, $|\psi\rangle$, are denoted $\breve{A}$ and $|\psi)$, respectively; $\breve{A} \in \mathbb{C}^{N \times N}$ and $|\psi) \in \mathbb{C}^N$ for some dimension $N$. Expectation values for finite-dimensional approximations of quantum observables are denoted $\langle \breve{A} \rangle := (\psi|\breve{A}|\psi)$. $\breve{A}$ and $|\psi)$ will be represented in the position basis, unless otherwise specified. Finally, we introduce notation for a global parameterization of points on the manifold as $\iota(\mathbf{x}) = (\iota_1(\mathbf{x}), ..., \iota_n(\mathbf{x}))$.

For geometric data analysis learning, a natural choice for the quantized Hamiltonian is $\hat{H}' = h^2 \Delta_g$, where $\Delta_g$ is the LB operator on $\mathcal{M}$, since the principal symbol for this operator is $\mathcal{H}' = |\mathbf{p}|_g^2 = \sum_{i,j} g^{ij} p_i p_j$, which is the classical Hamiltonian for free motion on the manifold. Thus, the corresponding flow $\Phi_{\mathcal{H}'}^t$ projected onto $\mathcal{M}$ transports a particle along geodesics, however its *speed* depends on the initial momentum magnitude. On the other hand, if we demand that the quantized Hamiltonian for the system take the form

$$\hat{H} = h\sqrt{\Delta_g}, \tag{5}$$

then the corresponding classical Hamiltonian is $\mathcal{H} := |\mathbf{p}|_g$ and its flow, which we denote by $\Phi_t$, projected onto $\mathcal{M}$ transports particles along *unit speed* geodesics, regardless of the initial (non-zero) momentum. Motivated by recovering intrinsic distances and the exponential map and in support of further applications where this unit-speed regularization is fruitful, it is this flow we wish to approximate.

The form of quantum-classical correspondence given by Theorem 1 is particularly suited to data-driven approximations of the LB operator of a manifold. A quantum system governed by a Hamiltonian $\hat{Q}$ whose principal symbol $q_0$ is

localized in phase space and such that $q_0(\mathbf{x}, \mathbf{p}) \equiv |\mathbf{p}|_{g_\mathbf{x}}$ in a neighbourhood of $(\mathbf{x}_0, \mathbf{p}_0) \in T^*\mathcal{M}$ is immediately linked to the geodesic flow $\Phi_t(\mathbf{x}_0, \mathbf{p}_0)$ through Eq. (3). Diffusion operators related to *graph Laplacians* now have a good probabilistic convergence theory from finite samples and we rely on the results of [1] that convert them to propagators $\hat{U}_{q_0}^t := e^{\frac{i}{\hbar}\hat{Q}t}$. We now set the stage for implementing this procedure.

Consider the matrix

$$\breve{L}_{\epsilon,\lambda} := \frac{2c_0}{c_2 \epsilon}\left( \breve{I}_N - \breve{D}_{\epsilon,\lambda}^{-1}\breve{\Sigma}_\epsilon^{-\lambda}\breve{T}_\epsilon\breve{\Sigma}_\epsilon^{-\lambda} \right), \qquad (6)$$

where $[\breve{T}_\epsilon]_{i,j} = K(\|v_i - v_j\|^2/2\epsilon)$, for $K : \mathbb{R} \to \mathbb{R}$ is a smooth function that exponentially decays in its argument. The diagonal matrices $[\breve{\Sigma}_\epsilon]_{i,i} = \sum_{j=1}^N [\breve{T}_\epsilon]_{i,j}$ and $[\breve{D}_{\epsilon,\lambda}]_{i,i} = \sum_{j=1}^N [\breve{\Sigma}_\epsilon^{-\lambda}\breve{T}_\epsilon\breve{\Sigma}_\epsilon^{-\lambda}]_{i,j}$ are normalizations and $c_j = \int_{\mathbb{R}^n} K(\|y\|^2)y_1^j dy$ for $j = 0, 2$ are constants that are trivial to compute upon a choice for $K$. The scale parameter $\epsilon > 0$ regulates the order of approximation beyond the infinite sample limit and its optimal choice depends on the density of the samples $X_N$ on $\mathcal{M}$. While there is no universal choice for $\epsilon$ at a given sample size, the asymptotic rates of convergence in Proposition 2 lend a rule-of-thumb. As the construction of our propagator and its discrete approximation are rooted in diffusion processes, the intuitions for the $\epsilon$ parameter carry over from that literature [5]: one attempts to assign a value such that a $\sqrt{\epsilon}$-ball captures the local neighborhood of any data point. Thus, $\epsilon$ is a measure of degree of uncertainty or resolution in configuration space data, and below we will show how it is intimately related to the fundamental phase space uncertainty $h$ imposed by quantum mechanics. Finally, the $\lambda$ parameter that influences the asymptotic (in $N$) convergence properties of $\breve{L}_{\epsilon,\lambda}$, which we now discuss. Note that in the main text we presented the $\lambda = 0$ version of $\breve{L}_{\epsilon,\lambda}$, and it was referred to there as $\Delta_{\epsilon,N}$

The asymptotic properties of $\breve{L}_{\epsilon,\lambda}$ have been thoroughly characterized in recent years, including the important convergence result [6] that $\breve{L}_{\epsilon,\lambda} \xrightarrow{N \to \infty} \hat{\mathcal{L}}_{\epsilon,\lambda}$, where

$$\hat{\mathcal{L}}_{\epsilon,\lambda} = \left( \Delta_g + \frac{2(1-\lambda)}{\mathsf{p}}\langle \nabla \mathsf{p}, \nabla \rangle_g \right) + \mathcal{O}(\epsilon). \qquad (7)$$

The leading term in this operator is the LB operator on the manifold, except for a correction term that depends on the sampling density $\mathsf{p}$. The correction term is zero under uniform sampling, and its influence can be tuned by choice of normalization parameter $\lambda$. While the choice $\lambda = 1$ asymptotically removes the dependence on sampling density, this is not necessary for our approach because Theorem 1 only depends on the principal symbol of the Hamiltonian generating the dynamics and these sampling density dependent terms enter as lower-order contributions to the symbol for $h^2\hat{\mathcal{L}}_{\epsilon,\lambda}$. Nevertheless, we leave this option for regularization open.

Although $\hat{\mathcal{L}}_{\epsilon,\lambda}$ approximates the LB operator on the manifold as $\epsilon \to 0$, this approximation is subtle from a quantization perspective since the $\mathcal{O}(\epsilon)$ terms include differential operators of increasing order. Indeed, since the basis of the approximation (Eq. (7)) is a first-order difference approximation to the infinitesimal of a diffusion operator in the parameter $\epsilon$, the corresponding *symbol*, once viewed from the quantization perspective, only locally approximates $|\mathbf{p}|_g^2$ in phase space [1]. The domain and quality of approximation grow as $\epsilon \to 0$ and in order to apply semiclassical methods, this must increase at a certain rate with $h \to 0$ as well. Further, to use this approximation to obtain quantum dynamics with a well-characterized classical limit, we must view it as applied to coherent states, which as we discussed in the previous section are wave-packets localized in phase space. The action of the asymptotic operator $\hat{\mathcal{L}}_{\epsilon,\lambda}$ on a coherent state in fact uncovers this crucial relationship between $\epsilon$ and $h$, as we now see.

**Proposition 1.** *Let $\alpha \geq 1$, $h > 0$, $\zeta_0 := (\mathbf{x}_0, \mathbf{p}_0) \in T^*\mathcal{M}$ and let $\psi_{\zeta_0}^h$ be a coherent state localized at $\zeta_0$. Then with $\epsilon := h^{(2+\alpha)}$ we have,*

$$(h^2\hat{\mathcal{L}}_{\epsilon,\lambda})\psi_{\zeta_0}^h(\mathbf{x}_0) = |\mathbf{p}_0|_{g(\mathbf{x}_0)}^2 \psi_{\zeta_0}^h(\mathbf{x}_0) + \mathcal{O}(h). \qquad (8)$$

*Proof.* The expansion of $\hat{\mathcal{L}}_{\epsilon,\lambda}$ is pointwise a Taylor series and can be found in [6] to be, on application to $u \in C^\infty$ in local coordinates,

$$\hat{\mathcal{L}}_{\epsilon,\lambda}[u](\mathbf{x}) = \left( \Delta_g + \frac{2(1-\lambda)}{\mathsf{p}}\langle \nabla \mathsf{p}, \nabla \cdot \rangle_{g_\mathbf{x}} \right)[u](\mathbf{x}) + \sum_{j=2}^\infty \frac{\epsilon^{j-1}}{j!}G_j(\mathbf{x}, D)[u](\mathbf{x}), \qquad (9)$$

with each $G_j$ a partial differential operator (PDO) of order $2j$ with smooth coefficients in the variable $\mathbf{x}$. The issue with applying this to a coherent state is that we must let $h \to 0$ while also having $\epsilon \to 0$, but the applications of the

PDOs to $\psi_{\zeta_0}^h$ bring about terms of negative orders in $h$, which may lead to the effect of higher order PDOs becoming asymptotically dominant. To curtail this, we must scale $h$ appropriately according to $\epsilon$ and also rescale $\hat{\mathcal{L}}_{\epsilon,\lambda}$ (in a manner common in semiclassical analysis). For generality, we let $\epsilon := \epsilon(h)$ be a function of $h > 0$.

On letting $s(z) := \exp_{\mathbf{x}}(z)$ provide normal coordinates, we find that on application to a coherent state,

$$\hat{\mathcal{L}}_{\epsilon\lambda}[\psi_{\zeta_0}^h](\mathbf{x}) = h^{-2}\psi_{\zeta_0}^h(\mathbf{x})\langle \nabla_z, \nabla_z\rangle\big|_{z=0}\phi \circ s + \mathcal{O}(h^{-1})\psi_{\zeta_0}^h(\mathbf{x}) + \sum_{j=2}^{\infty}\frac{\epsilon^{j-1}}{j!}\mathcal{O}(h^{-2j}),$$

wherein $\phi$ denotes the complex phase of $\psi_{\zeta_0}^h = e^{\frac{i}{h}\phi(\mathbf{x};\zeta_0)}$ and $\mathcal{O}(h^{-k})$ denotes a polynomial in $1/h$ of degree $k$ with coefficients depending only on derivatives of $\phi$ and the smooth functions appearing in the coefficients of the operators $G_j$. We wish to recover the dominant term of the semi-classical application

$$h^2 \Delta_g[\psi_{\zeta_0}^h](\mathbf{x}) = \psi_{\zeta_0}^h(\mathbf{x})\langle\nabla_z,\nabla_z\rangle\big|_{z=0}\phi\circ s + \mathcal{O}(h)$$
$$= |d\phi|_{\mathbf{x}}^2 \psi_{\zeta_0}^h(\mathbf{x}) + \mathcal{O}(h),$$

which follows from $ds|_{z=0} = I$, as the dominant term from the $\hat{\mathcal{L}}_{\epsilon,\lambda}$ approximation. To do so, we must balance the asymptotic scaling in the terms of higher order: namely, on setting $\epsilon \in \mathcal{O}(h^{2+\beta})$ with $\beta \geq 0$, then we have $\epsilon^{j-1}\mathcal{O}(h^{-2(j-1)}) \in \mathcal{O}(h^\beta)$ for all $j \geq 2$, so that

$$(h^2 \hat{\mathcal{L}}_{\epsilon,\lambda})[\psi_{\zeta_0}^h](\mathbf{x}_0) = (h^2 \Delta_g)\psi_{\zeta_0}^h(\mathbf{x}_0) + \mathcal{O}(h) + \mathcal{O}(h^\beta)$$
$$= |\mathbf{p}_0|_{g(\mathbf{x}_0)}^2\psi_{\zeta_0}^h(\mathbf{x}_0) + \mathcal{O}(h) + \mathcal{O}(h^\beta),$$

since $\phi(\mathbf{x}_0;\zeta_0) = 0$ and $d\phi|_{\mathbf{x}=\mathbf{x}_0} = -\mathbf{p}_0$ by definition of the admissibility condition on $\phi$. If now we use $\epsilon = h^{2+\alpha}$ with $\alpha \geq 1$, then the remainder term is just $\mathcal{O}(h)$ as in the statement of the Proposition. □

Eq. (8) states that, in the sense of the discussion in the previous section, the *high-frequency data* in $h^2\hat{\mathcal{L}}_{\epsilon,\lambda}$ is the symbol of the semiclassical LB operator, $\hat{H}'$ when the *configuration space resolution parameter* $\epsilon$ decays as $h^{2+\alpha}$ for some $\alpha \geq 1$ with respect to the *phase space resolution parameter* $h$. In other words, $h^2\hat{\mathcal{L}}_{\epsilon,\lambda}$ acting on a coherent state approximates the action of the semiclassical LB operator on that coherent state, to order $O(h)$.

In [1], $h^2\hat{\mathcal{L}}_{\epsilon,\lambda}$ is shown to be a $\Psi$DO with this scaling of $\epsilon$ and $h$ and then, the action of Heisenberg dynamics governed by $\hat{V}_t := e^{-it\sqrt{\hat{\mathcal{L}}_{\epsilon,\lambda}}}$ on coherent states is reduced to the form satisfying Theorem 1. In practice we approximate expectations such as those on the left hand side of Eq. (3) by discrete inner products that are influenced by the sampling density $\mathsf{p}$. Therefore, we must account for this density as a multiplier here in the asymptotic regime. Altogether, we have the quantum-correspondence principle for graph Laplacians, in the asymptotic (continuum) limit:

**Theorem 2** (Ref. [1]). *Let $\hat{A}$ be an observable for a quantized system and $a \in h^0 S^0$ its corresponding symbol. Let the quantized system have Hamiltonian $\hat{G} = h\sqrt{\hat{\mathcal{L}}_{\epsilon,\lambda}}$, with $h = \epsilon^{1/(2+\alpha)}$ for some $\alpha \geq 1$. The corresponding propagator is $\hat{V}_t = e^{-\frac{i}{h}\hat{G}t}$. Let $\Phi_t$ denote the classical Hamiltonian flow generated by the principal symbol of $\hat{H}$: $\mathcal{H} = |\mathbf{p}|_g$ (i.e., the homogeneous geodesic flow) and let $|\psi_{\zeta_0}^h\rangle$ be a normalized coherent state localized at $\zeta_0 = (\mathbf{x}_0, \mathbf{p}_0)$. Then, there is an $h_0 > 0$ such that for all $h \in (0, h_0]$ and $|t|$ less than the injectivity radius at $\mathbf{x}_0$,*

$$\left|\langle\psi_{t,\zeta_0}|\hat{\mathsf{p}}\hat{A}|\psi_{t,\zeta_0}\rangle - a\circ\Phi_t(\mathbf{x}_0,\mathbf{p}_0)\right| = \mathcal{O}(h), \tag{10}$$

*where $|\psi_{t,\zeta_0}\rangle := \hat{V}_t|\psi_{\zeta_0}^h\rangle/\sqrt{\langle\psi_{\zeta_0}^h|\hat{V}_t^\dagger\hat{\mathsf{p}}\hat{V}_t|\psi_{\zeta_0}^h\rangle}$ is the normalized propagated state, and $\hat{\mathsf{p}}$ is a diagonal operator representing the sampling density $\mathsf{p}$.*

This theorem tells us that if we propagate coherent states for $|t| < \mathrm{inj}(\mathbf{x}_0)$ with the asymptotic data-driven propagator, we can approximate observables of the classical Hamiltonian flow, and this approximation improves as $h \to 0$ ($\epsilon \to 0$). Any features of the classical flow can be reconstructed by suitable choices for $\hat{A}$. This result is reminiscent of a semiclassical expansion of quantum dynamics around a classical trajectory; *i.e.*, for coherent state initial states and short times, time-propagated quantum observables are equal to corresponding classical observables, with $\mathcal{O}(h)$ quantum corrections. Although such expansions are known in specialized settings [7, 8], our result is (i) specific to the manifold learning context, where the generator of evolution is the data-driven approximation $\hat{G}$, and (ii) general in the sense that it applies to dynamics on any Riemannian manifold *satisfying the assumptions of* Section II.B. In practice, we define coordinates on the manifold, $u: \mathcal{M} \to \mathbb{R}^m$, and compute $\langle\hat{u}_j\rangle := \langle\psi_{t,\zeta_0}|\hat{u}_j\hat{\mathsf{p}}|\psi_{t,\zeta_0}\rangle = \int_{\mathcal{M}} u_j(\mathbf{x})|\psi_{t,\zeta_0}(\mathbf{x})|^2 \mathsf{p}(\mathbf{x})d\mathbf{x}$, where $\hat{u}_j = \mathrm{Op}_h(u_j)$, which as an operator is simply multiplication by $u_j$. The equality follows from the fact that $\hat{u}_j$ and $\hat{\mathsf{p}}$ are diagonal operators in position basis.

### II.C.1. The role of $h$ in manifold learning

What is the role of $h$ in the quantum dynamics formulation of manifold learning? This was discussed in the main text and we present another perspective here. Unlike in physical theories, $h$ is not a constant but a function of the sampled dataset $X_N$. According to the proof of Proposition 1, setting $h \in \omega(\sqrt{\epsilon})$, for $\epsilon$ the scale parameter in the graph Laplacian construction of the heat kernel, results in a data-driven propagator that approximates the action of the LB operator on coherent states. This $h(\epsilon)$ relation is intended to set the quantization scale of the quantum system to the resolution of the manifold provided by the sampled data $X_N$. The rationale for this is motivated by the following: the graph Laplacian construction approximates a diffusion process that is discretized to $\sqrt{\epsilon}$ balls on the manifold. This process determines the smallest scale to which we can resolve the manifold and in doing so, uses all of the available resolution provided by the sampled data on configuration space. With our choice of coherent states as initial states, we redistribute this resolution equally between the position and momentum degrees of freedom. Therefore, following the uncertainty principle, we must set the minimal phase space resolution for the quantum dynamical system to be above this, i.e., $h = \omega(\sqrt{\epsilon})$. From this perspective, the $h \to 0$ limit is both the classical limit of the quantum dynamics and also the limit where the sampled data fully covers the manifold. The practice we have opted for here is to set $h = \epsilon^{1/(2+\alpha)}$ with $\alpha \geq 1$ as a way to balance errors that arise from coherent state delocalization and from approximating the dynamics, as discussed in the proof of Proposition 1.

Another form of the uncertainty principle [9] dictates that the restriction of a $\Psi$DO to a region of phase space constitutes its restriction to a portion of the spectrum. The $h$ parameter thus simultaneously plays the role of setting an implicit *spectral resolution* scale: due to the localization of a coherent state, applying a $\Psi$DO to it effectively restricts this action to a portion of the spectrum governed by the momentum and the scale $h$. In the present context, [1, Sec. 4] shows that coherent states are implicitly driven, to $\mathcal{O}(h)$ error, by spectral truncations of the propagator $\hat{V}_t$. That is, $|\langle \psi_{t,\zeta_0}|\hat{A}|\psi_{t,\zeta_0}\rangle - \langle \psi_{t,\zeta_0,\sigma}|\hat{A}|\psi_{t,\zeta_0,\sigma}\rangle| = \mathcal{O}(h)$ with $|\psi_{t,\zeta_0,\sigma}\rangle = \Pi_\sigma \hat{V}_t \Pi_\sigma \psi_{t,\zeta_0}$ and $\Pi_\sigma$ the spectral projector onto the eigenspaces of $\mathcal{L}_{\epsilon,\lambda}$ with eigenvalues at most $\sigma \sim |\mathbf{p}_0|^2_{g(\mathbf{x}_0)}/h^2$. Thus, the coherent state evolves according to an energy scale set by $|\mathbf{p}_0|_{g(\mathbf{x}_0)}$ and $h$.

### II.D. Data-driven approximation of quantum dynamics

This section outlines how the program to extract geodesic distances through quantum propagation described in the previous section can be realized through approximation of the quantum propagator, initial states, and measurements using the point cloud dataset, $X_N$. Pseudocode describing the procedures in this section is given in Section VII.

### II.D.1. Propagator

In view of Theorem 2, we form the matrix $\breve{G} = \sqrt{h^2 \breve{L}_{\epsilon,\lambda}}$ with $h = \epsilon^{1/(2+\alpha)}$ for some $\alpha \geq 1$, and then the data-driven approximation of the quantum propagator $\hat{V}_t$ over a time $t$ is formed as $\breve{V}_t = e^{-\frac{i}{h}\breve{G}t}$. This matrix $\breve{V}_t$ is referred to as $U_{\epsilon,N}^t$ in the main text. $\breve{L}_{\epsilon,\lambda}$ is not a symmetric matrix, and therefore in order to compute the above functions of it, we perform a similarity transform of $\breve{L}_{\epsilon,\lambda}$ to a symmetrized version $\breve{D}_{\epsilon,\lambda}^{1/2} \breve{L}_{\epsilon,\lambda} \breve{D}_{\epsilon,\lambda}^{-1/2}$, perform the required computations spectrally and invert this similarity transformation afterwards.

Note that $\breve{G}$ is an approximation of the Hamiltonian in the position basis of the quantized dynamical system. We can then propagate initial states expressed in the position basis, $|\psi_0\rangle$, as usual, $|\psi_t\rangle = \breve{V}_t|\psi_0\rangle$. Integrating classical dynamics to get time evolved quantities requires working with phase space variables, both positions and momenta. In contrast, by quantizing we are able to work entirely in the position basis, where the data is sampled, an obvious advantage of quantization for our manifold learning application.

### II.D.2. Initial states

Using the samples from $\mathcal{M}$ in $X_N$ we will construct approximations of the phase space localized coherent states, in Eq. (2), when the expected position $\mathbf{x}_0$ corresponds to a sample point $v_0 := \iota(\mathbf{x}_0) \in V$. The definition of a coherent state requires only satisfying the *admissibilty* conditions given in [1] for its phase and as discussed there, such a state can be readily prepared by using extrinsic coordinates or a local coordinate patch about $\mathbf{x}_0$. In practice, the prescribed phase may have a differential vector lying outside of the tangent space of $\iota(\mathbf{x}_0)$. Nevertheless, the procedures that follow give a phase sufficiently close to admissible *with high probability* that this issue can be circumvented.

Since $v_i$ represents a point $i$ in a set of global, extrinsic coordinates, an approximation of a coherent state with expected position $\mathbf{x}_0$ is given by a vector with elements

$$\left[|\tilde{\psi}_{\zeta_0}^h\rangle\right]_\ell = e^{\frac{i}{h}(v_\ell - v_0)^\mathsf{T} p_0} e^{-\frac{\|v_\ell - v_0\|^2}{2h}}, \quad 1 \leq \ell \leq N, \tag{11}$$

wherein $p_0 = (v_n - v_0)/\|v_n - v_0\| \in \mathbb{R}^n$ is a vector in the direction of $v_n$, which is one of any fixed number $n_{\text{coll}}$ nearest neighbors of $v_0$. Here, the uncertainty parameter $h$ is determined by the choice of $\epsilon$ used in the data-driven construction of the propagator. This is a finite-dimensional approximation of $|\tilde{\psi}_{\tilde{\zeta}_0}^h\rangle := |e^{\frac{i}{h}\varphi(\cdot;\tilde{\zeta}_0)}\rangle$, with $\tilde{\zeta}_0 = (\mathbf{x}_0 = \iota^{-1}(v_0), \mathbf{p}_0)$ and wherein $p_0$ approximates $\mathbf{p}_0$, the tangent direction between $v_n$ and $v_0$. We denote the normalized version of $|\tilde{\psi}_{\zeta_0}^h\rangle$ by $|\psi_{\zeta_0}^h\rangle$.

Alternatively, one can also form a coherent state using local information derived through a data-driven approximation of the tangent space at $\mathbf{x}_0$. An approach to this is the local principal component analysis procedure (LPCA) as described in [10], which as shown there, yields an approximation to the tangent space up to second order in the size of the local neighborhood[1], *with high probability*. More specifically, the input to LPCA is the set $X_{0,N} \subset X_N$ of data points contained in the ball of radius $\sqrt{\varepsilon}_{\text{PCA}}$ centered at the point $v_0 := \iota(\mathbf{x}_0)$ and its ouput is a linear transformation $\breve{O} : \mathbb{R}^n \to \mathbb{R}^{\tilde{\nu}}$ that maps points in the extrinsic space to their approximate coordinates in the tangent space at $v_0$, with approximate dimension $\tilde{\nu}$. In accordance with [10, Theorem B.1], using a small neighborhood size $\sqrt{\varepsilon}_{\text{PCA}} =: \sqrt{\gamma \delta}_{\text{PCA}} \in \omega(N^{-\frac{2}{d+2}})$ with $\gamma \sim \varepsilon_{\text{PCA}}$ and $\delta_{\text{PCA}} = O(1)$ to construct $\breve{O}$ yields w.h.p. an approximation of tangent space coordinates to $O(\varepsilon_{\text{PCA}})$ error for samples within the $\sqrt{\delta}_{\text{PCA}}$-ball about $v_0$, namely $\mathcal{S}_0 := \{\ell \mid \|v_0 - v_\ell\|^2 \leq \delta_{\text{PCA}}\}$.

The dimension $\tilde{\nu}$ of the approximated tangent space is dictated by the decay of singular values of the LPCA covariance matrix, or can be set if the manifold dimension is known *a priori*. Let $\vartheta_\ell^{v_0} \in \mathbb{R}^{\tilde{\nu}}$ be the LPCA coordinates for all $v_\ell$ with $\ell \in \mathcal{S}_0$. Once the LPCA approximation to the tangent space is formed, the initial momentum can be chosen to be a vector in this space. Furthermore, we approximate geodesic distances with the Euclidean distances in LPCA space, and the inner product between tangent vectors that determines the phase of a coherent state is approximated by a Euclidean inner product.

The LPCA procedure gives rise to the following initial state:

$$\left[|\tilde{\psi}_{\zeta_0}^h\rangle\right]_\ell = \begin{cases} e^{\frac{i}{h}(\vartheta_\ell^{v_0} - \vartheta_0^{v_0})^\mathsf{T} p_0} e^{-\frac{\|\vartheta_\ell^{v_0} - \vartheta_0^{v_0}\|^2}{2h}}, & \text{if } \ell \in \mathcal{S}_0 \\ 0, & \text{otherwise,} \end{cases} \tag{12}$$

with $p_0$ a unit vector in $\mathbb{R}^{\tilde{\nu}}$. Altogether, this is a finite-dimensional, unnormalized LPCA approximation of $|\psi_{\tilde{\zeta}_0}^h\rangle$, with $\tilde{\zeta}_0 = (\mathbf{x}_0 = \iota^{-1}(v_0), \mathbf{p}_0)$, and $p_0$ approximating $\mathbf{p}_0$ which is an orthogonal projection of $p_0$ onto the tangent space at $v_0$.

As before, the uncertainty parameter $h$ is determined by the choice of $\epsilon$ used in the data-driven construction of the propagator, and we denote the normalized version of this vector by $|\psi_{\zeta_0}^h\rangle$. The propagation speed of this state is one due to the normalization of the momentum vector, and hence we often state distances as times in the following. An important point to note is that the formation of this initial state vector does not require computations with the original (possibly high-dimensional) samples.

### II.D.3. Measurements to extract geodesics

We now use the data-driven and computationally tractable initial states just described, as surrogates for coherent states, to approximate the geodesic flow through the *expected position* of the discrete propagation $|\psi_{t,\zeta_0}^h\rangle := \breve{V}_t|\psi_{t,\zeta_0}\rangle/\|\breve{V}_t|\psi_{t,\zeta_0}^h\rangle\|$. The notion of *expected position* is dependent on a choice of atlas (or local coordinates) for $\mathcal{M}$ and perhaps the most direct route to the expected position is through the global extrinsic coordinates (in a larger embedding space) provided by $\iota : \mathcal{M} \to \mathbb{R}^n$, which gives for each sample the coordinates $v_\ell = \iota(\mathbf{x}_\ell)$. A given smooth coordinate mapping is in fact a vector of quantum observables, for the quantizations of the coordinate functions are (up to lower-order symbols) just order zero $\Psi$DOs given by diagonal operators of multiplication by the corresponding functions themselves. Therefore, in view of Theorem 2, we can follow geodesics on $\mathcal{M}$, up to $O(h)$ errors, through

---

[1] An application of this feature is that by preparing the state in LPCA coordinates at $\mathbf{x}_0$ and propagating to time $|t| < \text{inj}(\mathbf{x}_0)$, one can extend the approximation of normal coordinates from an $O(\sqrt{h})$ neighborhood to the whole of the sampled data within an $O(1)$ radius ball inside the normal neighbourhood at $\mathbf{x}_0$.

these coordinates by the dynamics $\bar{\mathbf{x}}_t := \langle \psi^h_{t,\zeta_0} | \hat{\mathbf{p}} \hat{\iota} | \psi^h_{t,\zeta_0} \rangle$. In light of this, we define for the probability distribution $q_\ell := |[|\psi_{t,\zeta_0})]_\ell|^2$, $1 \leq \ell \leq N$ over data points determined by the propagated state, its *expected (mean) position in extrinsic coordinates* to be given by the sample point $\langle \check{\mathbf{x}}_t \rangle \in V$ that minimizes the Euclidean distance in $\mathbb{R}^n$ to the empirical mean $\bar{v} := (\psi^h_{t,\zeta_0} | [\iota|_{X_N}] | \psi^h_{t,\zeta_0}) = \sum_\ell q_\ell v_\ell$. Here, $[\iota|_{X_N}] := [\text{diag}(v_{\cdot,1}), \ldots, \text{diag}(v_{\cdot,n})]^T$ denotes a vector of diagonal matrices with the $j$'th coordinates ($1 \leq j \leq n$) of each data point on the diagonal and the corresponding inner product defining $\bar{v}$ is taken element-wise with respect to $\iota|_{X_N}$. The mean $\bar{v}$ is the data-driven computational surrogate to $\bar{\mathbf{x}}_t$ and we will establish below that *with high probability (w.h.p.), the expected position* $\langle \check{\mathbf{x}}_t \rangle$ is within an $O(h)$-ball on $\mathcal{M}$ about the point $\mathbf{x}_t$ that is at geodesic distance $t$ from $\mathbf{x}_0 := \iota^{-1}(v_0)$ in the direction $\mathbf{p}_0$ approximating $p_0$ that is used to construct the initial state $|\psi^h_{\zeta_0}\rangle$.

As in the formation of the initial state, the localization of the propagated wavepacket also enables approximation of the expected position without global computations involving the high-dimensional extrinsic coordinates if we proceed through an LPCA approximation of the local neighborhood of the *maximum* of the propagated state. Namely, we perform LPCA around datapoint $v_{\ell^*}$, where $\ell^* = \arg\max_\ell q_\ell$, using all data points in $\mathcal{S}_{\ell^*} = \{\ell \mid ||v_{\ell^*} - v_\ell||^2 \leq \delta_{\text{PCA}}\}$. This LPCA computation is decoupled from that of the initial state preparation, hence this neighbourhood size parameter $\delta_{\text{PCA}}$ and the resulting coordinate map $\check{O}$ may also be different from that of the previous section. The resulting linear transformation in fact gives a coordinate mapping for the whole $\sqrt{\delta_{\text{PCA}}}$-neighbourhood $\mathcal{N}_{\ell^*} := \{v \in \iota(\mathcal{M}) \mid ||v - v_{\ell^*}||^2 \leq \delta_{\text{PCA}}\}$, which can be viewed as the vector of functions given by $u := \check{O} \cdot \iota : \mathcal{N}_{\ell^*} \to \mathbb{R}^{\tilde{\nu}}$ (for $\tilde{\nu}$ the LPCA dimension, also decoupled from initial state preparation). The *expected (mean) position in LPCA coordinates* is then defined as the data point in LPCA coordinates, $\langle \check{\mathbf{x}}_t \rangle^{\text{LPCA}} \in \check{O}[V]$ that minimizes the Euclidean distance in LPCA space to the empirical mean position $\bar{\vartheta} := (\psi^h_{t,\zeta_0} | [u|_{\mathcal{S}_{\ell^*}}] | \psi^h_{t,\zeta_0}) = \sum_{\ell \in \mathcal{S}_{\ell^*}} q_\ell \vartheta^{v_{\ell^*}}_\ell$, wherein $[u|_{\mathcal{S}_{\ell^*}}]$ is a vector of diagonal matrices given by the coordinate elements of $u$ applied to the data points in $\mathcal{S}_{\ell^*}$ and $\vartheta^{v_{\ell^*}}_\ell$ are LPCA coordinates for the data point $\ell$.

The above procedures yield a data point corresponding to $\check{\mathbf{x}}_t \in \{\iota^{-1}(\langle \check{\mathbf{x}}_t \rangle), \iota^{-1} \circ \check{O}^\mathsf{T} \langle \check{\mathbf{x}}_t \rangle^{\text{LPCA}}\} \subset \mathcal{M}$. They are justified by the fact that for $h$ following a certain decay rate, we have with high probability that $\check{\mathbf{x}}_t$ is *intrinsically* within distance $h$ from the geodesic point $\mathbf{x}_t := \pi_\mathcal{M} \Phi_t(\mathbf{x}_0, \mathbf{p}_0)$: this is,

**Proposition 2.** *Let $\beta := (2 + \alpha)(\nu + 1) + 1$ and $\beta_0 := (2 + \alpha)(5\nu/2 + 4) + 2(\nu + 2)$. Then, $\exists\, h_0 > 0$ such that*

1. *if the initial state is given by Eq. (11) and $N^{-1/\gamma_0} \lesssim h < h_0$, then with high probability, $d_g(\check{\mathbf{x}}_t, \mathbf{x}_t) < h$ and*

2. *if the initial state is given by Eq. (12) and $N^{-1/\gamma_1} \lesssim h < h_0$ with[2], then with high probability, $d_g(\check{\mathbf{x}}_t, \mathbf{x}_t) < h$,*

*wherein $\gamma_0 := \max\{\nu(\nu/4 + 1 + \beta), \beta_0\}$ and $\gamma_1 := \max\{(\nu+2)(\nu/4 + 1 + \beta)/2, \beta_0\}$ whenever $\nu \geq 2$ and if $\nu = 1$ is possible, these maxima must be taken with $3\nu/2 + 2\beta$.*

This proposition is a statement about the accuracy of the calculated expected positions with respect to the position determined by the classical geodesic flow. There are in total four combinations of state preparation and expectation calculation to consider:

1. The state $|\psi^h_{\zeta_0}\rangle$ is prepared using extrinsic coordinates as per Eq. (11) and the mean is also computed in extrinsic coordinates, giving $\bar{v}$.

2. The state is still Eq. (11), while the mean is computed in LPCA coordinates, giving $\bar{\vartheta}$.

3. The state is $|\psi^h_{\zeta_0}\rangle$ prepared using LPCA coordinates as per Eq. (12) and the mean is also computed in LPCA coordinates, $\bar{\vartheta}$.

4. The state is still Eq. (12), while the mean is now computed in extrinsic coordinates, giving $\bar{v}$.

In all cases, we are using initial states whose phases are some *perturbation* away from being admissible (see Section II.B). When the initial states are true coherent states, it is shown in [1, §6] that the corresponding version of this proposition holds in the case of extrinsic means with a better convergence rate and similarly, a better convergence rate than stated here follows for the LPCA case upon ensuring that the projected coordinates are diffeomorphic in a neighborhood of the maximum of the propagated coherent state. In the proof of this proposition below, we extend these results to the case of initial states that are perturbations of sufficiently low order, of coherent states. Thus, the main argument rests on criteria for our constructed initial states $|\psi^h_{\zeta_0}\rangle$ to be such perturbations, which is the

---

[2] Recall from Section II.D.2 that $\gamma$ scales the neighborhood size over which the LPCA matrix $\check{O}$ is computed in order to approximate well the tangent space at $\mathbf{x}_0$ for the preparation of the initial state $|\psi^h_{\zeta_0}\rangle$ as per (12). $\gamma \sim h^{\frac{2\gamma_1}{\nu+2}}$

content of the following two preliminary lemmas: that is, we show that with high probability, both the extrinsic and LPCA formulations of the phases are sufficiently close to admissible such that the corresponding initial states and their propagations with respect to $\check{V}_t$ are $\mathcal{O}(h)$ perturbations of the propagations of coherent states. Then, the properties of localization and means of the propagations of coherent states given in [1] can be applied directly to yield Proposition 2.

We begin with a criterion for the phases of the initial states, which is where the approximations ultimately land and we record it as,

**Lemma 1.** *Let $-\bar{\varphi} \in C^\infty(\mathcal{M} \times T^*\mathcal{M})$ be an admissible phase and $\zeta_0 \in T^*\mathcal{M}$. Then, for $|\psi^h_{\zeta_0}\rangle$ a (normalized) coherent state localized at $\zeta_0$ with phase $\varphi$ and given $\phi \in C^\infty(\mathcal{M})$, if $\varepsilon > 0$ and there is a constant $C \geq 0$ such that $||\phi(\cdot) - \varphi(\cdot;\zeta_0)||_\infty \leq Ch^{\nu/4+1}\varepsilon$, then there is $\varepsilon_0 > 0$ so that for all $\varepsilon \in (0, \varepsilon_0]$, $\check{\psi}^h_{\zeta_0}(\cdot) := e^{\frac{i}{h}\phi(\cdot)}/||e^{\frac{i}{h}\phi(\cdot)}||_{L^2(\mathcal{M})} = \psi^h_{\zeta_0}(\cdot) + \mathcal{O}(\varepsilon)$.*

*Proof.* We denote by $\varphi$ the specification $\varphi(\cdot; \zeta_0)$ and $v := \phi - \varphi$. Then, a Taylor expansion of $e^{\frac{i}{h}z}$ yields that there is a constant $C' > 0$ such that $||e^{\frac{i}{h}\phi} - e^{\frac{i}{h}\varphi}||_\infty \lesssim ||e^{\frac{i}{h}v} - 1||_\infty \leq C'h^{\frac{\nu}{4}}\varepsilon$ and hence, $||e^{\frac{i}{h}\phi} - e^{\frac{i}{h}\varphi}||_{L^2} \leq ||e^{\frac{i}{h}v} - 1||_\infty ||e^{\frac{i}{h}\varphi}||_{L^2} \leq C'^2 h^{\frac{\nu}{2}}\varepsilon$, so $|\,||e^{\frac{i}{h}\phi}||_{L^2} - ||e^{\frac{i}{h}\varphi}||_{L^2}\,| \leq C'h^{\frac{\nu}{2}}\varepsilon$. Thus,

$$\begin{aligned}
|\check{\psi}^h_{\zeta_0} - \psi^h_{\zeta_0}| &= \left| \frac{e^{\frac{i}{h}\phi}||e^{\frac{i}{h}\varphi}||_{L^2} - e^{\frac{i}{h}\varphi}||e^{\frac{i}{h}\phi}||_{L^2}}{||e^{\frac{i}{h}\varphi}||_{L^2}||e^{\frac{i}{h}\phi}||_{L^2}} \right| \\
&\leq \frac{|e^{\frac{i}{h}\phi} - e^{\frac{i}{h}\varphi}|\,||e^{\frac{i}{h}\varphi}||_{L^2} + |e^{\frac{i}{h}\varphi}|(||e^{\frac{i}{h}\varphi}||_{L^2} - ||e^{\frac{i}{h}\phi}||_{L^2})|}{||e^{\frac{i}{h}\varphi}||_{L^2}||e^{\frac{i}{h}\phi}||_{L^2}} \\
&\leq C'\varepsilon \frac{||e^{\frac{i}{h}\varphi}||_{L^2}h^{\frac{\nu}{4}} + |e^{\frac{i}{h}\varphi}|h^{\frac{\nu}{2}}}{||e^{\frac{i}{h}\varphi}||_{L^2}(||e^{\frac{i}{h}\varphi}||_{L^2} - C'h^{\frac{\nu}{2}}\varepsilon)} \\
&\lesssim \varepsilon
\end{aligned}$$

whenever $\varepsilon$ is sufficiently small. $\square$

We wish to see that the states prepared in practice, as in Eqs. (11) and (12), are themselves perturbations of bonafide coherent states. That this is achievable with high probability is given in,

**Lemma 2.** *Let $|\psi^h_{\zeta_{\text{ext}}})$ be prepared as in Eq. (11) with $\zeta_{\text{ext}} := (v_0, p_0) \in \mathbb{R}^{2n}$ and $|\psi^h_{\zeta_{\text{LPCA}}})$ be prepared as in Eq. (12) with $\zeta_{\text{LPCA}} := (\vartheta^{v_0}_0, p_0) \in \mathbb{R}^{2\tilde{\nu}}$ and $\tilde{\zeta}_{\text{LPCA}} := (v_0, \tilde{p}_0) \in \mathbb{R}^{2n}$ its realization in the extrinsic space with $\tilde{p}_0$ emanating from $v_0$ and residing in the LPCA subspace approximation of $T_{v_0}\iota(\mathcal{M})$. Then, setting $\mathbf{x}_0 := \iota^{-1}(v_0)$ and denoting the Riemannian normal coordinates centered at $\mathbf{x}_0$ by $\eta^{\mathbf{x}_0}(\mathbf{x}) = (\eta^{\mathbf{x}_0}_1(\mathbf{x}), ..., \eta^{\mathbf{x}_0}_\nu(\mathbf{x}))$, we have for all $\beta \geq 0$ that*

1. *there is an $L^2$ normalized coherent state $|\psi^h_{\zeta_0}\rangle$ localized at $\zeta_0 := (\mathbf{x}_0, \eta^{\mathbf{x}_0}\iota^{-1}(v_n)) \in T^*\mathcal{M}$ and a constant $h_0 > 0$ such that for all $h \in (0, h_0]$,*

$$\Pr[||d\iota|_{\mathbf{x}_0}\zeta_0 - \zeta_{\text{ext}}||_{\mathbb{R}^n} > h^{\nu/4+1+\beta} \wedge ||[\psi^h_{\zeta_{\text{ext}}})] - [\psi^h_{\zeta_0})]||_\infty > h^\beta] \leq e^{-\Omega(Nh^{\nu(\nu/4+1+\beta)})} + e^{-\Omega(Nh^{3\nu/2+2\beta})}$$

   *and*

2. *there is an $L^2$ normalized coherent state $|\psi^h_{\zeta_0}\rangle$ localized at $\zeta_0 = (\mathbf{x}_0, \zeta_{\mathbf{p}_0}) \in T^*\mathcal{M}$ and a constant $h_0 > 0$ such that if $\delta > 0$ and $\delta N^{-\frac{2}{(\nu+2)(\nu/4+1+\beta)}} \leq h \leq h_0$, then with $\gamma \sim h^{\nu/4+1+\beta}$,*

$$\Pr[||d\iota|_{\mathbf{x}_0}\zeta_0 - \tilde{\zeta}_{\text{LPCA}}||_{\mathbb{R}^n} > h^{\nu/4+1+\beta} \wedge ||[\psi^h_{\zeta_{\text{LPCA}}})] - [\psi^h_{\zeta_0})]||_\infty > h^\beta] \leq e^{-\Omega(\delta)} + e^{-\Omega(Nh^{3\nu/2+2\beta})}.$$

*Proof.* The discussion in [1, §3.1] shows that $|\tilde{\psi}^h_{\zeta_{\text{ext}}})$ is an $\mathcal{O}(h^{\nu/4+\beta})$ perturbation of $|\tilde{\psi}^h_{\zeta_0}\rangle$ whenever $v_n$ is chosen such that $||v_0 - v_n||_{\mathbb{R}^n} \lesssim h^{\nu/4+1+\beta} < \kappa$. Then, $d_g(\iota^{-1}(v_0), \iota^{-1}(v_n)) \lesssim h^{\nu/4+1+\beta}$ and by a Chernoff bound, such a point belongs to our sample set with probability at least $1 - e^{-\Omega(Nh^{\nu(\nu/4+1+\beta)})}$. Further, by [1, Lemma 19], $|\,||\tilde{\psi}^h_{\zeta_{\text{ext}}})||^2_N - ||\tilde{\psi}^h_{\zeta_{\text{ext}}})||^2_{L^2}\,| \lesssim h^{\nu/2+\beta}$ with probability at least $1 - e^{-\Omega(Nh^{3\nu/2+2\beta})}$. Combining these events with Lemma 1 gives,

$$||\psi^h_{\zeta_{\text{ext}}}) - |\psi^h_{\zeta_0}\rangle| \leq \left||\psi^h_{\zeta_{\text{ext}}}) - \frac{|\tilde{\psi}^h_{\zeta_{\text{ext}}})}{||\tilde{\psi}^h_{\zeta_{\text{ext}}})||_{L^2}}\right| + \left|\frac{|\tilde{\psi}^h_{\zeta_{\text{ext}}})}{||\tilde{\psi}^h_{\zeta_{\text{ext}}})||_{L^2}} - |\psi^h_{\zeta_0}\rangle\right| \lesssim h^\beta$$

29

and with a union bound we arrive at the probabilistic bound in the statement of the first part of the Lemma.

Now, for the second part, let $\iota : \mathcal{M} \hookrightarrow \mathbb{R}^n$ denote the isometric embedding such that for each $1 \leq \ell \leq N$, there is $\mathbf{x}_\ell \in \mathcal{M}$ so that $v_\ell = \iota(\mathbf{x}_\ell) \in V$. Further, for each $\mathbf{x} \in \mathcal{M}$, let $\Pi_\mathbf{x} : \mathbb{R}^n \to T_{\iota(\mathbf{x})}\iota(\mathcal{M}) \subset \mathbb{R}^n$ denote the orthogonal projection onto the subspace of $\mathbb{R}^n$ tangent to the submanifold at $\iota(\mathcal{M})$ and set $\varphi(\mathbf{x}; \zeta_\mathbf{x}, \zeta_\mathbf{p}) := \langle d\iota|_{\zeta_\mathbf{x}} \zeta_\mathbf{p}, \Pi_{\zeta_\mathbf{x}}[\iota(\zeta_\mathbf{x}) - \iota(\mathbf{x})]\rangle + \frac{i}{2} \frac{\langle d\iota|_{\zeta_\mathbf{x}} \zeta_\mathbf{p}\rangle}{\langle p_0\rangle} ||\Pi_{\zeta_\mathbf{x}}[\iota(\zeta_\mathbf{x}) - \iota(\mathbf{x})]||_{\mathbb{R}^n}^2$ with $\zeta := (\zeta_\mathbf{x}, \zeta_\mathbf{p}) \in T^*\mathcal{M}$. Then, $\varphi \in C^\infty(\mathcal{M} \times T^*\mathcal{M})$ and satisfies,

$$d\varphi|_{\mathbf{x}=\zeta_\mathbf{x}} = -\zeta_\mathbf{p}, \qquad d^2 \Im(\varphi)|_{\mathbf{x}=\zeta_\mathbf{x}} = \frac{\langle d\iota|_{\zeta_\mathbf{x}} \zeta_\mathbf{p}\rangle}{\langle p_0\rangle} d\iota|_{\zeta_\mathbf{x}}^\mathsf{T} d\iota|_{\zeta_\mathbf{x}},$$

hence it is an admissible phase. The corresponding coherent state $|\psi_\zeta^h\rangle$ with phase $-\bar{\varphi}$ will serve as the *model* for the LPCA prepared state Eq. (12).

The LPCA procedure [10, §2] yields a linear map, $\breve{O} : \mathbb{R}^n \to \mathbb{R}^{\tilde{\nu}}$ for $1 \leq \tilde{\nu} \leq n$. Its matrix representation is given by $\breve{O} := [u_1, \ldots, u_{\tilde{\nu}}]^\mathsf{T}$ such that for each $1 \leq j \leq \tilde{\nu}$, the column vector $u_j \in \mathbb{R}^n$ is the left singular vector corresponding to $j$-th largest singular value of a local neighbourhood matrix at $v_0$. Suppose, without loss of generality, that the embedding $\iota(\mathcal{M})$ is appropriately translated and rotated such that $v_0 = 0 \in \mathbb{R}^n$ and the subspace $T_{v_0}\iota(\mathcal{M}) \subset \mathbb{R}^n$ is aligned so that the first $\nu$ vectors of the standard basis $\{e_1, \ldots, e_n\}$ (meaning $e_j$ has $j$-th entry equal to 1 and is otherwise zero) of $\mathbb{R}^n$ form an orthonormal basis for $T_{v_0}\iota(\mathcal{M})$ and there are $\zeta_{\mathbf{p}_1}, \ldots, \zeta_{\mathbf{p}_\nu} \in T^*_{\zeta_{\mathbf{x}_0}}(\mathcal{M})$ such that for each $1 \leq j \leq \nu$, $d\iota|_{\zeta_{\mathbf{x}_0}} \zeta_{\mathbf{p}_j} = e_j$. Now assume the event, call it $\mathcal{D}$, that $\tilde{\nu} = \nu$ and recall that the phase constructed in Eq. (12) is given by

$$\phi(\mathbf{x}) := \langle p_0, \breve{O}[\iota(\mathbf{x}) - \iota(\mathbf{x}_0)]\rangle_{\mathbb{R}^\nu} + \frac{i}{2} ||\breve{O}[\iota(\mathbf{x}_0) - \iota(\mathbf{x})]||_{\mathbb{R}^\nu}^2.$$

Since the singular vectors $u_1, \ldots, u_\nu$ are orthonormal, we have $\breve{O}\breve{O}^\mathsf{T} = I_\nu$, so

$$\phi(\mathbf{x}) = \langle \tilde{p}_0, \breve{\Pi}[\iota(\mathbf{x}) - \iota(\mathbf{x}_0)]\rangle_{\mathbb{R}^n} + \frac{i}{2} ||\breve{\Pi}[\iota(\mathbf{x}_0) - \iota(\mathbf{x})]||_{\mathbb{R}^n}^2$$

with $\tilde{p}_0 := \breve{O}^\mathsf{T} p_0$ and $\breve{\Pi} := \breve{O}^\mathsf{T} \breve{O}$.

Further suppose we are in the event, call it $\mathcal{A}_\varepsilon$ for $\varepsilon > 0$, that there are $w_1, \ldots, w_\nu \in \mathbb{R}^n$, which form an orthonormal basis for $T_{\iota(v_0)}\iota(\mathcal{M})$ such that for each $1 \leq j \leq \nu$, $||u_j - w_j||_\infty \leq \varepsilon$. Then, $||\breve{\Pi} - \Pi_{\zeta_{\mathbf{x}_0}}|| \lesssim \varepsilon$ and setting $\zeta_{\mathbf{p}_0} := d\iota|_{\zeta_{\mathbf{x}_0}}^\mathsf{T} \Pi_{\zeta_{\mathbf{x}_0}} \tilde{p}_0$, we have

$$|\phi(\mathbf{x}) + \bar{\varphi}(\mathbf{x}; \zeta_{\mathbf{x}_0}, \zeta_{\mathbf{p}_0})|$$
$$\leq |\langle \tilde{p}_0, (\breve{\Pi} - \Pi_{\zeta_{\mathbf{x}_0}}) \cdot \iota(\mathbf{x})\rangle + \langle (I_n - \Pi_{\zeta_{\mathbf{x}_0}})\tilde{p}_0, \Pi_{\zeta_{\mathbf{x}_0}} \cdot \iota(\mathbf{x})\rangle| + \frac{1}{2} \left| ||\Pi_{\zeta_{\mathbf{x}_0}} \iota(\mathbf{x})||^2 - ||\breve{\Pi}\iota(\mathbf{x})||^2 \right|$$
$$\leq ||p_0|| \cdot ||(\breve{\Pi} - \Pi_{\zeta_{\mathbf{x}_0}}) \cdot \iota(\mathbf{x})|| + ||(\breve{\Pi} - \Pi_{\zeta_{\mathbf{x}_0}})\tilde{p}_0|| \cdot ||\Pi_{\zeta_{\mathbf{x}_0}} \cdot \iota(\mathbf{x}_0)|| + \frac{1}{2} \left| ||\Pi_{\zeta_{\mathbf{x}_0}} \iota(\mathbf{x})|| + ||\breve{\Pi}\iota(\mathbf{x})|| \right| \cdot ||(\breve{\Pi} - \Pi_{\zeta_{\mathbf{x}_0}}) \cdot \iota(\mathbf{x})||$$
$$\lesssim \varepsilon,$$

wherein we have used that $\breve{\Pi}\tilde{p}_0 = \tilde{p}_0$.

We now wish to apply Lemma 1, so we require that $\varepsilon \lesssim h^{\nu/4+1+\beta}$. Further, if $\delta N^{-\frac{2}{(\nu+2)(\nu/4+1+\beta)}} \leq h$, then from the proof of [10, Theorem B.1], we find that the corresponding event $\mathcal{A}_\varepsilon$ happens with probability at least $1 - e^{-\Omega(\delta)}$. Moreover, within the same event, we find that the singular values resulting from LPCA and corresponding to $\breve{O}$ have the following *gap property*: if $\sigma_1 \geq \cdots \geq \sigma_n \geq 0$ are the singular values corresponding to $u_j, \ldots, u_n$, then for $1 \leq j \leq \nu$, $\sigma_j \sim N\varepsilon^{\nu/2+1}$, while for each $\nu + 1 \leq m \leq n$, $\sigma_m \lesssim N\varepsilon^{\nu/2+2}$. Therefore, for all $(j, m) \in \{1, \ldots, \nu\} \times \{\nu+1, \ldots, n\}$, we have $\sigma_j/\sigma_m \gtrsim h^{-(2+\nu/4)}$, so within the same probabilistic event we can determine the dimension $\nu$, viz., the event $\mathcal{D}$ occurs simultaneously with $\mathcal{A}_\varepsilon$. By the proof of Lemma 1, we so far have that $||e^{\frac{i}{h}\phi} - e^{\frac{i}{h}\varphi}||_{L^\infty} \lesssim h^{\nu/4+\beta}$ and $||e^{\frac{i}{h}\phi}/||e^{\frac{i}{h}\phi}||_{L^2} - |\psi_{\zeta_0}^h\rangle||_{L^\infty} \lesssim h^\beta$. An application of [1, Lemma 19] gives that $|\,||\tilde{\psi}_{\zeta_{\mathrm{LPCA}}}^h||_N^2 - ||e^{\frac{i}{h}\phi}||_{L^2}^2\,| \lesssim h^{\nu/2+\beta}$ with probability at least $1 - e^{-\Omega(Nh^{3\nu/2+2\beta})}$. Thus, proceeding as in the previous part we find that assuming also this event gives $|\,|\psi_{\zeta_{\mathrm{LPCA}}}^h) - |\psi_{\zeta_0}^h\rangle||_{L^\infty} \lesssim h^\beta$, whence upon taking a union bound over all considered events, we arrive at the probabilistic bound as stated in the second part of the present Lemma. $\square$

Now that we may associate to each prepared state a proper coherent state, we wish to run the convergence results of [1, §6] on the corresponding coherent state and pull the results back to the originally prepared state, up to perturbative terms. This entails knowing that the propagated *density* $||\psi_{t,\tilde{\zeta}_0}^h)|^2$, for $\tilde{\zeta}_0 \in \{\zeta_{\mathrm{ext}}, \zeta_{\mathrm{LPCA}}\}$, is also a

perturbation of $|\hat{V}_t|\psi^h_{\zeta_0}\rangle|^2$ with $\zeta_0$ related to $\tilde{\zeta}_0$ by the previous Lemma. As an intermediary step, we will bound $|||\breve{V}_t|\psi^h_{\zeta_0}\rangle|^2 - ||\psi^h_{t,\tilde{\zeta}_0}\rangle|^2||_\infty$ with the help of a *generic* $L^\infty$ bound on the propagation of $C^\infty$ functions as discussed in [1, §6.2]. Such an $L^\infty$ bound makes sense because $\breve{V}_t$ is a finite-dimensional operator that extends to the continuum by Nyström extension (discussed in detail in Sec. 6 of [1]):

$$\breve{V}_t[u](\mathbf{x}) = f(0)u(\mathbf{x}) + \breve{T}_{\epsilon,\lambda}[Df(\breve{T}_{\epsilon,\lambda})[u]](\mathbf{x}), \tag{13}$$

where $f(z) = e^{it\sqrt{\frac{2c_0}{c_2}\frac{(1-z)}{\epsilon}}}$, $Df(z) = (f(z)-f(0))/z$, and $\breve{T}_{\epsilon,\lambda}$ is the diffusion operator defined above. If $u \in C^\infty$, the generic $L^\infty$ bound mentioned above is

$$|\breve{V}_t[u](\mathbf{x})| \lesssim ||u||_\infty + \frac{||u||_{N,2}\, t\epsilon^{-\frac{\nu+1}{2}}||k||_\infty ||p_{N,\epsilon,\lambda}||_{N,\infty}^{1/2}}{(\inf p_{N,\epsilon}^\lambda)(\min_{j\in[N]} p_{N,\epsilon}^\lambda)\min_{j\in[N]} p_{N,\epsilon,\lambda}^{3/2}}, \tag{14}$$

wherein $||\cdot||_{N,2}^2 := \frac{1}{N}(\cdot|\cdot)$, $||\cdot||_{N,\infty}$ denotes the maximum of a function over the sample set and $p_{N,\epsilon}, p_{N,\epsilon,\lambda} \in C^\infty(\mathcal{M})$ are given by

$$p_{N,\epsilon}(\mathbf{x}) := \frac{1}{N}\sum_{j=1}^N k_\epsilon(\mathbf{x},\mathbf{x}_j), \quad p_{N,\epsilon,\lambda}(\mathbf{x}) := \frac{1}{N}\sum_{j=1}^N k_{N,\epsilon,\lambda}(\mathbf{x},\mathbf{x}_j),$$

$$k_{N,\epsilon,\lambda}(\mathbf{x},\mathbf{y}) := \frac{k_\epsilon(\mathbf{x},\mathbf{y})}{(p_{N,\epsilon}(x)p_{N,\epsilon}(y))^\lambda}, \quad k_\epsilon(\mathbf{x},\mathbf{y}) := \epsilon^{-\nu/2}k(||\iota(\mathbf{x})-\iota(\mathbf{y})||^2/\epsilon);$$

these are related to the matrix definitions given above via

$$[\breve{T}_\epsilon]_{i,j} = Nk_\epsilon(\iota^{-1}v_i, \iota^{-1}v_j),$$
$$[\breve{\Sigma}_\epsilon]_{j,j} = Np_{N,\epsilon} \circ \iota^{-1}(v_j).$$

Therefore, once we have a sampling based bound on the factors involving $p_{N,\epsilon}$ and $p_{N,\epsilon,\lambda}$ and consistency of the $||\cdot||_{N,2}$ norms for $|\psi^h_{\zeta_0}\rangle$ and $|\psi^h_{\tilde{\zeta}_0}\rangle$, we can establish the deviation of their corresponding propagated densities. This leads to the recovery of geodesic propagation on $\mathcal{M}$ up to errors decaying in $h$ and $N$ simultaneously, with high probability, which we now establish in,

*Proof of Proposition 2.* We proceed with the cases 1 - 4. Recall that $\langle \breve{\mathbf{x}}_t \rangle$ and $\langle \breve{\mathbf{x}}_t \rangle^{\text{LPCA}}$ denote the mean approximated geodesic point as computed in extrinsic or LPCA coordinates, respectively and that this is independent of how the corresponding initial state is prepared. We will denote by $\langle \mathbf{x}_t \rangle$ either one of these points and specify a particular one with its specific notation when needed. Regarding the initial state, let $\tilde{\zeta}_0 \in \{\zeta_{\text{ext}}, \zeta_{\text{LPCA}}\}$ and let $|\psi^h_{\zeta_0}\rangle$ be the coherent state corresponding to $|\psi^h_{\tilde{\zeta}_0}\rangle$ as constructed in Lemma 2. We will work with the approximations with respect to $|\breve{\psi}^h_{t,\zeta_0}\rangle := \breve{V}_t|\psi^h_{\zeta_0}\rangle$, so we denote

$$\breve{q}_\ell := |[|\breve{\psi}^h_{t,\zeta_0}\rangle]_\ell|^2,$$
$$\breve{v}_\ell := \sum_{\ell \in S_{\ell^*}} v_\ell \breve{q}_\ell, \quad \breve{\vartheta}_\ell := \sum_{\ell \in S_{\ell^*}} \vartheta_\ell^{v_{\ell^*}} \breve{q}_\ell$$

and we denote $\langle \mathbf{x}_t \rangle_{\zeta_0} \in \{\langle \breve{\mathbf{x}}_t \rangle_{\zeta_0}, \langle \mathbf{x}_t \rangle_{\zeta_0}^{\text{LPCA}}\}$ with $\langle \breve{\mathbf{x}}_t \rangle_{\zeta_0}$ the closest point, with respect to $||\cdot||_{\mathbb{R}^n}$, to $\breve{v}_\ell$ in $\iota(\mathcal{M})$ and likewise, by $\langle \mathbf{x}_t \rangle_{\zeta_0}^{\text{LPCA}}$ the closest point to $\breve{\vartheta}_\ell \in \mathbb{R}^{\tilde{\nu}}$ with respect to the Euclidean norm in $\mathbb{R}^{\tilde{\nu}}$. Further, we will use the notation from the proof of Lemma 2 to let $\breve{O} : \mathbb{R}^n \to \mathbb{R}^{\tilde{\nu}}$ denote the LPCA map. We will also need the continuum counterparts to the discretely constructed *degree functions* $p_{N,\epsilon,\lambda'}$ for $\lambda' \in \{0,\lambda\}$, namely: $p_\epsilon(\mathbf{x}) := \int_\mathcal{M} k_\epsilon(\mathbf{x},\mathbf{y})\mathsf{p}(\mathbf{y})d\mathbf{y}$, $k_{\epsilon,\lambda}(\mathbf{x},\mathbf{y}) := k_\epsilon(\mathbf{x},\mathbf{y})/(p_\epsilon(\mathbf{x})p_\epsilon(\mathbf{y}))^\lambda$ and $p_{\epsilon,\lambda}(\mathbf{x}) := \int_\mathcal{M} k_{\epsilon,\lambda}(\mathbf{x},\mathbf{y})\mathsf{p}(\mathbf{y})d\mathbf{y}$.

Now set $\beta, \delta_0 \geq 0$, $\varepsilon > 0$ and assume the occurrence of the following events:

$$\mathcal{A}_0(h^\beta; \tilde{\zeta}_0) : \quad |||\psi^h_{\tilde{\zeta}_0}\rangle - |\psi^h_{\zeta_0}\rangle||_{L^\infty} \lesssim h^\beta,$$
$$\mathcal{E}_{\lambda'}(\delta_0) \quad : \quad ||p_{N,\epsilon,\lambda'} - p_{\epsilon,\lambda'}||_\infty \lesssim \delta_0$$
$$\mathcal{G}_{\text{ext}} \quad : \quad d_g(\iota^{-1}(\langle \breve{\mathbf{x}}_t \rangle_{\zeta_0}), \mathbf{x}_t) \leq h \qquad \text{if } \langle \mathbf{x}_t \rangle = \langle \breve{\mathbf{x}}_t \rangle$$
$$\mathcal{G}_{\text{LPCA}}(\varepsilon) \quad : \quad \begin{cases} d_g(\iota^{-1}\breve{O}^\mathsf{T}\langle \mathbf{x}_t \rangle_{\zeta_0}^{\text{LPCA}}, \mathbf{x}_t) \leq h, \\ \tilde{\nu} = \nu \wedge ||\breve{\Pi} - \Pi|| \leq \varepsilon \end{cases} \quad \text{if } \langle \mathbf{x}_t \rangle = \langle \mathbf{x}_t \rangle^{\text{LPCA}}.$$

We now see that these events imply that the computed mean is intrinsically $\mathcal{O}(h)$ close to $\mathbf{x}_t$. Let $\underline{C}_{\lambda'} := \inf p_{\epsilon,\lambda'} > 0$ and note that if $\delta_0 = \min\{\underline{C}_0^{2\lambda}, \underline{C}_\lambda^{3/2}\}/2$, then by Eq. (14) and the events $\mathcal{A}_0(h^\beta; \tilde{\zeta}_0)$ and $\mathcal{E}_{\lambda'}(\delta_0)$,

$$|||\check{\psi}^h_{t,\zeta_0}\rangle|^2 - ||\psi^h_{t,\tilde{\zeta}_0}\rangle|^2||_\infty \leq |||\check{\psi}^h_{t,\zeta_0}\rangle| - ||\psi^h_{t,\tilde{\zeta}_0}\rangle|||_\infty (|||\check{\psi}^h_{t,\zeta_0}\rangle|||_\infty + |||\psi^h_{t,\tilde{\zeta}_0}\rangle|||_\infty) \\ \lesssim h^\beta \epsilon^{-\frac{\nu+1}{2}}(h^{-\nu/4} + \epsilon^{-\frac{\nu+1}{2}}). \tag{15}$$

Therefore, setting $\beta = (2+\alpha)(\nu+1)+1$ gives $|||\check{\psi}^h_{t,\zeta_0}\rangle|^2 - ||\psi^h_{t,\tilde{\zeta}_0}\rangle|^2||_\infty \lesssim h$. With this, we have that for any $u \in C^\infty$,

$$|\sum_{\ell \in S_{\ell^*}} u_\ell (q_\ell - \check{q}_\ell)| \lesssim h$$

and hence upon denoting

$$\bar{v}_\ell := \sum_{\ell \in S_{\ell^*}} v_\ell q_\ell, \quad \bar{\vartheta}_\ell := \sum_{\ell \in S_{\ell^*}} \vartheta_\ell^{v_{\ell^*}} q_\ell,$$

we have in particular,

$$||\bar{v}_\ell - \check{v}_\ell||_{\mathbb{R}^n} \lesssim h, \quad ||\bar{\vartheta}_\ell - \check{\vartheta}_\ell||_{\mathbb{R}^\nu} \lesssim h,$$

wherein the second inequality follows from event $\mathcal{G}_{\text{LPCA}}$. Hence, $||\langle \check{\mathbf{x}}_t \rangle - \langle \check{\mathbf{x}}_t \rangle_{\zeta_0}||_{\mathbb{R}^n} \lesssim h$ so under the event $\mathcal{G}_{\text{ext}}$ with $h < \kappa$, we have $d_g(\iota^{-1}\langle \check{\mathbf{x}}_t \rangle, \mathbf{x}_t) \lesssim h$. Likewise, $||\langle \mathbf{x}_t \rangle^{\text{LPCA}} - \langle \mathbf{x}_t \rangle^{\text{LPCA}}_{\zeta_0}||_{\mathbb{R}^n} \lesssim h$ and there is a neighbourhood $V \subset \mathbb{R}^\nu$ of the origin, a neighbourhood $\mathcal{O} \subset \mathcal{M}$ of $\iota^{-1}\check{O}^\intercal \langle \mathbf{x}_t \rangle^{\text{LPCA}}$ and a constant $\varepsilon_0 > 0$ such that for all $\varepsilon \in (0, \varepsilon_0]$, the event $\mathcal{G}_{\text{LPCA}}(\varepsilon)$ gives that $\iota^{-1}\check{O}^\intercal : V \to \mathcal{M}$ is a diffeomorphism. Therefore, we have also in the LPCA case that $d_g(\iota^{-1}\check{O}\langle \mathbf{x}_t \rangle^{\text{LPCA}}, \mathbf{x}_t) \lesssim h$.

The event $\mathcal{A}(h^\beta; \tilde{\zeta}_0)$ happens with probability bounded below by that given in Lemma 2. The event $\mathcal{E}_{\lambda'}$ happens with probability at least $1 - e^{-\Omega(N\epsilon^{\nu/2})}$, by the arguments in the proof of [1, Lemma 14]. Since $\varepsilon \in (0, \varepsilon_0]$ can be taken to be constant, following the proof of Lemma 2, the second event of $\mathcal{G}_{\text{LPCA}}(\varepsilon)$ happens with probability at least $1 - e^{-\Omega(N^{4/(\nu+2)})}$. The event $\mathcal{G}_{\text{ext}}$ and the first event of $\mathcal{G}_{\text{LPCA}}$ happen with probabilities bounded below by [1, Proposition 4]: namely, there is a constant $h_0 > 0$ such that if $N^{-1/\beta_0} \lesssim h \leq h_0$ with $\beta_0 := (2+\alpha)(5\nu/2+4) + 2(\nu+2)$, then

$$\Pr[d_g(u(\langle \mathbf{x}_t \rangle_{\zeta_0}), \mathbf{x}_t) \leq h] > 1 - e^{-\Omega(Nh^{\beta_0})} \tag{16}$$

with $u := \iota^{-1}$ for $\langle \mathbf{x}_t \rangle_{\zeta_0} = \langle \check{\mathbf{x}}_t \rangle_{\zeta_0}$ and $u := \iota^{-1}\check{O}^\intercal$ for $\langle \mathbf{x}_t \rangle_{\zeta_0} = \langle \mathbf{x}_t \rangle^{\text{LPCA}}_{\zeta_0}$. Then, a union bound for each of the cases among the combinations of values of $\tilde{\zeta}_0$ and $\langle \mathbf{x}_t \rangle$ gives the consistencies in the statement of Proposition 2, *with high probability*. $\square$

Finally, we mention that while the expected value of the position coordinate over the probability distribution determined by the propagated state is the best estimate of position along the geodesic path, due to the localization of the propagated coherent state the maximizer $\ell^* := \arg\max_\ell q_\ell$ is often also a good estimate of $\mathbf{x}_t$ [1]. A similar convergence rate as above (up to a quadratic loss in $h$) can be established along the same lines, based on the corresponding arguments for the propagation of true coherent states given in [1]. Moreover, we have found this to be useful in practical cases where $N$ is small (see Section III).

### II.D.4. Dual perspective: signals processing & geodesic spheres

We have so far considered the approximation of geodesics on $\mathcal{M}$ through propagations $\hat{V}_t|\psi^h_{\zeta_0}\rangle$ of coherent states $|\psi^h_{\zeta_0}\rangle$ localized at a fixed $\zeta_0 \in T^*\mathcal{M}$ and discrete sampling schemes, in turn, approximating these. As mentioned at the end of the previous section, it follows from results in [1] that as a function of $\mathbf{x} \in \mathcal{M}$, the density $\rho_t(\mathbf{x}; \zeta_0) := |\langle \delta_\mathbf{x}|\psi_{t,\zeta_0}\rangle|^2$, along with its discrete approximations discussed above, attain (*w.h.p.*) their maximum in an $O(\sqrt{h})$-ball about $\mathbf{x}_t$. If we now fix the position-space point to $\mathbf{x}^* \in \mathcal{M}$ and let the initial phase-space localization point vary, then $\rho_t$ describes a density in phase-space as a function of $\zeta \in T^*\mathcal{M}$ that we may write as,

$$\rho_t(\mathbf{x}^*; \zeta) = |\langle \delta_{\mathbf{x}^*}|\hat{V}_t|\psi^h_\zeta\rangle|^2 = |\langle \psi^h_\zeta|\hat{V}_t^\dagger|\delta_{\mathbf{x}^*}\rangle|^2. \tag{17}$$

The right-hand side connects us with the rich and classical field of *time-frequency analysis* and grounds our notions regarding the *signal in the data $X_N$* being the approximation of geodesic rays in $\hat{U}_t|\delta_{\mathbf{x}^*}\rangle$ with the data-driven approximations $\check{V}_t|\psi_h^\zeta\rangle$.

*Time-frequency analysis* is a field of harmonic analysis with a wide range of scientific applications, most prominent of which is signals processing. It also has well-known and deep connections with quantum mechanics and the excellent text [11] provides a detailed study into this area; here, we just give a brief sketch, sufficient to portray our perspective. The term *time-frequency* comes historically from the origins of the transforms under study, which are variations on the Fourier transform of signals; and in this sense, the *time* component is so named due to the traditional signals being functions of a single variable, called *time* and their Fourier transform representing their *frequency* components. In the modern setting, the signals may more generally be functions of spatial variables and likewise, the notion of Fourier transform is adapted to study multidimensional frequencies: namely, given a distribution $u \in \mathcal{D}(\mathbb{R}^n)$, its Fourier transform gives its frequency content as the function of frequencies $\mathbf{p} \in \mathbb{R}^n$ given by,

$$\mathcal{F}[u](\mathbf{p}) := \int_{\mathbb{R}^n} e^{-i\mathbf{p}\cdot\mathbf{y}} u(\mathbf{y}) \, d\mathbf{y}.$$

In connection with quantum mechanics, the Fourier transform converts the position-space representation of a state $|u\rangle$ to its momentum-space representation. Thus, while the Fourier transform recovers all frequency content of $u$, it loses all spatial information: this is a form of the uncertainty principle concerning the compatibility of the knowledge of position and momentum of a quantum state. However, it is not necessary to spend all spatial resolution of a distribution in order to gain frequency resolution and this is precisely the notion realized and studied in time-frequency analysis.

A primary tool for modulating between spatial and frequency resolutions is the *short-time Fourier transform* (STFT), which gives the frequency content of $u$ along neighbourhoods of each spatial point: namely, given a *window function* $\chi \in C^\infty(\mathbb{R}^n \times \mathbb{R}^n)$ such that $\chi_\mathbf{x}(\cdot) := \chi(\mathbf{x}, \cdot)$ is localized in a neighbourhood of each $\mathbf{x} \in \mathbb{R}^n$, the STFT of $u$ is a function of phase space points $\zeta = (\mathbf{x}, \mathbf{p}) \in T^*\mathbb{R}^n$ given by,

$$\mathcal{F}_\chi[u](\mathbf{x}, \mathbf{p}) := \int_{\mathbb{R}^n} e^{-i\mathbf{p}\cdot\mathbf{y}} \chi_\mathbf{x}(\mathbf{y}) u(\mathbf{y}) \, d\mathbf{y}.$$

This transform gives a *representation* of $u$ in *time-frequency* or *phase space* points $(\mathbf{x}, \mathbf{p}) \in T^*\mathbb{R}^n$, through $\mathcal{F}_\chi[u] \in \mathcal{D}(T^*\mathbb{R}^n)$. As per the uncertainty principle, as the window $\chi_x$ becomes more localized, the resulting representation gives stronger resolution of the spatial dependence in $u$ about $\mathbf{x}$ and weaker resolution of its frequency content; vice versa, as $\chi_x$ is wider, the spatial resolution of $u$ about $\mathbf{x}$ is weaker and $\mathcal{F}_\chi[u]$ displays more of its frequency dependence, with the extreme being $\chi_x \equiv 1$, which gives only the frequency content in $u$ via the classical Fourier transform.

In addition to studying the combined spatial and frequency contents of distributions, we also wish to study them at *high* frequencies, which is where the nature of their singularities becomes apparent (see [3, 12]). One approach to this is by way of a *semiclassical rescaling* of the Fourier transform that scales the frequency space by a factor $1/h$ for $0 < h \leq 1$: namely,

$$\mathcal{F}^h[u](\mathbf{p}) := \int_{\mathbb{R}^n} e^{-\frac{i}{h}\mathbf{y}\cdot\mathbf{p}} u(\mathbf{y}) \, d\mathbf{y}$$

gives a frequency space representation of $u$ such that studying $\mathcal{F}^h[u]$ in a fixed region $\mathcal{R}$ of frequency space is equivalent to studying the $\mathcal{F}[u]$ at frequencies $\mathbf{p} \in \mathcal{R}/h$. Furthermore, since at each position $\mathbf{x} \in \mathbb{R}^n$, the frequency scales ought to depend only on $|\mathbf{p}|$ and $h$, we also reposition the spatial origin to $\mathbf{x}$ by modulating frequencies with a factor $e^{i\mathbf{x}\cdot\mathbf{p}}$. Under this rescaling and modulation, the STFT takes the form,

$$\mathcal{F}_\chi^h[u](\mathbf{x}, \mathbf{p}) := \int_{\mathbb{R}^n} e^{-\frac{i}{h}(\mathbf{y}-\mathbf{x})\cdot\mathbf{p}} \chi_\mathbf{x}(\mathbf{y}) u(\mathbf{y}) \, dy,$$

which we call the *h-STFT* of $u$. This h-STFT rescales a $1 \times 1/h$ hyperrectangle in phase (time-frequency) space for $\mathcal{F}_\chi[u]$ to a hypercube of unit length for the representation $\mathcal{F}_\chi^h[u]$.

Among the choices for window functions, the Gaussian gives an *optimal* phase space portrait of a distribution in the sense that it *saturates* the uncertainty principle, for it is equally localized in both, its spatial and frequency contents. The corresponding STFT is called the *Gabor transform*, which with respect to the semiclassical rescaling, has the window function $\chi(\mathbf{x}, \mathbf{y}; h) := e^{-\frac{1}{2h}|\mathbf{x}-\mathbf{y}|^2}$ and thus (after modulation as for the h-STFT), it is the map $T_h : \mathcal{D}(\mathbb{R}^n) \to C^\infty(T^*\mathbb{R}^n)$ defined by,

$$T_h[u](\mathbf{x}, \mathbf{p}) := \int_{\mathbb{R}^n} \overline{\psi_h^\zeta(\mathbf{y})} u(\mathbf{y}) \, d\mathbf{y} \tag{18}$$

with $\psi_h^\zeta(\mathbf{y}) := e^{-\frac{1}{2h}|\mathbf{x}-\mathbf{y}|^2 - \frac{i}{h}(\mathbf{x}-\mathbf{y})\cdot\mathbf{p}}$ and the phase space point being denoted by $\zeta := (\mathbf{x}, \mathbf{p})$. This is also known as the FBI transform and $\psi_h^\zeta$ is a *coherent state* centered at $\zeta$ in phase space. Since the definition of $T_h$ involves integration only locally about $x$ (modulo $O(h^\infty)$ errors), the Gabor transform can be naturally generalized to smooth, compact manifolds by taking $\psi_h^\zeta(\mathbf{y}) = e^{-i\overline{\phi}(\mathbf{y};\zeta)}$ with complex-valued $\phi$ satisfying certain *admissibility* conditions such that the function resembles its Euclidean counterpart in local coordinates in a neighbourhood of $\{(\mathbf{x}, \mathbf{x}, \mathbf{p}) \in \mathcal{M} \times T^*\mathcal{M}\}$. For details around this, see [13].

In many signals analysis applications, the amplitude of the STFT alone is sufficient: indeed, the *spectrogram* $\zeta \mapsto |\mathcal{F}_\chi[u](\zeta)|^2$ is used for signal analysis and detection in applications as varied as seismology [14], speech processing [15], radar detection [16] and gravitational wave detection [17]. A spectrogram of $u$ yields a density plot depicting its concentration in phase space at a resolution given by the corresponding window function. This provides a heatmap representation of signals, which along with the noted applications, further connects image processing tools and deep neural networks to analysis of data originating from such a variety of areas. Fundamentally, the spectrogram allows to associate frequency bands with the spatial regions at which they are most prominent, thereby identifying the signal-containing regions of space and the frequencies present there.

Coming back to our application and comparing (17) with (18) gives that in fact,

$$\rho_t(\mathbf{x}^*; \zeta) = |T_h[\hat{V}_t^\dagger|\delta_{\mathbf{x}^*}\rangle](\zeta)|^2,$$

that is, the phase space density $\rho_t(\mathbf{x}^*; \cdot)$ gives the *Gabor* spectrogram of the distribution $\hat{V}_t^\dagger|\delta_{\mathbf{x}^*}\rangle$. We have that $\rho_t(\mathbf{x}^*; \zeta) = |T_h[\hat{U}_t^\dagger|\delta_{\mathbf{x}^*}\rangle](\zeta)|^2 + O(h)$ due to $|T_h[\hat{U}_t^\dagger|\delta_{\mathbf{x}^*}\rangle](\zeta)|^2 \equiv |\langle\psi_\zeta^h|\hat{U}_t^\dagger|\delta_{\mathbf{x}^*}\rangle|^2$ and the results in [1], so it follows straightforwardly[3] that the approximation holds uniformly over compact subsets of phase space. Furthermore, it is also shown in [1] that $\mathbf{x} \mapsto \langle\delta_\mathbf{x}|\hat{V}_t|\psi_h^\zeta\rangle$ is $h$-localized[4] over $\mathbf{x} \in \mathcal{M}$ to an $O^*(h^{\frac{1}{2}})$-neighbourhood[5] of $\mathbf{x}_t := \pi_\mathcal{M}\Phi^t(\zeta)$. Therefore, by the dual perspective, we have that, *the spectrogram $\rho_t(\mathbf{x}^*; \zeta)$ observed over a bounded region of phase space containing $\mathcal{N}_{\mathbf{x}^*}^t := \{\zeta \in \mathcal{N} \mid \pi_\mathcal{M}\Phi^t(\zeta) = \mathbf{x}^*\}$ in its interior is $h$-localized in an $O^*(h^{\frac{1}{2}})$-neighbourhood of $\mathcal{N}_{\mathbf{x}^*}^t$*. This implies that by observing the spectrogram over a neighbourhood $\mathcal{N}$ of the unit cosphere bundle[6] $S^*\mathcal{M}$, we can recover the geodesic flow at time $t$ initialized at $\zeta_\mathbf{p}^* := (\mathbf{x}^*, \mathbf{p})$ for any momentum $\mathbf{p} \in S_{\mathbf{x}^*}^*\mathcal{M}$.

In practice, we have access only to the *discrete spectrogram*

$$\breve{\rho}_t(\mathbf{x}^*; \breve{\zeta}) := |(\psi_{\breve{\zeta}}^h|\breve{V}_t^\dagger|\delta_{\mathbf{x}^*})|^2$$

with $|\psi_{\breve{\zeta}}^h\rangle$ constructed as in Section II.D.2 using either LPCA or extrinsic coordinates. With high probability, this discrete picture is a faithful approximation to the spectrogram of $\hat{U}_t^\dagger|\delta_{\mathbf{x}^*}\rangle$ such that we can recover the geodesic flow in all directions emanating from $\mathbf{x}^*$; more precisely, we have,

**Proposition 3.** *If initial states $|\psi_{\breve{\zeta}_{\ell,p}}^h\rangle$ are constructed using LPCA with $\breve{\zeta}_{\ell,p} := (v_\ell, \breve{\Pi} \cdot p)$ for $p \in \mathcal{T}_\delta := \{\delta < ||p||_{\mathbb{R}^n} \leq 1 + \delta\}$ and $0 < \delta < 1$, then under the same conditions on the rates of $h$ and $\gamma$ as given in Proposition 2, we have that w.h.p.,*

$$\sup_{\substack{(\mathbf{x}^*; \ell, p) \in \\ \mathcal{M} \times (X_N \times \mathcal{T}_\delta)}} \sup_{d_g(\mathbf{x}, \mathbf{x}_\ell) \lesssim h^\beta} |\breve{\rho}_t(\mathbf{x}^*; \breve{\zeta}_{\ell, p}) - \rho_t(\mathbf{x}^*; \mathbf{x}, \Pi \cdot p)| = O(h).$$

*Thus, the discrete spectrogram concentrates around $\{\Phi^t(\mathbf{x}^*, \mathbf{p}) \mid \gamma - \delta \leq ||\mathbf{p}||_{\mathbf{x}^*} \leq 1 + \gamma + \delta\}$.*

**Remarks.**

1. The supremum over an $h^\beta$-neighborhood of $\mathbf{x}_\ell$ above tells that with sufficiently many samples and *w.h.p.*, our discrete perspective of the spectrogram extends to the neighborhood of each point along the manifold component of phase space and matches the continuum picture well.

2. The proposition is stated for initial states constructed using LPCA, but a similar result holds also for initial states constructed with extrinsic coordinates. We deal only with LPCA here because the statement is less cumbersome, for in the case of extrinsic coordinates we need to discretely consider the possible directions given by neighboring points of each $\mathbf{x}_\ell$, which then also complicates a bit the argument, though the core reasoning is essentially the same.

---

[3] We use that $\rho_t(\mathbf{x}^*; \zeta)$ and $\zeta \mapsto \langle\delta_{\mathbf{x}^*}|\hat{U}_t^\dagger|\psi_h^\zeta\rangle$ are smooth in $\zeta$, which is due to the smoothing property of $T_h$ applied to distributions.
[4] We say that $u_h \in C^\infty$ is *h-localized* in a neighbourhood $\mathcal{N}$ if $|u_h| \lesssim h$ in the complement of $\mathcal{N}$.
[5] We use the notation $O^*(\varepsilon) := O(\varepsilon^{1-\delta})$ for any $\delta > 0$.
[6] The unit cosphere bundle $S^*\mathcal{M}$ consists of the points in phase space with unit cotagent vectors.

*Proof sketch.* If the state is prepared using LPCA according to (12) with $\check{\zeta} = (v_\ell, \check{p}) \in X_N \times \mathbb{R}^{\tilde{\nu}}$, then there is a corresponding $\zeta = (\mathbf{x}_\ell, \mathbf{p}) \in T^*\mathcal{M}$ with $\mathbf{x}_\ell := \iota^{-1}(v_\ell)$ and $\mathbf{p}$ given by the orthogonal projection of $\check{p}$ onto the tangent space at $v_\ell$. The proof of Lemma 2 readily shows that with $\gamma \in \omega(N^{-\frac{2}{d+2}})$, given $\beta > 0$ and $h \sim \gamma^{\frac{4}{\nu+4(1+\beta)}}$, we have w.h.p. that $\nu = \tilde{\nu}$, $\|\check{\Pi} - \Pi_{\mathbf{x}_\ell}\| \leq \gamma$ and $\sup_{(\mathbf{x}^*;\ell,p) \in \mathcal{M} \times X_N \times \mathcal{T}_\delta} |\langle \delta_{\mathbf{x}^*} | \psi^h_{\check{\zeta}_{\ell,p}} \rangle - \langle \delta_{\mathbf{x}^*} | \psi^h_{\zeta_{\ell,p}} \rangle| \lesssim h^\beta$ for $\check{\zeta}_{\ell,p} := (v_\ell, \check{\Pi} \cdot p)$ and $\zeta_{\ell,p} := (\mathbf{x}_\ell, \Pi \cdot p)$. By application of the inequality (15) to our case and noting that the underlying inequality (14) is independent of the phase space localization parameter $\zeta$, we then have that $\sup_{(\mathbf{x}^*;\ell,p) \in \mathcal{M} \times X_N \times \mathcal{T}_\delta} ||\psi^h_{t,\check{\zeta}_{\ell,p}}\rangle|^2 - ||\check{\psi}^h_{t,\zeta_{\ell,p}}\rangle|^2| \lesssim h$. Following the methods in the proof of [1, Lemma 14] and using smoothness of $p \mapsto |\psi^h_{\zeta_{\ell,p}}\rangle$, we have that *w.h.p.*, $\sup_{(\mathbf{x}^*;\ell,p) \in \mathcal{M} \times X_N \times \mathcal{T}_\delta} ||\psi^h_{t,\check{\zeta}_{\ell,p}}\rangle|^2 - ||\check{\psi}^h_{t,\zeta_{\ell,p}}\rangle|^2| \lesssim h$. Further, another application the same methods gives that $\sup_{d_g(\mathbf{x},\mathbf{x}_\ell) \lesssim h^\beta} ||\psi^h_{t,\zeta_{\ell,p}}\rangle|^2 - ||\psi^h_{t,\zeta_{\mathbf{x},p}}\rangle|^2| \lesssim h$. Thus, combining these inequalities gives the desired bound as in the statement of the Proposition. □

Figure 3 in the main text illustrates the ability of the discrete spectrogram $\check{\rho}(\mathbf{x}^*; \check{\zeta})$ to resolve geodesic information with the proper choice of $h$.

### II.D.5. Using geodesic distances to embed

As mentioned in the main text, the geodesic distances extracted using quantum dynamics can be used to embed datasets in Euclidean space for visualization and recovery of reduced order coordinates. To do so, for each sample point $v_\ell$ in the dataset, we propagate $n_{\text{coll}}$ coherent states whose position coordinates are initially localized on $v_\ell$. The coherent states are constructed with momenta in orthogonal directions in the tangent space at $v_\ell$, estimated using LPCA, as detailed above. Each state in the collection will propagate along a geodesic. The propagation is done in steps of $\Delta t$, by applying $\check{U}_{\Delta t}$, for $n_{\text{prop}}$ steps. For each step, this enables the extraction of a *geodesic spray* consisting of $n_{\text{coll}}$ sample points that are approximately a geodesic distance $i \times \Delta t$ for $1 \leq i \leq n_{\text{prop}}$ from the initial state. We populate a geodesic distance matrix $\mathcal{G}$ using this data[7], which defines an adjacency matrix for a graph $\mathcal{X}_N$ on $X_N$. This adjacency matrix captures intrinsic manifold distances between the data points in $X_N$, and we can perform emebeddings of $\mathcal{X}_N$ to achieve tasks such as recover reduced-dimensional coordinates, clustering, classification, *etc.*, as we shall demonstrate below. The matrix $\mathcal{G}$ can be considered a sparsified version of the original distance matrix $\check{T}_\epsilon$ based on Euclidean distances, which typically has no strictly zero entries.

### II.D.6. Choice of parameters

The computational procedure we have outlined in this section requires choice of several parameters, the most important of which are $\epsilon$ and $\alpha$. Appropriate choice of both of these parameters is critical to accurate estimation of geodesic distances, and consequently, manifold learning. This is a familiar situation in many manifold learning approaches and often data-driven heuristics are used to guide the choice of optimal parameters; *e.g.,* see Ref. [18] for an approach to choosing the $\epsilon$ scale parameter in graph Laplacian applications. We now show that Proposition 1 can be used to define a procedure for identifying regions in $(\epsilon, \alpha)$ space that lead to good performance for a given dataset.

Proposition 1 and the semiclassical representation of $\hat{\mathcal{L}}_{\epsilon,\lambda}$ imply the following about its expectation under a coherent state $h^2 \langle \psi^h_{\zeta_0} | \hat{\mathcal{L}}_{\epsilon,\lambda} | \psi^h_{\zeta_0} \rangle = |\mathbf{p}_0|^2 + \mathcal{O}(h)$ [1]. If we choose normalized momenta for initial states, this expectation $\approx 1$. Therefore, a heuristic for choosing $\epsilon$ and $\alpha$ is to compute in the finite-dimensional setting, the deviation $\mathcal{D} = |h^2 \langle \psi^h_{\zeta_0} | \check{L}_{\epsilon,\lambda} | \psi^h_{\zeta_0} \rangle - 1|$ and identify the viable $(\epsilon, \alpha)$ pairs as the ones where this quantity is small. Note that this only requires construction of the matrix $\check{L}_{\epsilon,\lambda}$ and initial state vector, no spectral computations or propagations. Given that $|\psi^h_{\zeta_0}\rangle$ is a restriction of a coherent state to the samples, this deviation condition is effectively evaluating the closeness of matrix elements elements of $\check{L}_{\epsilon,\lambda}$ (in the $\mathcal{O}(h)$ orthonormal coherent state basis) to the asymptotic quantity $\hat{\mathcal{L}}_{\epsilon,\lambda}$. In practice, we recommend computing the average of this deviation over initial states centered at several sampled data points or to fix the initial state location ($\mathbf{x}_0$) and to compute $\check{L}_{\epsilon,\lambda}$ using several samples from the dataset.

---

[7] The matrix $\mathcal{G}$ is a function of $N, \epsilon, \alpha, \Delta t, n_{\text{prop}}, n_{\text{coll}}, \delta_{\text{PCA}}$, but we do not explicitly notate this dependence for conciseness.

## III. APPLICATIONS

In this section we demonstrate geodesic distance extraction using quantum propagation, and its application for visualization, extraction of reduced order coordinates, and clustering.

We approximate the unitary propagator $\hat{V}_t$ through the graph Laplacian specified in Eq. (6). We choose an exponential function for $K(\cdot)$, resulting in a Gaussian kernel $[\check{T}_\epsilon]_{i,j} = e^{-||v_i-v_j||^2/2\epsilon}$ (in which case, $c_0 = \sqrt{\pi}$ and $c_2 = \sqrt{\pi}/2$) and in the following, we will assume $\lambda = 1$ normalization to reduce the error terms appearing in Eq. (7), for in practice we find it to perform slightly better than $\lambda = 0$, although the results are quite comparable. For each application, we have used the quantities $\langle \check{\mathbf{x}}_t \rangle$ or $\langle \check{\mathbf{x}}_t \rangle^{\text{LPCA}}$ (see Section II.D.3) computed directly from the data.

As detailed in Section II.D.6 the deviation measure $\mathcal{D}$ can be used to select parameters $\epsilon$ and $\alpha$. We present plots of $\mathcal{D}$ for all the datasets presented in this section in Section VI.

### III.A. Model examples

We begin with a demonstration of our approach in the case of two model manifolds, for which we are able to compute to high precision the true geodesics, spectrum and eigenfunctions. This allows us to evaluate not only the accuracy of our methods to the geodesic approximations, but also the effects of the spectral deviation between the graph Laplacian and Laplace-Beltrami operators on these methods.

We also give experimental evidence that the quantum propagation of initial states from Section II.D.2 *has more global knowledge* of geodesics than straight line projections from LPCA. This is a natural comparison since the constructions of our initial states, whether using extrinsic or LPCA coordinates in the phases, are directly related to linear approximations of local neighborhoods of the manifold, which locally resolve normal coordinates. Nevertheless, the neighbourhood sizes for which LPCA coordinates centered at a point $\mathbf{x}_0$ resolve normal coordinates are asymptotically small in $N$, while the quantum propagations resolve geodesics within the geodesic ball of radius $\text{inj}(\mathbf{x}_0)$ about $\mathbf{x}_0$. Therefore, we expect to see a deviation between the two forms of geodesic approximations: specifically, given LPCA coordinates about $v_0 := \iota(\mathbf{x}_0)$ and a direction vector $p_0$ in these coordinates, we include in our comparisons the point $\mathbf{x}_t^E$ that is closest in Euclidean norm to $\vartheta_0^{v_0} + p_0 t$.

#### III.A.1. 2-sphere

We uniformly sample $N = 8000$ points from the unit 2-sphere embedded in $\mathbb{R}^3$ ($n = 3, \nu = 2$). For this example, we use LPCA-based constructions of initial states and position expectations. The quantum propagator $\check{V}_{\Delta t}$ is formed with $\Delta t = 0.1, \epsilon = e^{-4.7}$. Fig. 2(a) shows an example propagation of an initial coherent state ($\alpha = 1.6$) localized a point (black square) with some initial momentum. The red curve shows the true unit-speed geodesic path, obtained by solving the geodesic equation for the 2-sphere, with the given initial conditions, and the magenta dots show $\langle \check{\mathbf{x}}_t \rangle^{\text{LPCA}}$ for $t = i\Delta t$ for integer $i \leq 10$. The color of the point cloud encodes the propagated coherent state $|\psi_{t=1,\zeta_0}\rangle$, with the data points colored according to the probability mass at the point, $q_\ell$. The state remains localized as it propagates along the geodesic, and it is concentrated on points a geodesic distance $t$ from the initial point. In Fig. 2(b) we show a snapshot of the propagated state at $t = 1$, with the black circles highlighting sample points $v_\ell$ satisfying $d_g(v_\ell, v_0) < h$ – where $d_g$ is the geodesic distance on the sphere and $v_0$ is the sample the initial state is localized on – which clearly shows that the delocalization of $|\psi_{t,\zeta_0}\rangle$ is set by $h$.

To quantify how well the propagated coherent states track geodesics, we repeat such propagations $m = 50$ times with uniformly sampled initial points and initial momenta toward the nearest neighboring point in each case. At each $t = i\Delta t$, we compute $\langle \check{\mathbf{x}}_t \rangle^{\text{LPCA}}$ and $\mathbf{x}_t^E$. Fig. 2(c) shows the average and one standard deviation error bars for the geodesic distance of these positions to the initial point. We see that $\langle \check{\mathbf{x}}_t \rangle^{\text{LPCA}}$ corresponds to points that are approximately a geodesic distance $t$ away on average, with very small variation. In contrast, $\mathbf{x}_t^E$, the points computed by the naïve local Euclidean approximation of propagation along geodesics begin to deviate from geodesic distance $t$ for $t > 0.6$. The inset to Fig. 2(c) also shows the average error in direction of propagation for $\langle \check{\mathbf{x}}_t \rangle^{\text{LPCA}}$ and $\mathbf{x}_t^E$. For each of the random propagations, this error is computed as $\Delta_{\text{dir}}(t) = |1 - p_0^T g_{v_0} p_0'(t)|$, where $p_0$ is the initial momentum dictating propagation direction, $g_{v_0}$ is the metric tensor for the 2-sphere at the initial point $v_0$, and $p_0'(t)$ is the initial momentum that results in propagation to the point $\langle \check{\mathbf{x}}_t \rangle^{\text{LPCA}}$ or $\mathbf{x}_t^E$. This quantity is just the misalignment between the true momentum vector at $t = 0$ and the momentum vector required to propagate to the point determined by either procedure. We evaluate $\Delta_{\text{dir}}(t)$ away from $t = 0$ since at very short times because of the poor resolution of the manifold given by the finite number of samples, this error can be large; *i.e.,* there may not be a data point

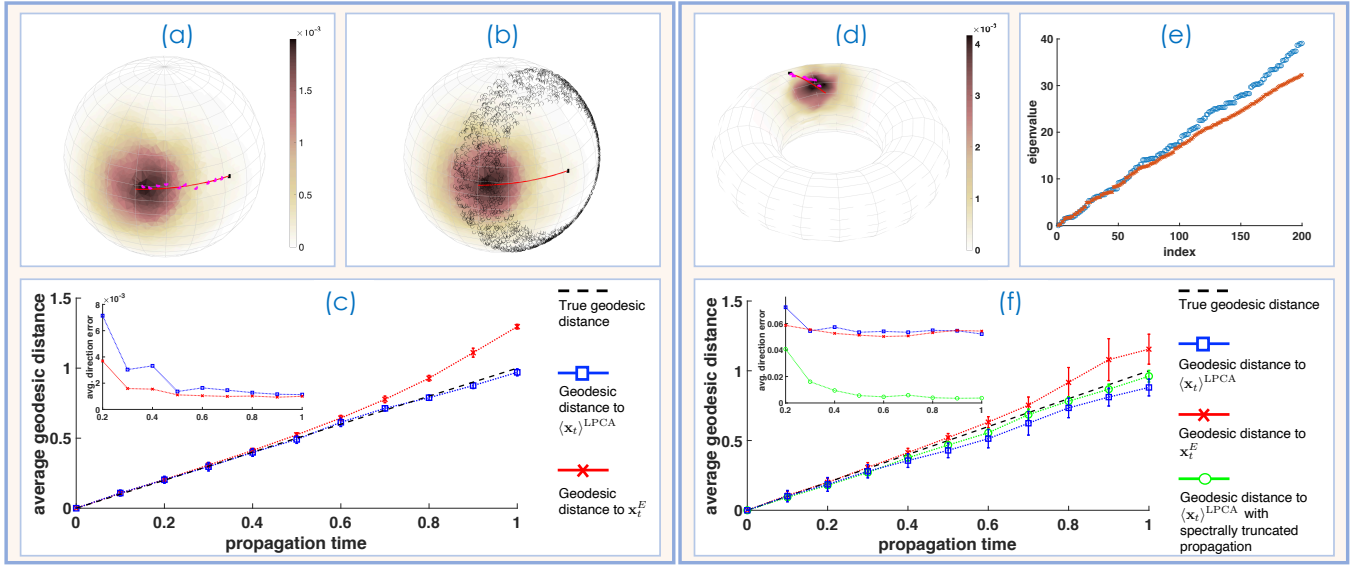

FIG. 2: Illustration of geodesic extraction through quantum dynamics on two sample manifolds: the sphere ((a)-(c)) and torus ((d)-(f)). In both cases, LPCA is used to construct initial states and compute expected position. (a) Example propagation of a coherent state on the unit sphere with $N = 8000$ sample points and $\epsilon = e^{-4.7}, \alpha = 1.6, \delta_{\text{PCA}} = 1.5$. Initial expected position is indicated by the black square. The point cloud is colored according to $q_\ell$, and triangulated using Delaunay triangulation for visualization purposes. The red curve is the true unit-speed geodesic curve up to $t = 1$ with the specified initial state, and the magenta dots indicate mean positions determined through data-driven coherent state propagation for ten steps of $\Delta t = 0.1$. (b) The propagated state at $t = 1$, with the black circles indicating sample points $v_\ell$ such that $d_g(v_\ell, v_0) \leq h$, to show that the propagated state is localized in a region dictated by $h$. (c) The average geodesic distance of the propagated state from the initial point as a function of propagation time. The average is taken over $m = 50$ propagations and the spread around the average is shown using one standard deviation error bars. The blue squares correspond to statistics of the geodesic distance to the point extracted through quantum dynamics and the red crosses correspond to Euclidean propagation. The inset shows the average error in propagation direction as a function of time. (d) Example propagation of a coherent state on a torus with radii $(r = 0.8, R = 2)$ with $N = 12000$ sample points and $\epsilon = e^{-3.8}, \alpha = 1.8, \delta_{\text{PCA}} = 2.0$. The point cloud is colored according to $q_\ell$, and an $\alpha$-shape triangulation of the point cloud is used for visualization. The red curve is the true unit-speed geodesic curve up to $t = 1$ with the specified initial state, and the magenta dots indicate mean positions determined through data-driven coherent state propagation for ten steps of $\Delta t = 0.1$. (e) A comparison of the first 200 eigenvalues of the true LB operator on the torus (blue circles) and the data-driven reconstruction (red crosses). (f) The average geodesic distance of the propagated state from the initial point as a function of propagation time. The average is taken over $m = 50$ propagations and the spread around the average is shown using one standard deviation error bars. The blue squares correspond to statistics of the geodesic distance to the point extracted through quantum dynamics and the red crosses correspond to Euclidean propagation. The green circles correspond to statistics of geodesic distance calculated using a spectrally truncated quantum propagator. The inset shows the average error in propagation direction on the torus.

directly on the geodesic curve and the closest point requires a significantly different initial momentum to propagate to with small $t$. This is an artifact of finite sampling that becomes negligible at larger $t$, as shown by Fig. 2(c).

### III.A.2. Torus

The sphere is a relatively simple manifold with constant curvature. We next illustrate our method on a more complex manifold with positive and negative curvature, the torus embedded in $\mathbb{R}^3$ ($n = 3, \nu = 2$). We uniformly sample $N = 12000$ points on a torus with radii $r = 0.8, R = 2$ and form the quantum propagator $\check{V}_{\Delta t}$ with $\Delta t = 0.1, \epsilon = e^{-3.8}$. Fig. 2(d) shows an example propagation of an initial coherent state (constructed using LPCA, $\alpha = 1.8$). The red curve shows the true unit-speed geodesic path, obtained by solving the geodesic equation for the torus, with the given initial conditions, and the magenta dots show mean positions determined through data-driven coherent state propagation for ten steps of $\Delta t$. The color of the point cloud encodes the propagated coherent state $|\psi_{t=1,\zeta_0}\rangle$, with the data points

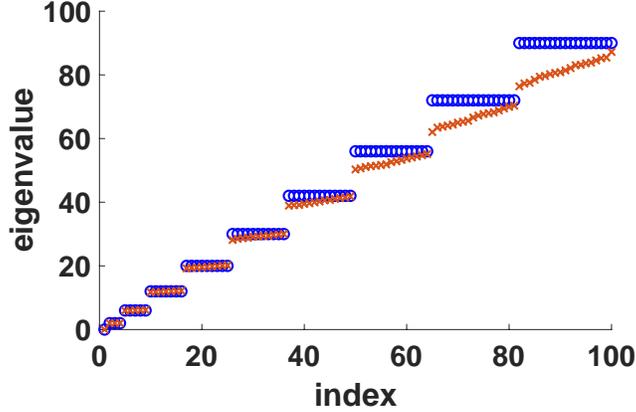

**FIG. 3:** Eigenvalues of the LB operator on the sphere (blue circles) compared to the eigenvalues of the data-driven approximation of the LB operator constructed from $N = 8000$ samples from the sphere with $\epsilon = e^{-4.8}$.

colored according to the probability mass at the point, $q_\ell$. To understand more systematically the quality of geodesic distance extraction, we repeat $m = 50$ such propagations from uniformly sampled initial points (with initial momenta toward the nearest neighboring point in each case) and show the statistics of the geodesic distance to the extracted point, $\langle \check{\mathbf{x}}_t \rangle^{\text{LPCA}}$, as a function of propagation time $t$ in Fig. 2(f). The blue squares show the average and one standard deviation variation of the geodesic distance extracted up to $t = 1$. For comparison we also show the statistics of geodesic distance to the local Euclidean approximation of geodesic propagation $\mathbf{x}_t^E$. The inset in Fig. 2(f) also shows the error in propagation direction (quantified in the same way as for the sphere example) for the two propagation methods.

We see that the quantum propagation incurs slightly more error for the torus than for the sphere, however, the average error is still $\mathcal{O}(h)$. The discussion in Section II.C.1 implies that the propagation of a coherent state with scale set by $h$ is dictated by a restricted portion of the spectrum of $e^{\frac{i}{\hbar}\hat{H}t}$. Thus, to see the potential of the approach with larger number of data samples, it makes sense to compare to propagation under a more accurate spectral truncation of $\hat{H}$. In Section V we use intrinsic information about the torus to directly construct the LB operator for the torus by numerical diagonalization. We truncate the spectral data at 800 total eigenvalues [8] (including degeneracies) and eigenvectors and form a *spectrally truncated propagator*, $\tilde{V}_{\Delta t}$, defined over a regular discretization of the torus over $N' = 51300$ points. Fig. 2(e) compares the first 200 eigenvalues of the data-driven LB reconstruction, and the more accurate reconstruction through numerical diagonalization, and clearly shows the disagreement between spectra after $\sim 100$ eigenvalues. We then propagate initial coherent states (with $h = e^{-1}$ as above), extract points closest to the expected position of the propagated states, $\langle \check{\mathbf{x}}_t \rangle^{\text{LPCA}}$, using the same approach described above, and the statistics of the geodesic distance to these points are shown in Fig. 2(f) as green circles. The results show extremely low error in geodesic distance of $\langle \check{\mathbf{x}}_t \rangle^{\text{LPCA}}$, and low errors in propagation direction (green circles in the inset of Fig. 2(f)). We note that the reduction in error in direction is also a result of a better estimation of the tangent space at a point due to the larger number of data samples. Since the phase-space localization $h$ is fixed throughout, this is an example where truncation to a better spectral approximation yields tighter geodesic approximation. By the spectral convergence of the graph Laplacian to the LB operator [19], we conclude that with more data the extremely low error geodesic extraction demonstrated with $\tilde{V}_{\Delta t}$ could be reproduced by our method.

For completeness we also present spectrum for the graph Laplacian for the sphere in Fig. 3 (blue circles) and the eigenvalues of the data-driven approximation constructed from $N = 8000$ points (uniformly) sampled from the sphere (red crosses). We see again that these quantities begin to disagree at higher eigenvalues – of particular note is the inability of the data-driven construction to match the degeneracies of the high eigenvalues.

### III.B. Euclidean embeddings

In support of typical applications of manifold learning in data analysis, we show the ability of the embedding procedure described in the main text and Section II.D.5 to capture the geometry of model datasets. Namely, in Fig. 4 we

---

[8] The magnitude of the largest eigenvalue used is 156.5.

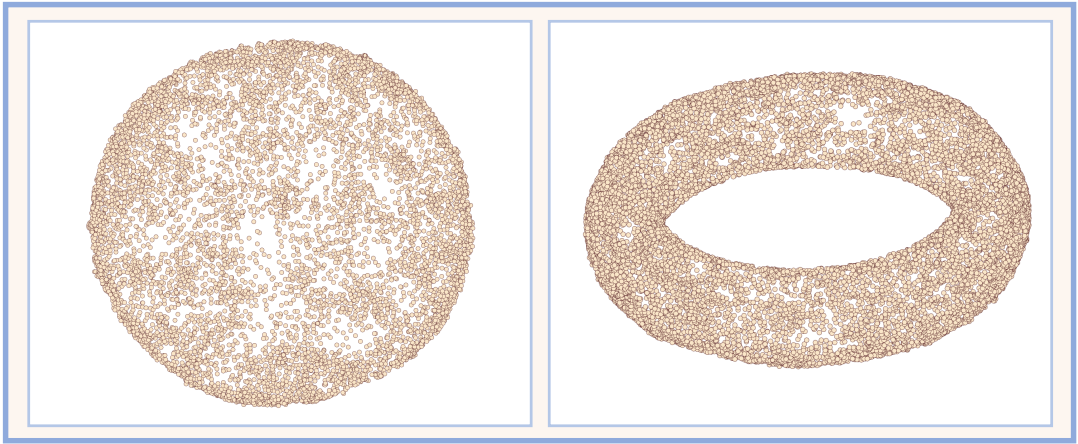

**FIG. 4:** 3D embeddings of samples from a sphere and torus, using force-based graph layout based on geodesic distances extracted using our approach. The data is described in Sec. III. The parameters used are: (**sphere:** $\epsilon = e^{-4.7}, \alpha = 1.6, \delta_{\text{PCA}} = 1.5, \Delta t = 0.1, n_{\text{prop}} = 5, n_{\text{coll}} = 40$), (**torus:** $\epsilon = e^{-3.8}, \alpha = 1.8, \delta_{\text{PCA}} = 2.0, \Delta t = 0.4, n_{\text{prop}} = 1, n_{\text{coll}} = 40$).

display the embedding of the sample points of the sphere and torus data sets (Section III.A.1 and Section III.A.2, respectively) in three dimensions using force-directed layout [20] on the matrix $\mathcal{G}$ extracted from quantum propagations, with parameters as described in the figure's caption.

In both cases, the reconstruction is a good representation of the original manifold. Of course, there is no dimensional reduction in this case, however the reconstructed embedding clearly reproduces the geometry of the original point cloud dataset.

### III.C. SafeGraph COVID-19 mobility data

In this section we present more details of the visualization and clustering of COVID-19 mobility data discussed in the main text.

The global COVID-19 pandemic has resulted in massive and unprecedented disruptions to patterns of daily life. Analyzing changes in mobility during the past year has led to insights into social distancing patterns and infection modalities that can inform public health policy and management [21–24]. Geolocation information from mobile phones provides a wealth of mobility data, and has been the basis of most of the previous studies on mobility changes during the pandemic. This data can be complex and extremely high dimensional, and traditional manifold learning techniques have been shown to be informative for visualization and clustering of this data [22].

We analyze aggregated mobility data in the Social Distancing Metric dataset from SafeGraph Inc. [25]. This is a fine-grained dataset that collects geolocation information from mobile devices, aggregates it at the census block group (CBG) level in the United States and records it daily for a period of over a year. It contains a wealth of information and we extract a simple metric to gauge the mobility patterns of citizens: we compute a daily stay-at-home (SAH) fraction for each CBG that is the ratio of the `completely_home_device_count` to the `device_count` values. The latter is the number of devices seen during the date whose home is determined to be in the CBG being considered, and the former is the number of these devices that were determined to not have left the home CBG during the day [25]. The SAH fraction for a date provides a measure of how curtailed mobility was within a CBG. To allow comparison to an earlier study with the same dataset [22] we limit the data to the 117-day time period from Feb. 23, 2020 to Jun. 19, 2020, which provides a snapshot of mobility patterns during the first three months of the pandemic.

In the following, we present results for analysis of mobility data for the state of Georgia (GA). After removing 17 CBGs with poor quality data, there are 5509 CBGs within the state. Therefore, our dataset has $N = 5509$ samples, each of dimension 117. We minmax normalize the data samples so that the minimum SAH fraction is zero and maximum is one. We performed quantum manifold learning on this dataset with parameters $\epsilon = e^{-1/2}, \alpha = 1.4, \Delta t = 0.1, n_{\text{prop}} = 20, n_{\text{coll}} = 60$. See Section VI for a plot of the deviation measure $\mathcal{D}$ for this dataset that informs these parameter choices. We compute initial states and $\langle \breve{\mathbf{x}}_t \rangle$ using the original extrinsic coordinates and do not employ LPCA for simplicity and since the data dimension is not large enough to prohibit this. After extracting a geodesic distance matrix $\mathcal{G}$, we embed the data in three dimensions using a force-directed layout using $\mathcal{G}$. The data is then clustered into 5 clusters an in Ref. [22]. We use the k-means algorithm for clustering with the coordinates of the 3D

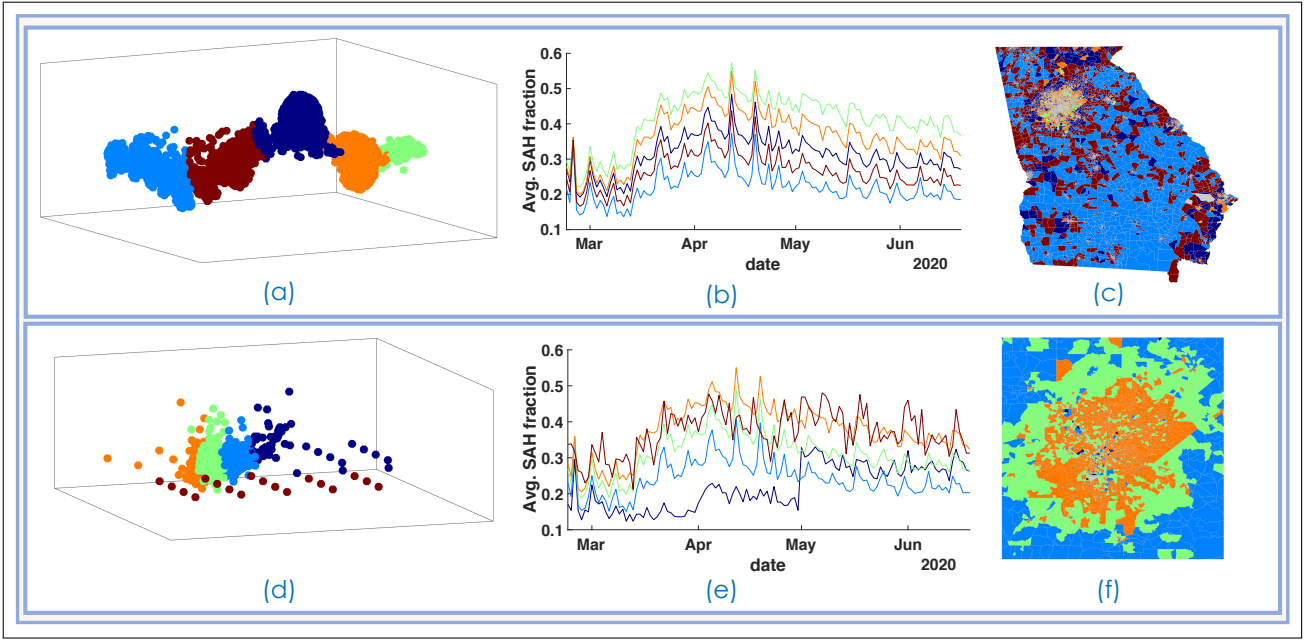

**FIG. 5:** An application of geodesic extraction through quantum dynamics on mobility data from the US state of Georgia (GA). **(a)** Force-based graph layout in $\mathbb{R}^3$ of CBGs based on extracted geodesic distance matrix, where the location of the propagated states is calculated using the mean position $\langle \mathbf{\ddot{x}}_t \rangle$. The 3D coordinates from this embedding are used to cluster the CBGs using k-means clustering, and the five clusters are marked by color. **(b)** The average SAH fraction time series for each of the clusters in (a), with the same color-coding. **(c)** Geographic location of the clusters in GA. CBGs colored in gray are ones for which there was missing or poor quality data – these were excluded from the analysis. **(d)** An $\mathbb{R}^3$ embedding of the same data as in (a), but with the location of the propagated state calculated using the maximum of the associated distribution over data points. Again, clustering is performed using k-means on the embedded coordinates. Poorly connected outlier CBGs are indicated by the brown and dark blue points. **(b)** Average SAH fraction time series for each of the clusters in (d), with the same color-coding. **(f)** Geographic location of the CBG clusters in the Atlanta metropolitan area. We present a zoom in into this area because some of the outlier CBGs identified in (d) are located in this area.

embedding, as opposed to Gaussian mixture model (GMM) clustering as was done in Ref. [22], although the results are similar with GMM clustering.

Fig. 5(a) shows the embedding in 3D with the clusters indicated by colors. In addition, the average SAH fraction time series for each cluster are shown in Fig. 5(b), which shows that the clustering is meaningful; *i.e.,* although almost all CBGs exhibit a similar mobility pattern, the clusters reflect different amounts of overall SAH fraction. The average mobility traces of the clusters are clearly separated. Finally, Fig. 5(c) shows a map of the CBGs, with the same color coding of clusters as in the other subfigures. This geographic representation clearly shows the rural-urban divide in degree of mobility change during the pandemic; the higher SAH fraction clusters are associated with the urban centers in GA, while the majority of the state exhibited behavior consistent with lower SAH fraction curves (light blue and brown clusters).

It is instructive to compare the results in Fig. 5 to those in Ref. [22], where a similar clustering analysis was done with Laplacian eigenmaps. In that work, the authors needed to embed into 14 dimensions in order to obtain good clustering, whereas we obtained meaningful clusters that represent distinct classes of SAH degree through a greater dimensional reduction to three dimensions. An explanation for this might be offered upon looking at the asymptotic situation for manifolds: then, generally speaking, the Laplacian eigenmaps embedding places an eigenfunction of $\hat{\mathcal{L}}_{\epsilon,\lambda}$ at each coordinate axis directly and hence relies on the clustering method to pick up on the geometry exposed by the selected eigenfunctions. It is by now well understood [26] that one may need to look deep into the higher order eigenfunctions with sufficiently small $\epsilon$ to recover non-trivial features of the geometry; *a priori*, this would therefore involve using a large number of the smallest eigenfunctions, especially if there are degeneracies. On the other hand, the method we have presented assigns neighbours based directly upon the intrinsic geodesic distances, which implicitly involve higher order eigenfunctions [27] so that the dependence on $N$ and $\epsilon$ is only in the quality of the application of Egorov's theorem and hence, of geodesic approximation. Therefore, our method potentially enables a smaller embedding dimension to identify the same geometric features of data.

Finally, to illustrate another feature of our approach to manifold learning via quantum dynamics, in Fig. 5(d) we show 3D embedding and clustering using the same methods described above (and same parameters), except using $\ell^*$, the maximum of the probability distribution dictated by a propagated coherent state, as the estimate of the position of the propagated state. While the mean position $\langle \check{\mathbf{x}}_t \rangle$ enables extraction of clusters with consistent average behavior, using the maximum to determine points a certain geodesic distance away results in a procedure that is sensitive to data outliers. This is illustrated by the brown and dark blue clusters in Fig. 5(d), which are disconnected or poorly connected to the remaining data points. When examining the average SAH fraction time series for these clusters in Fig. 5(e) we see that they are distinct from the other time series. Particularly remarkable is the dark blue cluster, which shows a marked change in SAH fraction around May 1st 2020. We note that the number of these outliers is small (18 (62) data points in the brown (dark blue) cluster) and identifying them is a challenging task in such a large dataset. These outlier CBGs are distributed across the whole state, but particularly concentrated in urban areas; *e.g.,* in Fig. 5(f) we show a zoom into the Atlanta metropolitan area, which contains some of the outlier CBGs.

In order to understand the anomalous behavior of the SAH fraction time series for the dark blue cluster we examine the 62 CBGs in this cluster in more detail. Remarkably, the majority of these CBGs contain universities and other institutions of higher learning. Using this as a clue, we can search for markers of changes in social distancing policy at schools and universities in Georgia. There were two significant policy changes for institutions of higher education in GA during this time period. First, most universities transitioned to full online instruction in late-March [28], and second, the cancellation of final examinations and commencement ceremonies at most universities effectively made the end of the Spring semester in late-April [29]. These two changes correlate with the jumps in average SAH fraction seen in dark blue cluster, and thus could provide explanations for these anomalous trends seen in early-April and early-May. It is remarkable that such weak signals present in only 62 data points (1.2% of the 5509 data points) are identified by our algorithm.

The application presented in this section and subsequent analysis demonstrate the applicability of our manifold learning approach to real-world data sets, even ones that are moderate in size ($N \sim 5000$ points). Remarkably, we have empirically found that our approach is also useful in smaller datasets (see next two subsections), where the theoretical properties derived in the large $N$ limit are not necessarily guaranteed.

### III.D. Global COVID-19 mobility patterns

Here we consider another dataset with mobility information during the COVID-19 pandemic, but this time analyze it at a much more coarse-grained geographic scale.

We consider the COVID-19 Community Mobility Reports dataset from Google [30] that aggregates mobile phone location information to capture mobility trends. The dataset charts movement trends in almost every country, including categorizing according to subregions/states in several countries, and covers dates from Feb. 15, 2020 to present day (as of the writing of this manuscript). In addition, the mobility information catalogued across six different categories of places: Retail and Recreation, Grocery and Pharmacy, Parks, Transit Stations, Workplaces and Residences. The key relevant quantity in the dataset is the *percentage change in mobility* from a baseline value for each category and region calculated during the period Jan. 3rd, 2020 to Feb. 6, 2020. The baseline value is calibrated according to day of the week.

In the following, we analyze the dataset over the time period Feb. 15, 2020 to Jan. 24, 2021 (345 days). We first preprocess the data in a couple of steps. First, we coarse-grain the dataset to country-level information; *i.e.,* we consider mobility change patterns for each country. Out of the total 132 countries in the dataset we drop 6 countries with poor quality or significant missing data, leaving a dataset with $N = 126$ countries. Second, we ignore the Residential category in the dataset since (i) it has different units than the other categories and the documentation specifically says it cannot be treated on the same footing as the other categories [30]. This leaves 5 categories across which mobility changes are tracked for 345 days per country, which defines a time series for each category. We concatenate all 5 time series for each country yielding data vectors of dimension $n = 345 \times 5 = 1725$. We normalize each data sample by the maximum $L^2$ norm of the $N = 126$ data samples so that each sample can be viewed as a vector in the unit ball in $R^{1725}$.

We calculate geodesic distances between countries in this dataset and then use the geodesic distance matrix to embed the data in $\mathbb{R}^3$ using the force-directed graph layout embedding technique. Due to the small size of the dataset, we form initial states and calculate expected positions using extrinsic coordinates (no local PCA) and using the maximum of the propagated state as the estimate of the expected position. The parameters used are: $\epsilon = e^{-0.6}, \alpha = 1, dt = 0.2, n_{\text{prop}} = 4, n_{\text{coll}} = 5$. After this embedding (which is a reduction from 1725 dimensions to 3) we perform k-means clustering using the embedding coordinates to cluster the countries into 6 clusters (we experimented with varying numbers of clusters and empirically found that 6 was optimal in terms of grouping clusters according to mobility patterns). The resulting embedding and color-coded clusters are shown in Fig. 6.

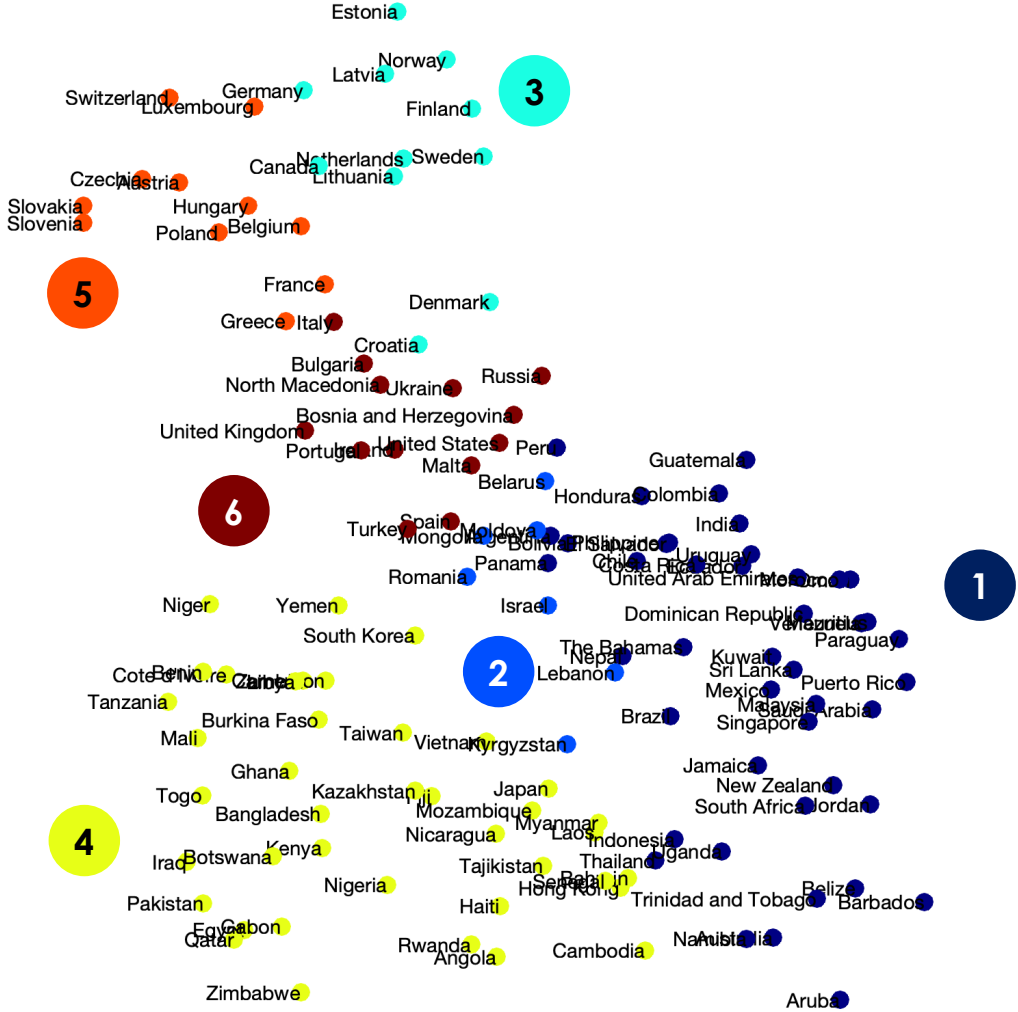

FIG. 6: Force-based graph layout of countries based on geodesic distance matrix extracted using our approach applied to mobility pattern data in these countries during the COVID-19 pandemic. The colors indicate clusters found using k-means clustering based on the 3D coordinates determined by the embedding.

The geodesic distance based embedding is meaningful is several ways. It clearly distinguishes different geographic regions, with countries in a region (*e.g.,* northern Europe) being close to each other. This is reasonable since social distancing policies and the timing of their introduction during the pandemic varied across geographic regions. In addition, the mobility patterns are highly correlated for countries within a cluster. In Fig. 7 we show mobility patterns for some sample countries within each cluster (top) and also the average percentage change in mobility for the countries in each cluster (bottom). Both of these are shown for each of the five categories considered. Each cluster captures a distinct mobility pattern over the 345 days and the countries within a cluster show remarkably similar mobility patterns. Both of these features indicate a meaningful clustering of this high-dimensional dataset.

### III.E. NBA player performance

Another small $N$, high-dimensional dataset we consider is the NBA Players dataset on Kaggle [31], which contains over two decades of statistics on basketball players in the National Basketball Association (NBA). The data was scraped and collated by Justinas Cirtautas using the NBA Stats API.

We first preprocess the dataset and remove any players without at least five consecutive seasons of data, which leaves $N = 920$ samples in the dataset. Then we focus on three key statistics for each player, over the first five years of their career in the NBA (or first five years for which there are records in the NBA Players dataset): average points

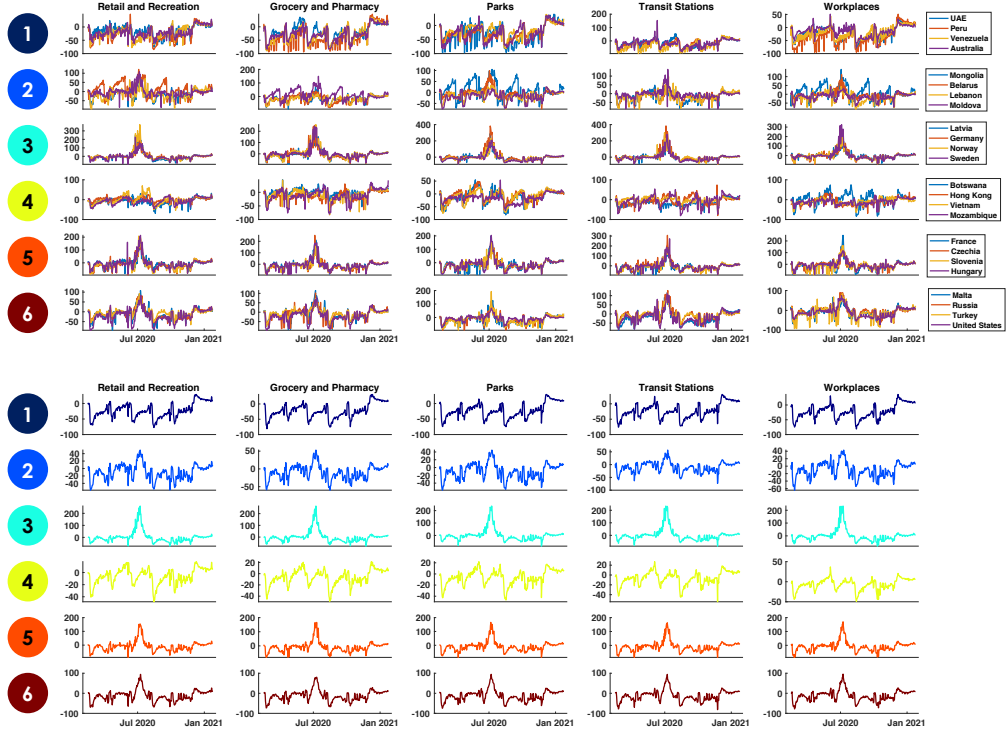

**FIG. 7: (Top)** Percentage mobility changes over the monitored time period (Feb 15, 2020 to Jan 24, 2021) for four randomly sampled countries in each cluster (rows), in each of the categories collected in the Google Community Mobility Reports (columns). The countries are listed to the right of each row. Note the similarity of the mobility patterns for countries in the same cluster. **(Bottom)** The average percentage mobility change for all countries within each cluster, and separated across the categories (columns). Each cluster shows a distinct average mobility pattern.

per game, average rebounds per game, and average assists per game. This yields $3 \times 5 = 15$ features per data point. The data samples are minmax normalized so that the minimum statistic is zero and maximum is one.

We calculate geodesic distances between players in this dataset and then use the geodesic distance matrix to embed the data in $\mathbb{R}^3$ using the force-directed graph layout embedding technique. Due to the small size of the dataset, we form initial states and calculate expected positions using extrinsic coordinates (no local PCA) and using the maximum of the propagated state as the estimate of the expected position. The parameters used are: $\epsilon = e^{-1.3}, \alpha = 1.2, \Delta t = 0.1, n_{\text{prop}} = 6, n_{\text{coll}} = 40$. As with the previous example, we perform k-means clustering based on the coordinates from the 3D embedding. In Fig. 8 we show the embedding and clustering for five clusters. Some of the nodes are also labeled with the player names.

The embedding and clustering shown in Fig. 8 groups players according to performance and playing style. In Fig. 9 shows the average statistics over five years for each of the clusters. Each cluster clearly shows a distinct pattern in terms of average statistic over the five years. The highest performers, or superstars who excel in all three categories (points, rebounds, assists), are grouped in the light blue cluster, while the orange cluster contains specialist centers and forwards who excel at rebounding and also score points, and the brown cluster contains specialist guards who excels at assists and scoring points.

As with the global COVID-19 mobility pattern dataset, the low-dimensional embedding and subsequent clustering of this dataset using our approach demonstrates its utility on even small datasets.

## IV. TANGENT SPACE ESTIMATION VIA LOCAL PCA

Local principal component analysis (LPCA) is utilized to formulate the initial coherent state and to calculate the expected position of the propagated state using only local data. Both of these procedures rely on the fact that LPCA provides an estimate of the tangent space of a manifold from sampled data, and LPCA coordinates match normal

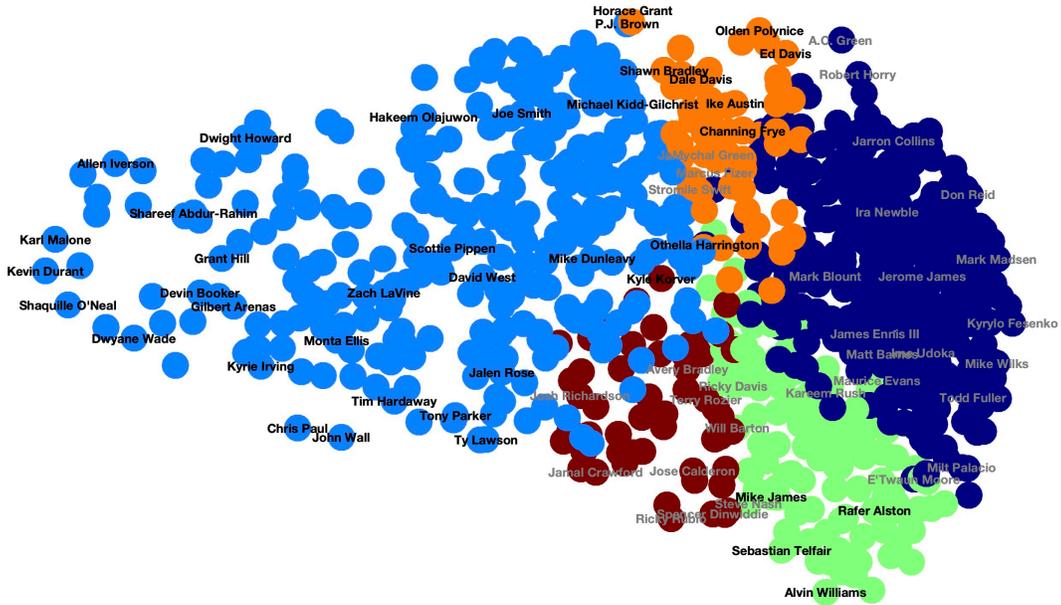

**FIG. 8:** Force-based graph layout of 920 NBA players based on playing statistics over five years. The colors indicate clusters found using k-means clustering based on the 3D coordinates determined by the embedding. Some of the player names are listed to show representative players from each cluster (listing all names would make the plot illegible)

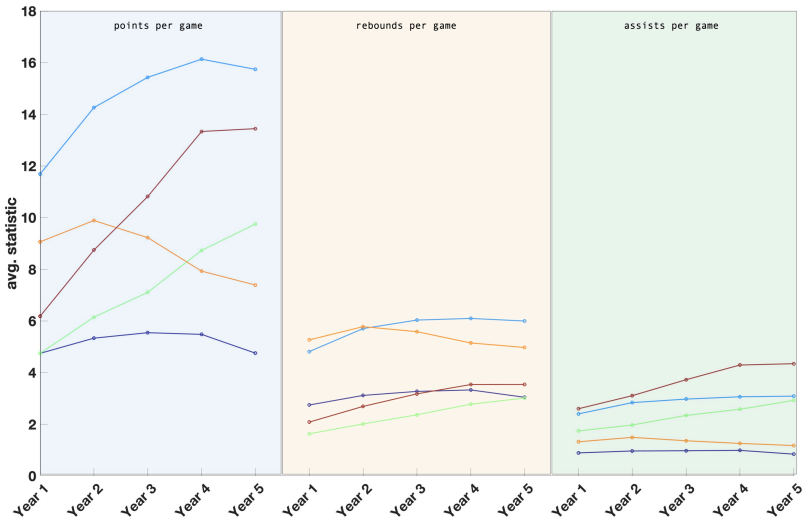

**FIG. 9:** Average statistics, when averaged over all players in a given cluster. The colors of the averages correspond to the cluster color-codes in Fig. 8. The panels show from left to right, the average points per game, average rebounds per game, and average assists per game, all over five years.

coordinates on the manifold up to third order in the size of the local neighborhood [32].

In this section we illustrate the quality of LPCA constructed tangent spaces for the sphere and torus model manifolds for which intrinsic global coordinates are known and analytical expressions for all geometric quantities can be derived. We parameterize both two-dimensional manifolds by angular coordinates $(\theta, \phi)$ (explicit forms of the coordinatizations given below).

We quantify the accuracy of LPCA tangent space estimates as a function of the LPCA neighborhood size, $\delta_{\text{PCA}}$ using two error metrics. First, we compare the tangent space estimated by LPCA to the true tangent space at a point on the manifold using the subspace angle error metric:

$$\Delta_{TS} = \arccos(|V_1^\mathsf{T} V_2|), \tag{19}$$

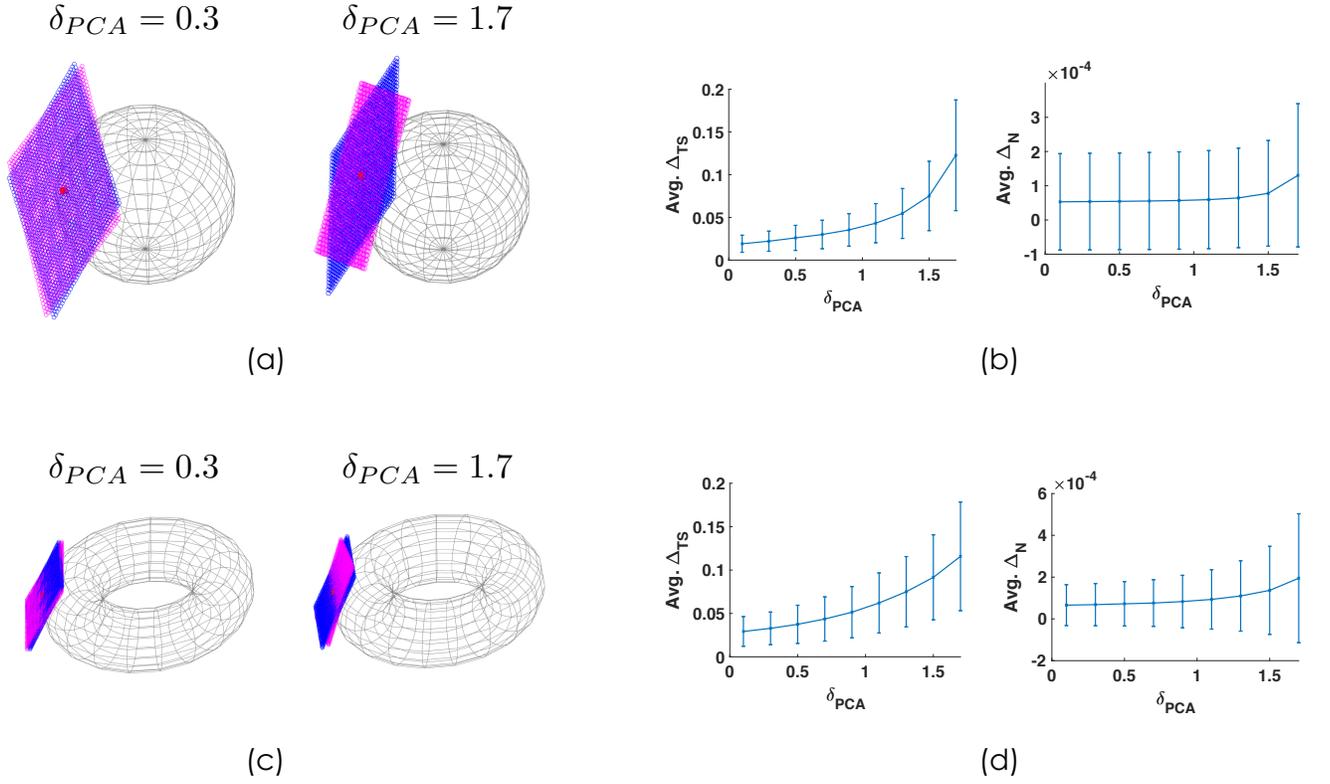

**FIG. 10:** Error metrics for LPCA-based tangent space estimation for two model manifolds.

where $V_1(V_2)$ is a matrix whose columns are vectors that span the LPCA estimated tangent space (true tangent space). The second error metric we employ quantifies the error in the estimated norm of vectors in the tangent space at a point $(\theta_0, \phi_0)$. We compute $|\mathbf{p}| = |\vartheta_0 - \vartheta_j|$, where $\mathbf{p}$ is a vector between the point of interest $(\theta_0, \phi_0)$ with LPCA coordinates $\vartheta_0$, and some nearby point $\vartheta_j$. This quantity is compared against the tangent vector norm $\hat{n} = \sqrt{[d\theta, d\phi] g_{\theta,\phi}(\theta_0, \phi_0) [d\theta, d\phi]^\mathsf{T}}$, with $d\theta = \theta_j - \theta_0$ and $d\phi = \phi_j - \phi_0$, as:

$$\Delta_N = |||\mathbf{p}| - \hat{n}|| \qquad (20)$$

This is an important error to quantify because the momenta for initial coherent states are normalized in LPCA coordinates, and in order to ensure unit speed propagation we require $|\mathbf{p}| = 1$.

### IV.A. Sphere

Consider a sphere unit radius parameterized by angles $0 \leq \theta < \pi$, $0 \leq \phi < 2\pi$ with radii $R, r$ ($r < R$) and parameterized by angles $\theta, \phi$, with Euclidean coordinates in a 3D embedding given by:

$$(x, y, z) = (\sin(\theta)\cos(\phi), \sin(\theta)\sin(\phi), \cos(\theta)) \qquad (21)$$

The tangent space for this manifold at a point $\theta_0, \phi_0$ is a plane in the 3D embedding space spanned by the two vectors

$$e_1 = [-r\sin(\theta_0)\cos(\phi_0), -r\sin(\theta_0)\sin(\phi_0), r\cos(\theta_0)]$$
$$e_2 = [-(R + r\cos(\theta_0))\sin(\phi_0), R + r\cos(\theta_0)\cos(\phi_0), 0] \qquad (22)$$

We uniformly sample $N = 8000$ points from a 3D embedding of the sphere and evaluate the error metrics $\Delta_{TS}$ and $\Delta_N$ at all points, as a function of $\delta_{\text{PCA}}$, the Euclidean distance cutoff that dictates the LPCA neighborhood size. Fig. 10(a) shows two examples of the estimated versus true tangent space at a point on sphere for $\delta_{\text{PCA}} = 0.3, 1.7$. Then in Fig. 10(b) we show the average value of these error metrics (averaged over all 8000 sampled points) and one standard deviations error bars.

## IV.B. 2-torus

Consider a 2-torus with radii $R, r$ ($r < R$) and parameterized by angles $0 \leq \theta < 2\pi$, $0 \leq \phi < 2\pi$, with Euclidean coordinates in a 3D embedding given by:

$$(x, y, z) = ((R + r\cos(\theta))\cos(\phi), (R + r\cos(\theta))\sin(\phi), r\sin(\theta)) \tag{23}$$

The tangent space for this manifold at a point $\theta_0, \phi_0$ is a plane in the 3D embedding space spanned by the two vectors

$$\begin{aligned} e_1 &= [-r\sin(\theta_0)\cos(\phi_0), -r\sin(\theta_0)\sin(\phi_0), r\cos(\theta_0)] \\ e_2 &= [-(R + r\cos(\theta_0))\sin(\phi_0), R + r\cos(\theta_0)\cos(\phi_0), 0] \end{aligned} \tag{24}$$

We uniformly sample $N = 12000$ points from a 3D embedding of a 2-torus with $R = 2, r = 0.8$ and evaluate the error metrics $\Delta_{TS}$ and $\Delta_N$ at all points, as a function of $\delta_{\text{PCA}}$, the Euclidean distance cutoff that dictates the LPCA neighborhood size. Fig. 10(c) shows two examples of the estimated versus true tangent space at a point on the 2-torus for $\delta_{\text{PCA}} = 0.3, 1.7$. Then in Fig. 10(d) we show the average value of these error metrics (averaged over all 12000 sampled points) and one standard deviations error bars.

As is clear the statistics of these error metrics for the sphere and torus examples, the LPCA estimate of the tangent space is best for small $\delta_{\text{PCA}}$. This creates a trade-off for our application: to get the best estimate of the tangent space a small $\delta_{\text{PCA}}$ is preferred, but in order to obtain a good construction of an initial coherent state, we require a local neighborhood ($\mathcal{S}_0$ in Section II.D.2) that is not too small. To resolve this trade-off we fix $\delta_{\text{PCA}}$ to be large enough to construct good fidelity coherent states, but construct those states using LPCA coordinates extracted from an LPCA computed in a neighborhood of size $\gamma\delta_{\text{PCA}}$, for a scale factor $\gamma \leq 1$. In the illustrations shown in Section III, we used $\gamma = 0.1$ for the sphere, and $\gamma = 0.05$ for the torus.

Note that in general, a different choice of $\delta_{\text{PCA}}$ can be made for the initial state construction and the expectation calculation (if both are done using LPCA). In all our illustrations in this work however, we choose the same value for both.

## V. SPECTRALLY TRUNCATED PROPAGATOR FOR THE 2-TORUS

In Section III.A.2 we presented the quantum evolution of coherent state by a spectrally truncated propagator for the 2-torus, as a way of illustrating the performance of our approach to extracting geodesics when the amount of data allows very accurate approximation of the Laplace-Beltrami (LB) operator on the data manifold. Here we present details of how the spectrally truncated propagator is constructed and additional data on propagation for long times with this propagator.

We construct the spectrally truncated propagator by numerically solving for the spectrum and eigenvectors of the LB operator on the 2-torus. Consider the explicit parameterization of a 2-torus with radii $r, R, r < R$ embedded in $\mathbb{R}^3$ given in Eq. (23). In these coordinates, the LB operators can be be explicitly written as:

$$\Delta = -\frac{1}{r^2 \rho(\theta)} \frac{\partial}{\partial \theta} \rho(\theta) \frac{\partial}{\partial \theta} - \frac{1}{\rho(\theta)^2} \frac{\partial^2}{\partial \phi^2} \tag{25}$$

We wish to solve for the eigenfunctions and eigenvalues of this operator, *i.e.*, solve

$$-\frac{1}{r^2 \rho(\theta)} \frac{\partial}{\partial \theta} \rho(\theta) \frac{\partial}{\partial \theta} \varphi_n(\theta, \phi) - \frac{1}{\rho(\theta)^2} \frac{\partial^2}{\partial \phi^2} \varphi_n(\theta, \phi) = \lambda_n \varphi_n(\theta, \phi), \tag{26}$$

for $\lambda_n$ and $\varphi_n(\theta, \phi)$. Taking into account the symmetries of the torus, we can separate variables and write the eigenfunctions as $\varphi_n(\theta, \phi) = \psi_n(\theta) e^{ik_n \phi}$, for $k_n \in \mathbb{N}_0$, and consequently Eq. (26) can be simplified to an equation for $\psi_n$ as

$$\begin{aligned} &-\frac{1}{r^2 \rho(\theta)} \frac{d}{d\theta} \rho(\theta) \frac{d}{d\theta} \psi_n(\theta) + \frac{k_n^2}{\rho(\theta)^2} \psi_n(\theta) = \lambda_n \psi_n(\theta), \\ &\Rightarrow \left[ \frac{-\rho'(\theta)}{r^2 \rho(\theta)} \frac{d}{d\theta} - \frac{1}{r^2} \frac{d^2}{d\theta^2} + \frac{k_n^2}{\rho(\theta)^2} \right] \psi_n(\theta) = \lambda_n \psi_n(\theta) \end{aligned} \tag{27}$$

with boundary conditions $\psi_n^{(j)}(0) = \psi_n^{(j)}(2\pi)$ on $\psi_n$ and all of its derivatives.

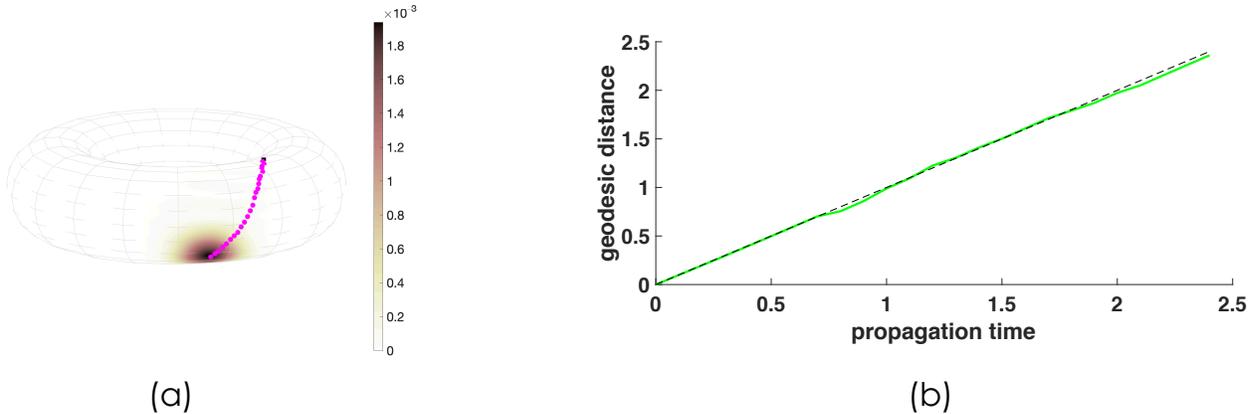

FIG. 11: An example of long time propagation with a spectrally truncated propagator on the 2-torus. **(a)** The path of a coherent state propagated for $t = 2.4$ that traverses a significant portion of the torus. Note that the coherent state remains localized and of roughly Gaussian profile even after such a long propagation. **(b)** The geodesic distance of $\langle \check{\mathbf{x}}_t \rangle^{\text{LPCA}}$, the mean position of the propagated state at time $t$ from the initial point. We see that the mean remains a geodesic distance $t$ away even for long times.

We can approximate solutions to Eq. (27) by iterating over integers $k_n$, and for each value of $k_n$, discretizing the $\theta$ domain into $N_\theta$ points, and approximating the quantity in the square brackets by a matrix, $M(k_n)$. Then numerically solving a matrix eigenproblem of the form $M(k_n)\Psi = \lambda_n \Psi$ yields discrete approximations of the LB eigenfunctions and eigenvalues, with error scaling as $\mathcal{O}(1/N_\theta^2)$. We choose a fine enough discretization of the $\theta$ domain and take enough iterations of $k_n$ so that the numerically approximated spectral data is converged. The discretization used for the $r = 0.8$ torus studied in Section III.A.2 was 1023 (uniformly spaced) points in the domain $[0, 2\pi)$. We apply this procedure to approximate the first 400 eigenvalues, $\lambda_n$, and associated partial eigenvectors $\psi_n(\theta)$, of the 2-torus LB operator. This data can be used to construct full eigenvectors, $\varphi_n(\theta, \phi)$, by discretizing the $\phi$ coordinate and evaluating the phase $e^{ik_n\phi}$ at each point, for each $n$. Note that for each $k_n$ that generates a solution to the eigensystem in Eq. (27), so does $-k_n$. Therefore, for each $\lambda_n, \psi_n(\theta)$ pair, we construct two $\varphi_n(\theta, \phi)$, with $\pm k_n$ phase factors. Using this eigendata we form approximate unitary propagator as $\tilde{V}_{\Delta t} = X \exp(i\sqrt{\Lambda}\Delta t)X^\dagger$, where $\Lambda$ is a matrix of the first 800 eigenvalues (with degeneracies) and the columns of $X$ are the corresponding eigenvectors, $\varphi(\theta, \phi)$. In order to have a numerically tractable propagator $\tilde{V}_{\Delta t}$, we downsample the 1023 points in $\theta$ to 171 and discretize $\phi$ to 300 (uniformly spaced) points, yielding an approximation of the propagator evaluated on $N' = 51300$ total points on the torus.

### V.A. Propagation details

Propagation with the spectrally truncated propagator is performed in almost the same manner as described in Section II.D. Initial states are formed using LPCA coordinates with $\delta_{\text{PCA}} = 2.0$, and to allow comparison to the data-driven propagation, we fix the coherent state uncertainty parameter at the value dictated by the $\epsilon, \alpha$ choice made in Section III.A.2 for the data-driven propagator with $N = 12000$ points; i.e., $h = e^{-1}$. An important difference to the procedure described in Section II.D is that for the spectrally truncated propagation, the initial momentum $p_0$ is chosen to be a uniformly random 2-vector in LPCA coordinates, instead of being a vector towards the nearest neighbor to any point. By the proof of Proposition 2 in Section II.D.3 this is valid because the LPCA coordinates approximate the tangent space when there is sufficient density of local sampling.

### V.B. Long-time propagation

To support the 2-torus numerics in Section III.A.2, in Fig. 11 we show an example of long-time propagation with this spectrally truncated propagator. The propagation is remarkably accurate – the coherent state remains localized and the propagation is along a unit-speed geodesic curve as the distance propagated plotted in Fig. 11(b) shows.

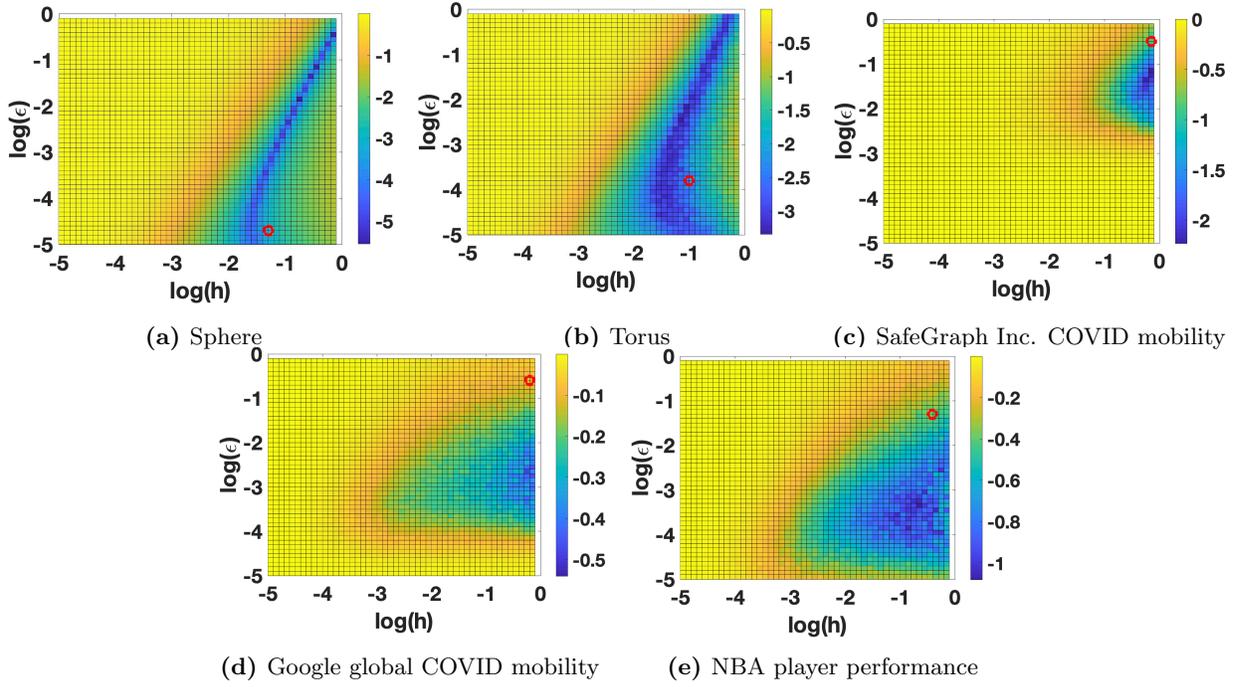

**FIG. 12:** Plots of log of the average deviation measure $\mathcal{D}$ (averaged over 28 initial points on each dataset) as a function of $\log(\epsilon)$ and $\log(h)$ for the five datasets studied in this work. The red circle in each figure indicates the parameter values used in the analysis presented in Section III.

## VI. DEVIATION MEASURE $\mathcal{D}$ FOR DATASETS

In Section II.D.6 we defined a useful quantity, $\mathcal{D} = |h^2(\psi_{\zeta_0}^h|\breve{L}_{\epsilon,\lambda}|\psi_{\zeta_0}^h) - 1|$, that can be used a heuristic for choosing good values of $\epsilon$ and $\alpha$ for a dataset. For ideal constructions of the propagator and coherent states, this quantity should be $\mathcal{O}(h)$. In this section we plot this deviation measure as a function of $\epsilon$ and $\alpha$ for all the datasets analyzed in this work. We average this quantity over 28 choices of initial point, $\mathbf{x}_0$, on each dataset to avoid sampling-related bias around any particular point.

## VII. ALGORITHMS

In this section we provide pseudocode that implements the core components of the quantum manifold learning program described above. The algorithms presented are effectively those used throughout nearly all of our aforementioned applications. They have not been optimized for fast implementation on large datasets, though there are are several areas where this is possible and we leave this to future work. Furthermore, we have made choices in the algorithms that make them less general than they could potentially be: some examples are the choices of the kernels $K$ and $K_{\text{PCA}}$ (see the parameters list below for explanations). Slightly more inconspicuous is that when LPCA is used, we chose to use it for both, the initial state preparation and mean computation simultaneously and with the same parameters (see Algorithms PrepInitialState and GeodesicFlowFromQuantumMean). As noted in Section II.D.3, these two procedures are independent of one another so that it is possible to use LPCA for one and not the other and also to change the corresponding parameters for either. Nevertheless, the presented algorithms have sufficed to effectively analyze the variety of datasets we have presented above, ranging from model manifolds to real-world (and thus, imperfect) statistics.

We also note that in the applications presented in Sections III.D and III.E, where the real-world data is sparse, we have used the *maximum* measurement of the distribution $|\check{V}_t|\psi_\zeta^h\rangle|^2$ for the propagated quantum state, rather than the mean measurement given in Algorithm GeodesicFlowFromQuantumMean. Since this is a trivial change, we have left out the corresponding modifications in preference for simplicity.

The primary inputs to our algorithms are the parameters:

1. $X_N = \{v_1, \ldots, v_N\} \subset \mathbb{R}^n$ is the dataset consisting of $N$ vectors, each of dimension $n$.

2. $K \in C^\infty(\mathbb{R}_+)$ is a non-negative, monotonically decreasing function with exponentially decaying tail (see [1] for precise conditions) used to define $\check{T}$, which in turn gives the graph Laplacian. In our applications, we have used the exponential kernel $K(s) = e^{-s}$, but other local kernels are also applicable (*e.g.,* bump functions).

3. $\epsilon > 0$ is the scale parameters used in constructing the graph Laplacian.

4. $\alpha > 0$ is the scaling parameters that determines the relationship between $h$ and $\epsilon$.

5. $0 \leq \lambda \leq 1$ determines the graph Laplacian normalization. For all the illustrations in this paper, $\lambda = 1$.

6. $\Delta t > 0$ is the time step used in constructing the unitary quantum propagator.

7. $n_{\text{prop}} \in \mathbb{N}$ is the number of time steps to propagate for. Geodesic distances are determined in steps of $\Delta t$ up to a distance $n_{\text{prop}} \Delta t$.

8. $n_{\text{coll}} \in \mathbb{N}$ is the number of distinct momenta to propagate from each data point on the manifold.

9. `PCA` is a boolean that flags whether LPCA is used in the algorithm, or states and measurements are constructed through extrinsic coordinates in $X_N$. The following parameters determining LPCA are then also needed:

    (a) $\delta_{\text{PCA}} > 0$ is the Euclidean distance bound that determines the points used to construct the LPCA space around each data point.

    (b) $0 < \gamma < 1$ is the scale factor that determines the LPCA projection (see discussion in Section II.D.2).

    (c) $\sigma_{\text{thresh}} > 0$ is a threshold determining the gap size to check for among the singular values of the covariance matrix corresponding to LPCA. This parameter is used in LPCA to determine the dimension $\tilde{\nu}$ of the LPCA space.

    (d) $K_{\text{PCA}} \in C^2([0,1])$ is a positive, monotonically decreasing function with support on $[0,1]$ as per [10]. **We have used $K_{\text{PCA}}(s) = \chi_{[0,1]}(u)$ in all of our applications**, but we simply include this parameter to indicate that other choices commonly used in the statistical literature are also permissible (*e.g.,* the Epanechnikov kernel $K_{\text{PCA}}(s) = (1-s^2)\chi_{[0,1]}(u)$).

**Algorithm** COMPUTEGEODESICDISTANCEMATRIX: Compute geodesic distance matrix via quantum dynamics.

**Input:** $X_N, \epsilon, \alpha, \lambda, \Delta t, n_{\text{prop}}, n_{\text{coll}}, \text{PCA}, \delta_{\text{PCA}}, \gamma, \sigma_{\text{thresh}}$
**Output:** Geodesic distance matrix $D$
1: $\breve{V}_{\Delta t} \leftarrow$ COMPUTEPROPAGATOR$(V, s \mapsto e^{-s}, \epsilon, \lambda, \Delta t)$ ▷ Compute discrete approximation of quantum propagator from data
2: $h \leftarrow \epsilon^{\frac{1}{(2+\alpha)}}$
3: $D \leftarrow \texttt{ZeroMatrix}(N, N)$ ▷ Initialize $N \times N$ geodesic distance matrix with zeros
4: **for** $v_j \in X_N$ **do**
5:     **for** $m = 1 : n_{\text{coll}}$ **do**
6:         $|\tilde{\psi}^h_{\Delta t}) \leftarrow$ PREPINITIALSTATE$(v_j, m, \text{PCA}, \delta_{\text{PCA}}, \gamma, \sigma_{\text{thresh}}, h)$ ▷ Initial state to be propagated $n_{\text{prop}}$ times
7:         **for** $n = 1 : n_{\text{prop}}$ **do**
8:             $|\tilde{\psi}^h_{\Delta t}) \leftarrow \breve{V}_{\Delta t} |\tilde{\psi}^h_{\Delta t})$ ▷ Efficiently apply $\breve{V}^n_{\Delta t}$ to initial state
9:             $\ell^d \leftarrow$ GEODESICFLOWFROMQUANTUMMEAN$(|\tilde{\psi}^h_{\Delta t}), X_N, \text{PCA}, \delta_{\text{PCA}}, \gamma, \sigma_{\text{thresh}})$
10:            **if** $[D]_{j,\ell^d} == 0$ or $[D]_{j,\ell^d} > n\Delta t$ **then**
11:                 $[D]_{j,\ell^d}, [D]_{\ell^d,j} \leftarrow n\Delta t$ ▷ Set geodesic distance from $j$ to $\ell^d$ to be $n\Delta t$ unless it is already closer
12: **return** $D$

---

**Algorithm** COMPUTEPROPAGATOR: Construct the propagator $\breve{V}_t$ as per Section II.D.1 for a single time-step.

**Input:** $X_N, s \mapsto K(s), \epsilon, \lambda, \Delta t$
**Output:** Propagator $\breve{V}_{\Delta t}$
1: **for** $(v_j, v_k) \in X_N \times X_N$ **do** ▷ Compute kernel matrix for Markov operator
2:     $\left[\breve{T}_\epsilon\right]_{j,k} \leftarrow K\left(\|v_j - v_k\|^2 / 2\epsilon\right)$ ▷ In our applications $K(s) = e^{-s^2}$
3: $\breve{D}_\epsilon \leftarrow$ diagonal matrix with diagonal elements $\sum_{k=1}^{N} [\breve{T}_\epsilon]_{j,k}$ for $1 \leq j \leq N$
4: $\breve{D}_{\epsilon,\lambda} \leftarrow$ diagonal matrix with diagonal elements $\sum_{k=1}^{N} [\breve{D}_\epsilon^{-\lambda} \breve{T}_\epsilon \breve{D}_\epsilon^{\lambda}]_{j,k}$ for $1 \leq j \leq N$
5: $\breve{L}_{\epsilon,\lambda} \leftarrow \frac{4}{\epsilon}(\breve{I}_N - \breve{D}_{\epsilon,\lambda}^{-1} \breve{D}_\epsilon^{-\lambda} \breve{T}_\epsilon \breve{D}_\epsilon^{-\lambda})$ ▷ Compute normalized graph Laplacian
6: **return** $\breve{V}_{\Delta t} \leftarrow \exp\left(-i\Delta t \sqrt{\breve{L}_{\epsilon,\lambda}}\right)$ ▷ Compute quantum propagator

---

**Algorithm** PREPINITIALSTATE: Initial state preparation, implementing Section II.D.2

**Input:** $v_j \in X_N, m \in \{1, \ldots, n_{\text{coll}}\}, \text{PCA} = \text{bool}(0,1), \delta_{\text{PCA}}, \gamma, \sigma_{\text{thresh}}, h,$
**Output:** Initial (unnormalized) state $|\tilde{\psi}^h_{\zeta_j})$ localized about $v_j$
1: **if** PCA **then**
2:     $\vartheta^j, \mathcal{N}_j \leftarrow$ LOCALPCA$(v_j, \chi_{[0,1]}, \delta_{\text{PCA}}, \gamma, \sigma_{\text{thresh}})$
3:     $k \leftarrow$ choose $\ell \in \mathcal{N}_j$ s.t. $\#\{\ell' \in \mathcal{N}_j \mid \|\vartheta^j_{\ell'} - \vartheta^j_j\| < \|\vartheta^j_\ell - \vartheta^j_j\|\} = m - 1$ ▷ $\vartheta^j_\ell$ is the $\ell$'th column of the matrix $\vartheta^j$
4:     $p_0 \leftarrow \vartheta^j_k - \vartheta^j_j$
5:     **for** $\ell = 1 : N$ **do** ▷ Compute elements of $|\tilde{\psi}^h_{\zeta_j})$
6:         $[\tilde{\psi}^h_{\zeta_j}]_\ell \leftarrow \begin{cases} e^{\frac{i}{h}(\vartheta^j_\ell - \vartheta^j_j)^\mathsf{T} \cdot p_0 / \|p_0\|} e^{-\frac{\|\vartheta^j_\ell - \vartheta^j_j\|^2}{2h}} & \text{if } \ell \in \mathcal{N}_j \\ 0 & \text{otherwise} \end{cases}$
7: **else**
8:     $p_0 \leftarrow v_k - v_j$ for point $v_k$ that is $m$'th closest to point $v_j$ in original $n$-dimensional coordinates
9:     **for** $\ell = 1 : N$ **do**
10:        $[\tilde{\psi}^h_{\zeta_j}]_\ell \leftarrow e^{\frac{i}{h}(v_\ell - v_j)^\mathsf{T} \cdot p_0 / \|p_0\|} e^{-\frac{\|v_\ell - v_j\|^2}{2h}}$
11: **return** $|\tilde{\psi}^h_{\zeta_j}) \equiv [\tilde{\psi}^h_{\zeta_j}]$

**Algorithm** GEODESICFLOWFROMQUANTUMMEAN: Measurement of mean of propagated state, implementing Section II.D.3

**Input:** $|\tilde{\psi}^h_{\Delta t}\rangle, X_N, \mathtt{PCA} = \mathrm{bool}(0,1), \delta_{\mathrm{PCA}}, \gamma, \sigma_{\mathrm{thresh}},$
**Output:** Index of sample point in $X_N$ closest to the mean of the probability distribution given by $|\tilde{\psi}^h_{\Delta t}\rangle := \breve{V}_{\Delta t}|\tilde{\psi}^h\rangle)$
1: $|\psi^h_{\Delta t}\rangle \leftarrow |\tilde{\psi}^h_{\Delta t}\rangle/|||\tilde{\psi}^h_{\Delta t}\rangle||$
2: $q \leftarrow ||\psi^h_{\Delta t}\rangle|^2$   ▷ $q$ is the probability distribution over position space for the state $|\tilde{\psi}^h_{\Delta t}\rangle$
3: **if** PCA **then**
4:   $\ell^* \leftarrow \arg\max_\ell q_\ell$
5:   $\vartheta^{\ell^*}, \mathcal{N}_{\ell^*} \leftarrow \text{LOCALPCA}(v_{\ell^*}, \chi_{[0,1]}, \delta_{\mathrm{PCA}}, \gamma, \sigma_{\mathrm{thresh}})$
6:   $\bar{v} \leftarrow \sum_{\ell \in \mathcal{N}_{\ell^*}} \vartheta^{\ell^*}_\ell q_\ell$   ▷ $\vartheta^{\ell^*}_\ell$ denotes the $\ell$'th column of the matrix $\vartheta^{\ell^*}$
7:   $\ell^d \leftarrow \arg\min_{\ell \in \mathcal{N}_{\ell^*}} ||\vartheta^{\ell^*}_\ell - \bar{v}||$
8: **else**
9:   $\bar{v} \leftarrow \sum_{\ell=1}^N v_\ell q_\ell$
10:   $\ell^d \leftarrow \arg\min_{1 \leq \ell \leq N} ||\bar{v} - v_\ell||$
11: **return** $\ell^d$

---

**Algorithm** LOCALPCA: Local PCA as per the discussion in Section II.D.2 & proof of Lemma 2 (based on [10])

**Input:** $v_j \in X_N, K_{\mathrm{PCA}}, \delta_{\mathrm{PCA}}, \gamma, \sigma_{\mathrm{thresh}}$
**Output:**
  $\mathcal{N}_j$   subset of indices of samples in a $\sqrt{\delta_{\mathrm{PCA}}}$ nbd. of $v_j$
  $\mathcal{S}^{\mathrm{PCA}}_j$   matrix with columns consisting of Local PCA coordinates for each sample with index in $\mathcal{N}_j$
1: $\varepsilon_{\mathrm{PCA}} \leftarrow \gamma \delta_{\mathrm{PCA}}$
2: $\breve{\Xi}_j \leftarrow \sum_{k=1}^N K_{\mathrm{PCA}}\left(||v_k - v_j||/\sqrt{\varepsilon_{\mathrm{PCA}}}\right)(v_k - v_j)(v_k - v_j)^\mathsf{T}$   ▷ In our applications, $K_{\mathrm{PCA}}(s) = \chi_{[0,1]}(s)$
3: $[\breve{U}_j, \breve{\Sigma}_j, \breve{V}^\mathsf{T}_j] \leftarrow \mathrm{SVD}(\Xi_j)$   ▷ Singular value decomposition: $\breve{\Xi}_j = \breve{U}_j \breve{\Sigma}_j \breve{V}^\mathsf{T}_j$
4: $[\sigma_1 \cdots \sigma_{N_j}] \leftarrow$ singular values on the diagonal of $\breve{\Sigma}_j$   ▷ $N_j := \#\{v_k \in X_N \mid K_{\mathrm{PCA}}(||v_j - v_k||/\sqrt{\varepsilon_{\mathrm{PCA}}}) \leq 1\}$
5: $\tilde{\nu} \leftarrow \min \ell$ s.t. $(\forall) 1 \leq r \leq \ell, \ell + 1 \leq s \leq N, \sigma_r/\sigma_s > \sigma_{\mathrm{thresh}}$   ▷ dimension estimate, as in the proof of Lemma 2
6: $\breve{O}_j \leftarrow \mathbb{R}^{\tilde{\nu} \times D}$ matrix with first $\tilde{\nu}$ rows of $U^\mathsf{T}_j$
7: $\mathcal{N}_j \leftarrow \{k \in [1, N] \mid ||v_j - v_k|| \leq \sqrt{\delta_{\mathrm{PCA}}}\}$   ▷ Indices of samples from $X_N$ in a $\sqrt{\delta_{\mathrm{PCA}}}$ nbd. about $v_j$
8: $\mathcal{S}_j \leftarrow [v_k]_{k \in \mathcal{N}_j}$   ▷ Matrix with columns given by samples indexed by $\mathcal{N}_j$
9: $\mathcal{S}^{\mathrm{PCA}}_j \leftarrow \breve{O} \cdot \mathcal{S}_j$
10: **return** $\mathcal{S}^{\mathrm{PCA}}_j, \mathcal{N}_j$

## VIII. RELATED WORK

Manifold learning has a long history with much of the early work dedicated to approximating geodesics or the distance map via discretization of the Hamilton-Jacobi or eikonal equations in one form or another [33, 34]. Even Dijkstra's algorithm and its various cousins have their justifications rooted in these non-linear PDEs on configuration space [2]. These methods rely either on intrinsic knowledge of the manifold such as a parameterization or some form of regularized meshing such as triangulation; in the case of Dijkstra's algorithm, which remains closer to the *inverse problem* roots of manifold learning, convergence is not always guaranteed [2]. More recently, the field has received a resurgence of interest and many new techniques have emerged since discrete graph structures were shown to approximate operators of the form $\Delta + O(\partial^1)$ on manifolds, with high probability [6, 35]. These works to date, to the best of the authors' knowledge, owe in significant part to the highly local properties of differential operators and the corresponding diffusion processes they generate [6, 35–37]. In this work, we have gone *against the grain* in this sense: we have relied on the highly local nature of the diffusion process to approximate its generator, which *semi-classically* has the form $\Delta + O(h\partial^1)$ and proceeded to drive localized initial conditions with *non-local* dynamics. In doing so, we have developed an entirely linear framework for approximating the highly non-linear geodesic equations, eschewed the need for short-time asymptotics and any gradient-type computations (both of which have stability issues) as needed, for example in [38] and traded the search for the correct set of eigenfunctions – a difficult problem in the context of manifold learning applications [39] – with the optimization of a continuous parameter, $h$ through the coherent state, for which we provide a region to narrow and guide the choice of optimal parameters. Additionally, while most manifold learning methods are sensitive to the sampling density, p, as discussed in Section II.C, our approach has reduced sensitivity because p contributes at low order to the principal symbol of the data-driven Hamiltonian.

The methods we employ combine diffusion dynamics on manifolds [6, 40], which provide a necessary step to essentially *bootstrap* from discrete graph structures to differential operators $\hat{\mathcal{L}}_{\epsilon,\lambda}$ that encode geometric data, with

semiclassical analysis [3] that informs the sense in which these operators encode configuration-space dynamics, the appropriate dynamical operators $\hat{V}_t$ and initial conditions $|\psi_{\zeta_0}^h\rangle$ with which to access this and the length scales $h$ at which this correspondence is valid. Related to the latter aspect is the large body of work from microlocal analysis [12, 41–43], which uses the concept of the wavefront set to identify propagation of singularties along the classical dynamics encoded in the driving operator. In essence, the spectral function $\hat{V}_t$ of $\hat{\mathcal{L}}_{\epsilon,\lambda}$ is approximating the Fourier Integral Operator (FIO) $\hat{U}_t := e^{it\sqrt{\Delta_g}}$ and the application of the coherent state $|\psi_{\zeta_0}^h\rangle$ is in fact resolving the wavefront set of the Schwartz kernel of $\hat{U}_t$ to resolution governed by $h$ and for the point $\zeta_0$ in phase space $T^*\mathcal{M}$. It is conceivable that one could devise other FIOs with the appropriate canonical relation to resolve geodesics, or other function classes to dissect the phase space in search for the wavefront set, or canonical relation of an FIO. In general, both are hard to deploy, especially in the presence of anisotropy; such a program has been carried out in some part for the case of data on Cartesian grids in [44–46], hence this is in a rather different setting to ours. To the best of the authors' knowledge, ours is the first work to employ such methods in the sense of the manifold learning inverse problem.

One aspect we merely touch upon here is that of embedding the manifold. The problem of isometrically embedding a manifold in low dimensions goes back at least to Nash [47], who showed with great technique how to do this constructively. In doing so, the proof methods also show the inherent difficulty of this task: in that setting, all the details of a manifold are assumed – indeed, Nash's theorems solve a *direct* problem – and still, this requires highly intractable (to implement) recursive methods. Approximation methods from the *inverse* problem perspective range from minimizing certain *stress* functionals [48, 49] to aligning approximate tangent spaces [50–52] to using the spectral data of an approximated diffusion operator [35, 40, 53] or more recently, employing geometric controllability of the wave equation [54]. In essentially all cases, a key role is played by the distance map on the manifold: this is approximated to some order by each method, typically chaining local Euclidean distances from the given, possibly (very) high-dimensional embedding, to global approximations of geodesics. The quality of these approximations, going from local to global essentially controls the order to which a given embedding will be isometric and close to the original dimension of the manifold. Thus, the task of approximating geodesics well is key to solving the often-sought embedding problem.

Recently, new deterministic methods have been proposed to resolve the embedding problem with significant improvement in dimension reduction, in a more tractable fashion than Nash's algorithm [54, 55]. These stray somewhat from the present context, but there is nevertheless some intersection with [54] in that geometric control of the solution to a spectral approximation to the wave equation is used to embed the manifold from data and in doing so, also to recover the distance map. In that case, it is the unique extension property of smooth solutions that plays a key role, rather than the specific nature of propagation and phase-space localization properties that we employ. Other than these cases, most of the effort in manifold embedding coming from dynamics on function spaces has been rooted in diffusion processes: eigenfunction embedding methods such as [40, 53] rely heavily on the asymptotically small-time limit of the heat equation to recover geodesics, as per Varadhan's formula [38]. To compare, on the one hand, our proposed method is only little more than the very original 0-1 graph Laplacian operator as proposed in [35]. On the other hand, when this operator marks geodesic neighbours, globally rather than only locally, we can interpret it as an approximation to the *spherical means* operator [27, 56]. In that sense, when using single neighbour (the 0-1 case), this is a sparse representation of the canonical relation in $\cos(t\sqrt{\Delta_g})$ that encodes geodesics and simultaneously it provides a nearest neighbours approach to embedding.

A common problem in the real-world application of spectral methods is the search for eigenfunctions: a typical use-case is to embed high dimensional data in low dimensions in order to separate the data into meaningful clusters. The heuristic is the existence of a spectral gap in the diffusion matrix $\check{T}_{\epsilon,\lambda} := \check{D}_{\epsilon,\lambda}^{-1}\check{\Sigma}_\epsilon^{-\lambda}\check{T}_\epsilon\check{\Sigma}_\epsilon^{-\lambda}$ among the top eigenvalues (which are the low-lying eigenvalues of graph Laplacian), both to identify the number of clusters and the dimension of embedding in which to identify them with lower dimensional techniques [57]. However, such a gap may be dampened or effectively non-existent due to various features of data manifold, such as large variances in cluster sizes or a dominatingly large background cluster, which absorbs most of the diffusion process and thus localizes many of the leading eigenvectors to itself [26, 39]. This is of course particularly problematic for outlier detection [39, 58]. In contrast, our approach works with the high-energy eigenfunctions of the graph Laplacian by the nature of $\hat{V}_t$ that drives the dynamics: indeed, the propagation of localized states, or in the microlocal setting, singularities, by the half-wave propagator it approximates is governed by the high-energy eigenfunctions of $\Delta_g$ [27]. These eigenfunctions are, by their very nature, highly *delocalized* [27] and tend to concentrate on geodesics – which are global – rather than concentrating on dense regions, either due to sampling concentration or geometry, as would be the case for slow mixing of diffusion processes. Moreover, the use of coherent states $|\psi_{\zeta_0}^h\rangle$ has another effect from the clustering perspective: the parameter $h$ sets that on propagation with this state, the operator is implicitly projected onto eigenvalues to order $1/h$. We have suggested a simple test to give a regime in which the choice of $(\epsilon, h)$ yields good embeddings for all of the examples we have considered here. Thus, we are employing a large range of the spectrum at once and $h$ provides an *implicit* parameter for spectral truncation, with the ability to *a priori* ascertain optimal regions. This

may explain the applicability of our approach to the various tests provided here, both in large and small data settings: in particular, we note that in Figure 5 (d-f) we witness the existence of an outlier statistic with accompanying cluster even with a very rudimentary embedding with extremely low dimension (three), which is not seen to be present with the diffusion based approaches until at least eight eigenvectors are used and clustering is done in eight dimensions.

### ACKNOWLEDGMENTS



\end{document}